\documentclass[12pt,a4paper]{article}
\usepackage{a4wide}
\usepackage{latexsym}
\usepackage{epsf}
\usepackage{amssymb}
\makeatletter
\@addtoreset{equation}{section}
\makeatother

\begin{document}

\begin{flushright}
\small
IFT-UAM/CSIC-98-3\\
{\bf hep-th/9806120}\\
June  $16$th, $1998$
\normalsize
\end{flushright}

\begin{center}


\vspace{.7cm}

{\Large {\bf An $Sl(2,\mathbb{Z})$ Multiplet of Nine-Dimensional Type~II 
Supergravity Theories}}

\vspace{.7cm}


{\bf\large Patrick Meessen}${}^{\diamondsuit\clubsuit}$
\footnote{E-mail: {\tt meessen@martin.ft.uam.es}}
{\bf\large and  Tom\'as Ort\'{\i}n}${}^{\diamondsuit,\spadesuit}$
\footnote{E-mail: {\tt tomas@leonidas.imaff.csic.es}}
\vskip 0.4cm

${}^{\diamondsuit}$\ {\it Instituto de F\'{\i}sica Te\'orica, C-XVI,
Universidad Aut\'onoma de Madrid \\
E-28049-Madrid, Spain}

\vskip 0.2cm

${}^{\clubsuit}$\ {\it Departamento de F\'{\i}sica Te\'orica, C-XI,
Universidad Aut\'onoma de Madrid \\
E-28049-Madrid, Spain}

\vskip 0.2cm

${}^{\spadesuit}$\ {\it I.M.A.F.F., C.S.I.C., Calle de Serrano 113\\ 
E-28006-Madrid, Spain}

\vspace{.7cm}


{\bf Abstract}

\end{center}

\begin{quotation}

\small

We show that only by performing generalized dimensional reductions all
possible brane configurations are taken into account and one gets the
complete lower-dimensional theory. We apply this idea to the reduction
of type~IIB supergravity in an $SL(2,\mathbb{R})$-covariant way and
establish T~duality for the type~II superstring effective action in
the context of generalized dimensional reduction giving the
corresponding generalized Buscher's T~duality rules.

The full (generalized) dimensional reduction involves all the S~duals
of D-7-branes: Q-7-branes and a sort of composite 7-branes. The three
species constitute an $SL(2,\mathbb{Z})$ triplet. Their presence
induces the appearance of the triplet of masses of the 9-dimensional
theory.

The T~duals, including a ``KK-8A-brane'', which must have a compact
transverse dimension have to be considered in the type~IIA side.
Compactification of 11-dimensional KK-9M-branes (a.k.a.~M-9-branes) on
the compact transverse dimension give D-8-branes while
compactification on a worldvolume dimension gives KK-8A-branes. The
presence of these KK-monopole-type objects breaks translation
invariance and two of them given rise to an
$SL(2,\mathbb{R})$-covariant {\it massive 11-dimensional supergravity}
whose reduction gives the massive 9-dimensional type~II theories.

\end{quotation}

\newpage

\pagestyle{plain}


\section{Introduction}

During the last few years, the study of the low-energy string
effective action has shown itself to be most profitable.  It has
helped us to establish duality between many pairs of string theories
(duality of the effective actions being a necessary condition) and has
provided us with semiclassical solutions describing the long-range
fields of perturbative and non-perturbative states of string theories.
In particular, the now popular relation between M~theory and type~IIA
string theory was first suggested by the relation between
11-dimensional and type~IIA supergravity \cite{kn:Wi}. 11-dimensional
supergravity is our best source of knowledge about M~theory. On the
other hand, in the stringy microscopic explanation of the origin of
black-hole entropy (see e.g.~Ref.~\cite{kn:M}) it is fundamental to
have a semiclassical solution describing the black hole to calculate
the area of the horizon.

Superstring effective actions are nothing but the actions of
supergravity theories. Thus, there is much to be learned from the old
techniques used to compactify them if we translate them to string
language. For instance, compactifying the action of $N=1,d=10$
supergravity on a circle, it is easy to recover Buscher's T~duality
rules \cite{kn:Bu} as a global discrete symmetry of the 9-dimensional
theory that interchanges two vector fields associated to momentum
modes' charges and winding modes' charges \cite{kn:MS,kn:BKO2}. In the
type~II context, only through the use of the effective action it was
possible to derive the generalization of Buscher's T~duality rules
\cite{kn:BHO}.

In Ref.~\cite{kn:SS} Scherk and Schwarz proposed the method of
``generalized dimensional reduction'': In standard dimensional reduction
it is required that all fields are independent of the coordinates of the
compact dimensions. However, when there are global symmetries (always
present in the internal dimensions), in order to guarantee that the
lower-dimensional theory is independent of them, it is enough to require
that the fields depend on them in a certain way. The terms depending
on the internal directions generate mass terms on lower dimensions.

The authors of Ref.~\cite{kn:BRGPT} first used the idea of
Scherk-Schwarz generalized dimensional reduction \cite{kn:SS} with
global symmetries of no (known) geometrical origin. They applied it to
the symmetry under constant shifts of the type~IIB RR scalar and
obtained a massive 9-dimensional type~II theory, precisely the one one
obtains through standard dimensional reduction from Romans' massive
type~IIA supergravity \cite{kn:Ro2}. Another important result of
Ref.~\cite{kn:BRGPT} is that they identified the presence of
D-7-branes in the background as the origin of the 9-dimensional mass.
By T~duality arguments the mass parameter of Romans' theory was
identified with the presence of D-8-branes\footnote{This
  identification had already been made in Ref.~\cite{kn:PW}.}, (more
precisely as a ``$(-1)$-form RR potential'') and Buscher-type
T~duality rules could then be derived.

In this theory $SL(2,\mathbb{R})$ (the classical S~duality
group\footnote{$SL(2,\mathbb{R})$ is broken to $SL(2,\mathbb{Z})$ by
  quantum effects such as charge quantization. Most of our
  considerations throughout this paper will be purely classical and
  thus we will mostly talk about $SL(2,\mathbb{R})$. However, at
  specific places the restriction to the discrete $SL(2,\mathbb{Z})$
  will be important and we will deal with it in full detail.})  was
broken. This was to be expected since the S~duals of D-7-branes were
not present. In other words, the global symmetry chosen was only a
subgroup of the full $SL(2,\mathbb{R})$ global symmetry available.

In Ref.~\cite{kn:LLP2} a systematic way of performing the generalized
dimensional reduction associated to the type~IIB S~duality group was
proposed. Again, the full S~duality group was not used and the
dimensional reduction was not finished due to the complications
introduced by the self-dual 5-form. The resulting  9-dimensional massive
type~II theory was then incomplete and not $SL(2,\mathbb{R})$-invariant.

In fact, the breaking of S~duality seems unavoidable. However,
S~duality is supposed to be an exact symmetry of type~IIB superstring
theory. The solution to this puzzle is that one has to take into
account the transformation of the mass parameters, which should be
considered ``$(-1)$-form'' potentials, and include all the S~duality
related mass parameters. Then, the method of Ref.~\cite{kn:LLP2}
suggests that, by making use of the full $SL(2,\mathbb{R})$ group in
the generalized dimensional reduction a 9-dimensional massive
$SL(2,\mathbb{R})$-{\it covariant} type~II theory should be
obtainable. If the mass parameters are considered fixed constants of
the theory, then what one obtains can be considered as an
$SL(2,\mathbb{R})$ multiplet of 9-dimensional massive type~II
theories.

The first goal of this paper is to find this theory and interpret it
in terms of 10-dimensional 7-branes. We also want to get a better
understanding of the method of generalized dimensional reduction and 
we will propose two alternative methods giving the same results.

Since (as we will argue after we present a toy model of generalized
dimensional reduction in Section~\ref{sec-example}) this 9-dimensional
theory should be considered {\it the} 9-dimensional type~II
theory\footnote{A slightly more general massive 9-dimensional type~II
  theory may be constructed, though. This will be explained in the
  Conclusion Section.}, one expects T~duality to hold in this context.

Our second goal in this paper will be to establish T~duality in the
context of generalized dimensional reduction of the type~IIB theory.
To achieve this goal one faces an important problem that can be expressed
in two different ways:

\begin{enumerate}

\item It is clear, from the above discussion that one has to identify 
the S~duals of the D-7-brane and then one has to identify their T~duals.
The T~dual theory will be type~IIA theory in a background containing
these objects.

\item One has to find the generalization of type~IIA supergravity which
gives rise to the 9-dimensional masses we get from the type~IIB side.
However it does not seem possible to further generalize Romans' massive
type~IIA supergravity.

\end{enumerate}

A clue to the resolution of this problem is the fact that the type~IIB
$SL(2,\mathbb{R})$ symmetry is identical to the $SL(2,\mathbb{R})$
symmetry of 11-dimensional supergravity compactified on $T^{2}$ which
acts on the internal manifold
\cite{kn:BHO,kn:BJO,kn:Be,kn:As}. S~duality then interchanges the
11-dimensional theory that gives rise to Romans' theory with the
10-dimensional theory associated to the T~duals of the S~duals of
D-7-branes.

The 11-dimensional origin of Romans' theory is somewhat mysterious
because 11-dimensional supergravity cannot be deformed to include a
mass or a cosmological constant according to the no-go theorem of
Refs.~\cite{kn:BDHS}. In Refs.~\cite{kn:L,kn:O,kn:BLO}, though, a
modification of 11-dimensional supergravity that breaks 11-dimensional
covariance (one of the hypothesis of the no-go theorem) was proposed.
This theory gives Romans' type~IIA upon dimensional reduction.  The
original motivation for the construction of this theory was the
realization that the worldvolume effective actions of the extended
solitons of Romans' type~IIA can be derived from 11-dimensional gauged
$\sigma$-models\footnote{The effective action of the Kaluza-Klein (KK)
monopole is also a gauged $\sigma$-model \cite{kn:BJO2}. However, in
this case the gauging is not associated to any mass parameter.
Actually, to describe the KK in massive background the gauging of a
second isometry is necessary \cite{kn:BEL}.}. The gauging of an
isometry breaks the gauge invariance of the Wess-Zumino term. To
restore it, it is not enough to modify this term. One has to modify
the gauge transformations o the 11-dimensional fields. The massive
11-dimensional theory constructed in Ref.~\cite{kn:BLO} is precisely
the one which is invariant under these modified gauge transformations.

The reason for the explicit breaking of 11-dimensional covariance was
not sufficiently explained. It has been suggested recently in
Ref.~\cite{kn:BS} that it is due to the presence of some objects,
M-9-branes which we call KK-9M-branes. Now we are saying that, due to
S~duality, a 10-dimensional type~IIA theory with broken Poincar\'e
covariance is needed in order to make contact with the general massive
9-dimensional type~II theory that we are about to construct.
Furthermore, our previous arguments show that this breaking of
covariance is due to the presence of certain objects: The T~duals of
the S~duals of D-7-branes, which we call KK-8A-branes and which can be
obtained by dimensional reduction of KK-9M-branes.  This theory can be
obtained by the compactification of the massive 11-dimensional
supergravity of Ref.~\cite{kn:BLO} along a different coordinate.

These objects must be somewhat similar to type~IIA Kaluza-Klein (KK)
monopoles, which are the T~dual of the S~dual of D-5-branes (solitonic
NS-NS S-5-branes): A dimension transverse to their worldvolume has to
be compactified on a circle and there is an isometry associated to it.
The presence of this isometry is the reason for the breaking of
10-dimensional (and 11-dimensional) covariance.  Compactification of
the KK-9M-brane along a worldvolume direction gives the 10-dimensional
type~IIA KK-8A-brane while compactification along the isometry
direction gives the D-8-brane \cite{kn:BS}.  Compactification of the
KK-8A-brane along the isometry gives a 9-dimensional Q-7-brane, the
S~dual of a 9-dimensional D-7-brane. This is similar to what happens
with KK-monopoles: In eleven dimensions the KK-monopole can be called
KK-7M-brane.  Compactification along a worldvolume direction gives the
type~IIA KK-monopole that we can call KK-6A-brane and compactification
along the isometry direction gives the D-6-brane.  Further
compactification of the KK-6A-brane along the isometry gives a
9-dimensional NS-NS S-5-brane. These, and more relations, are depicted
in Figure~\ref{fig:kkduals} and will be explored in
Section~\ref{sec-KK789}.

The worldvolume theory of the KK-8A- and the KK-9M-brane must also be
given by a gauged $\sigma$-model\footnote{The corresponding action for
the KK-9M-brane as been written in Ref.~\cite{kn:BS}.}, where the
symmetry gauged is associated to the isometry that these objects must
have in the compact dimension, just as happens with the usual KK
monopoles \cite{kn:BJO2}.

Thus, we are lead to the following picture which solves our problem:
There is a massive 11-dimensional supergravity theory associated to
the presence of {\it two} KK-9M-branes and thus has 11-dimensional
covariance broken in two directions: the isometries of the
KK-9M-branes. This theory has $SL(2,\mathbb{R})$-covariance in those
two special isometric directions. One can eliminate one isometry by
compactifying along it obtaining a modification of Romans' type~IIA
theory with covariance broken in the other isometry direction. This is
type~IIA in the presence of a D-8-brane and a KK-8-brane. Reducing
further along the other isometry direction gives the desired
9-dimensional massive type~II theory, which is type~II theory in
presence of D-7-branes and and their S~dual Q-7-branes.

We will show that this is the right picture and we will comment on its
possible generalizations in the Conclusion.

At this point it is perhaps convenient to study a simple example to
illustrate some of our ideas.


\subsection{Generalized Dimensional Reduction of the 
Einstein-Dilaton Theory}
\label{sec-example}

We consider the following toy model which exhibits the general
features of generalized dimensional reduction associated to global
symmetries with no geometrical origin\footnote{In this section we use
hats for $d$-dimensional objects and no hats for $(d-1)$-dimensional
objects.}:

\begin{equation}
\hat{S} = \int d^{d}\hat{x} \sqrt{|\hat{g}|}\ \left[\hat{R} 
+{\textstyle\frac{1}{2}}\left(\partial\hat{\phi} \right)^{2} \right]\, .
\label{eq:toymodel}
\end{equation}

This action is invariant under constant shifts of the scalar
$\hat{\phi}$, the reason being that $\hat{\phi}$ only occurs through
its derivatives. The presence of this global symmetry allows us to
extend the general Kaluza-Klein Ansatz (i.e.~all fields, and in
particular $\hat{\phi}$, are independent of some coordinate, say $z$)
to a more general Ansatz in which $\hat{\phi}$ depends on $z$ in a
particular way:

\begin{equation}
\label{eq:genAnsatz}
\hat{\phi}(x,z) = \hat{\phi}^{\rm b}(x) +mz\, ,
\hspace{1cm}  
\hat{x}^{\hat{\mu}}=(x^{\mu},z)\, ,
\end{equation}

\noindent where the superscript ${}^{b}$ stands for {\it bare}, 
or $z$-independent.

This dependence on $z$ can be produced by a {\it local} shift of
$\hat{\phi}(x)$ with a parameter linear in $z$.  The invariance
of the action under constant shifts ensures that the action will not
depend on $z$. 

This is only a practical recipe to write a good Ansatz. To understand
better what one is doing, one has to recall that $z$ is a coordinate
on a circle $S^{1}$ subject to the identification $z\sim z+2\pi l$.
In standard Kaluza-Klein reduction one only considers single-valued
fields, so that the needed Fourier decomposition of the fields living
on ${\cal M}\otimes S^{1}$, reads

\begin{equation}
\hat{\phi}\left( \hat{x}\right) \;=\; 
\sum_{n\in\mathbb{Z}}e^{2\pi nz/l}\, \phi^{(n)}(x) \; .
\end{equation}

\noindent Dimensional reduction then means keeping the massless modes,
{\it i.e.}  $\phi^{(0)}$, only.  Some fields can be multivalued,
however.  If the scalar $\hat{\phi}$ is such that
$\hat{\phi}=\hat{\phi}+2\pi m$, the above Fourier expansion is
enhanced to

\begin{equation}
\hat{\phi}\left( \hat{x}\right) \;=\; \frac{mNz}{l} \,+\,
\sum_{n\in\mathbb{Z}}e^{2\pi nz/l}\, \phi^{(n)}(x) \; ,
\end{equation}

\noindent where $N\in\mathbb{Z}$ labels the different {\it topological
sectors}.  Now, the action for a field living on an $S^{1}$ is always
invariant under arbitrary shifts of the field, even if the field is to
be identified under discrete shifts. This then ensures that the lower
dimensional theory does not depend on $z$, the dimensional reduction,
if only $\frac{mNz}{l}+\phi^{(0)}$ is kept.  Each topological sector
is characterized by the charge

\begin{equation}
N\;=\; \lim_{x\rightarrow\infty}\, \frac{1}{2\pi lm}\,
\oint d\hat{\phi} \; ,
\end{equation}

\noindent which is nothing but the {\it winding number}.

A more physical interpretation of the technical description of the
generalized dimensional reduction recipe will be given later on.

Making use of the standard KK Ansatz for the Vielbein

\begin{equation}
\left( \hat{e}_{\hat{\mu}}{}^{\hat{a}} \right) = 
\left(
\begin{array}{cc}
e_{\mu}{}^{a} & kA_{(1)\ \mu} \\
&\\
0             & k        \\
\end{array}
\right)
\, , 
\hspace{1cm}
\left(\hat{e}_{\hat{a}}{}^{\hat{\mu}} \right) =
\left(
\begin{array}{cc}
e_{a}{}^{\mu} & -A_{(1)\ a}   \\
& \\
0             &  k^{-1}  \\
\end{array}
\right)\, , 
\label{eq:KKbasis}
\end{equation}

\noindent we readily obtain the  $(d-1)$-dimensional action

\begin{equation}
S=\int d^{d-1}x\sqrt{|g|}\ k\ \left[ R 
-{\textstyle\frac{1}{4}} k^{2}F_{(2)}^{2}
+{\textstyle\frac{1}{2}} \left({\cal D}\phi \right)^{2}
-{\textstyle\frac{1}{2}}m^{2} k^{-2} \right]\, ,
\label{eq:toymodelreduced1}
\end{equation}

\noindent where the field strengths are defined by

\begin{equation}
\left\{
\begin{array}{rcl}
F_{(2)\ \mu\nu} & = & 2\partial_{[\mu}A_{(1)\ \nu]}\, ,\\
& & \\
{\cal D}_{\mu} \phi & = & \partial_{\mu}\phi -mA_{(1)\ \mu}\, ,\\
\end{array}
\right.
\end{equation}

\noindent and 

\begin{equation}
\phi \equiv \hat{\phi}^{\rm b}\, .  
\end{equation}

\noindent A further rescaling of the metric 

\begin{equation}
g_{\mu\nu}  \rightarrow k^{-2/(d-3)}g_{\mu\nu}\, ,
\end{equation}

\noindent brings us to the final form of the action:

\begin{equation}
S=\int d^{d-1}x\sqrt{|g|}\ \left[ R 
+{\textstyle\frac{1}{2}}\left(\partial\varphi \right)^{2}
-{\textstyle\frac{1}{4}}e^{-a\varphi}F_{(2)}^{2}
+{\textstyle\frac{1}{2}} \left({\cal D}\phi \right)^{2}
-{\textstyle\frac{1}{2}}m^{2} e^{a\varphi} \right]\, ,
\label{eq:toymodelreduced2}
\end{equation}

\noindent where

\begin{equation}
k = e^{-\varphi/2a}\, ,
\hspace{1cm}
a=-\sqrt{\frac{2(d-2)}{(d-3)}}\, .
\end{equation}

This action and the field strengths are invariant under the following
massive gauge transformations:

\begin{equation}
\left\{
\begin{array}{rcl}
\delta \phi & = & m\chi\, ,\\
& & \\
\delta A_{(1)\ \mu} & = & \partial_{\mu} \chi\, .\\
\end{array}
\right.
\end{equation}

\noindent These transformations correspond in the $d$-dimensional
theory to the $z$-independent reparametrizations of $z$:

\begin{equation}
\delta z = -\chi (x)\, .  
\end{equation}

This is the theory resulting from the standard recipe for generalized
dimensional reduction \cite{kn:BRGPT}.

There is another way of getting the same result in this toy model:
We gauge the translation $\hat{\phi}\rightarrow\hat{\phi}+m$ and
impose that the gauge field is non-vanishing and constant in the
internal direction only (a Wilson line). Since the metric does not
transform, it is sufficient to demonstrate this on the kinetic term
for $\hat{\phi}$.

In order to gauge the translation invariance on $\hat{\phi}$ we introduce
the gauge field by minimal coupling

\begin{equation}
\partial_{\hat{\mu}}\hat{\phi}\,\rightarrow\,
 {\cal D}_{\hat{\mu}}\hat{\phi}\,=\, 
     \partial_{\hat{\mu}}\hat{\phi}\,+\,\hat{\cal E}_{\hat{\mu}} \, ,
\end{equation}

\noindent so that under a local transformation
$\hat{\phi}\rightarrow\hat{\phi} +\Lambda (\hat{x})$ the gauge field
transforms in an Abelian manner, i.e.

\begin{equation}
  \hat{\cal E}_{\hat{\mu}}^{\prime} \;=\;
      \hat{\cal E}_{\hat{\mu}}\,+\, \partial_{\hat{\mu}}\Lambda (\hat{x})\; .
\end{equation}

Making then the {\it standard} KK Ansatz and imposing that $\hat{\cal
  E}_{\hat{\mu}}$ is non-vanishing and constant, with value $m$, in
the compact direction only, one finds

\begin{equation}
\left\{
  \begin{array}{rcl}
 {\cal D}_{a} \phi & = & {e_{a}}^{\mu}\left( 
   \partial_{\mu}\phi \,-\, mA_{(1)\ \mu}\right) \,\equiv\,
       {e_{a}}^{\mu} {\cal D}_{\mu}\phi \; , \\
& & \\
 {\cal D}_{z} \phi & = & k^{-1}m \; ,\\
\end{array}
\right.
\end{equation}

\noindent leading to 

\begin{equation}
\int d^{d}x\sqrt{|\hat{g}|}\, 
{\textstyle\frac{1}{2}} \left( \partial\phi\right)^{2}
\,=\, \int d^{d-1}x\sqrt{|g|}\, k 
\left[ 
{\textstyle\frac{1}{2}} \left( {\cal D}\phi\right)^{2} \,-\,
{\textstyle\frac{1}{2}} k^{-2} m^{2}
\right] \, .
\end{equation} 

\noindent Comparing this result with Eq.~(\ref{eq:toymodelreduced1}),
one sees that, at least in this toy-model, generalized Scherk-Schwarz
reduction leads to the same result as the above algorithm.

We will also use this method in the context of type~IIB supergravity
and check that one gets the same results as well.

Observe that the field content looks the same as in the standard
dimensional reduction: There is a vector and two scalars (apart from
the metric). The symmetries and couplings are different, though. The
massive gauge symmetry allows us to eliminate one scalar (the
Stueckelberg field Ref.~\cite{kn:St}) and give mass to the vector
field. The number of degrees of freedom is exactly the same. So, what
is it we have done? To shed some light on the meaning of this
procedure we are going to perform the ``standard'' dimensional
reduction of the action (\ref{eq:toymodel}) but Poincar\'e-dualizing
first the scalar into a $(d-2)$-form potential\footnote{When indices
  are not explicitly shown we assume all indices to be antisymmetrized
  with weight one. This is slightly different from differential form
  notation.}  $\hat{A}_{(d-2)\ \hat{\mu}_{1}\cdots\hat{\mu}_{(d-1)}}$:

\begin{equation}
\partial\hat{\phi} ={}^{\star}\hat{F}_{(d-1)}\, .
\end{equation}

\noindent The dual action is

\begin{equation}
\tilde{\hat{S}} = \int d^{d}x \sqrt{|\hat{g}|}\ \left[\hat{R} 
+{\textstyle\frac{(-1)^{(d-2)}}{2\cdot (d-1)!}}
\hat{F}_{(d-1)}^{2} \right]\, .
\label{eq:dualtoymodel}
\end{equation}

{\it Standard} dimensional reduction with the same Vielbein Ansatz gives

\begin{equation}
\tilde{S}=\int d^{d-1}x\sqrt{|g|}\ k\ \left[ R 
-{\textstyle\frac{1}{4}} k^{2}F_{(2)}^{2}
+{\textstyle\frac{(-1)^{(d-2)}}{2\cdot (d-1)!}}
F^{2}_{(d-1)}
+{\textstyle\frac{(-1)^{(d-3)}}{2\cdot (d-2)!}}
k^{-2} F^{2}_{(d-2)} \right]\, ,
\label{eq:dualtoymodelreduced1}
\end{equation}

\noindent where

\begin{equation}
\left\{
\begin{array}{rcl}
F_{(d-1)} & = & (d-1)\partial A_{(d-2)} 
+(-1)^{(d-1)} A_{(1)}F_{(d-2)}\, ,\\
& & \\
F_{(d-2)} & = & (d-2)\partial A_{(d-3)}\, ,\\
\end{array}
\right.
\end{equation}

\noindent are the field strengths of the $(d-2)$- and $(d-3)$-form
potentials of the $(d-1)$-dimensional theory.

We can now dualize the potentials. A $(d-2)$-form potential in $(d-1)$
dimensions is dual to a constant that we call $m$. Adding the term

\begin{equation} 
-{\textstyle\frac{1}{(d-1)!}}
\int d^{d-1}x\ m\epsilon \left[F_{(d-1)} 
+(-1)^{d}(d-1)A_{(1)}F_{(d-2)} \right]\, ,
\end{equation}

\noindent to the action (\ref{eq:dualtoymodelreduced1}), and eliminating
$F_{(d-1)}$ using its equation of motion 

\begin{equation}
m =k{}^{\star}F_{(d-1)}\, ,
\end{equation}

\noindent in the action we get

\begin{equation}
\begin{array}{rcl}
\tilde{S} & = & \int d^{d-1}x\sqrt{|g|}\ 
\left\{ k\ 
\left[ R 
-{\textstyle\frac{1}{4}} k^{2}F_{(2)}^{2}
+{\textstyle\frac{(-1)^{(d-3)}}{2\cdot (d-2)!}}
k^{-2} F^{2}_{(d-2)} 
-{\textstyle\frac{1}{2}}m^{2}k^{-2}
\right] 
\right.
\\
& & \\
& & 
\left.
+{\textstyle\frac{1}{(d-2)!}}\frac{\epsilon}{\sqrt{|g|}}F_{(d-2)}
\left[
-mA_{(1)}
\right]
\right\}
\, . \\
\end{array}
\label{eq:dualtoymodelreduced2}
\end{equation}

\noindent Now we dualize into a scalar field the $(d-3)$-form potential:
We add to the above action the term

\begin{equation}
{\textstyle\frac{1}{(d-2)(d-2)!}}
\int d^{d-1}x\ \epsilon F_{(d-2)}\partial\phi\, ,
\end{equation}

\noindent and eliminate $F_{(d-2)}$ by substituting in the action 
its equation of motion

\begin{equation}
F_{(d-2)} = (-1)^{(d-2)} k\ {}^{\star}{\cal D}\phi\, ,
\end{equation}

\noindent obtaining, perhaps surprisingly,
Eq.~(\ref{eq:toymodelreduced1}).

\begin{figure}[!ht] 
\begin{center} 
\leavevmode 
\epsfxsize= 6cm
\epsffile{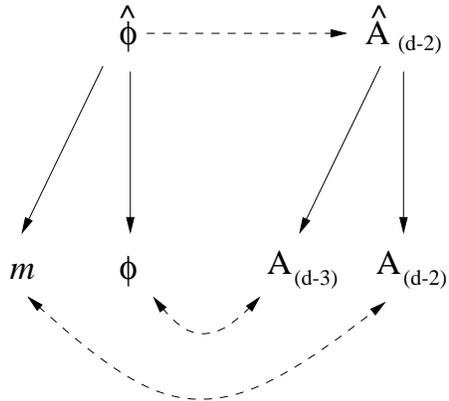} 
\caption{\footnotesize This diagram represents
two different ways of obtaining the same result: Generalized dimensional
reduction and ``dual'' standard dimensional reduction.\normalsize}
\label{fig:toygeneralized} 
\end{center} 
\end{figure}

What we have done is represented in figure~\ref{fig:toygeneralized}.

The translation to brane language is obvious: Generalized dimensional
reduction, which is essentially applied to scalars, is a way of
keeping track of the dual $(d-3)$- and $(d-4)$-branes which should
arise had we started with the dual of the scalar field.

Observe that in the generalized dimensional reduction Ansatz,
Eq.~(\ref{eq:genAnsatz}), the scalar is not single-valued in the
compact coordinate: $\hat{\phi} (z+1)=\hat{\phi}(z) +m$. The charge of
the $(d-3)$-brane can be associated to the monodromy of $\hat{\phi}$
and to the $(d-1)$-dimensional vector mass:

\begin{equation}
q \sim \int {}^{\star}\hat{F}_{(d-1)} \sim \int d\hat{\phi} \sim m\, .
\end{equation}

The implication of these results is obvious: The standard recipe for
generalized dimensional reduction is just a way of performing a
dimensional reduction taking into account all the possible fields
(i.e.~branes) that can arise in $(d-1)$ dimensions.  In particular,
the presence of $(d-3)$-branes is associated to the dependence on the
internal coordinate and the charge of the background $(d-3)$-branes is
proportional to the mass parameter. Generalized dimensional reduction
should, from this point of view, be considered the standard full
dimensional reduction, while the standard dimensional reduction is
incomplete and there is an implicit truncation. The reason why this
has not been realized before is that the missing fields only carry
discrete degrees of freedom. The mass parameters are to be considered
fields, although one can equally consider them as
expectation values of
those fields.

In the remainder of the paper we are going to perform a generalized
dimensional reduction in the, more complex, context of type~IIB
supergravity. The underlying physics is, however, the same.  The
upshot is that what we are going to do is to perform the full
dimensional reduction, without missing any fields as if we were able
to Poincar\'e-dualize the type~IIB scalars into 8-form potentials
which is technically complicated.

Since we know that the type~IIA and~IIB string theories are T~dual and
we know that this implies the same for their low-energy (supergravity)
theories, we expect T~duality to keep working in the generalized
dimensional reduction context. This poses several questions that we
will also try to answer.

Thus, the contents and structure of the paper are as follows: In
Section~\ref{sec-generalized} we perform the generalized dimensional
reduction of type~IIB supergravity in an $SL(2,\mathbb{R})$-covariant
way and obtain the massive 9-dimensional type~II theory which is
$SL(2,\mathbb{R})$-covariant. We analyze its global and local
symmetries.

In Section~\ref{sec-generalizedgauging} we obtain the same result using
a new recipe for generalized dimensional reduction which involves the
gauging of the global symmetry. 

\begin{figure}[!ht]
\begin{center}
\leavevmode
\epsfxsize= 6cm
\epsffile{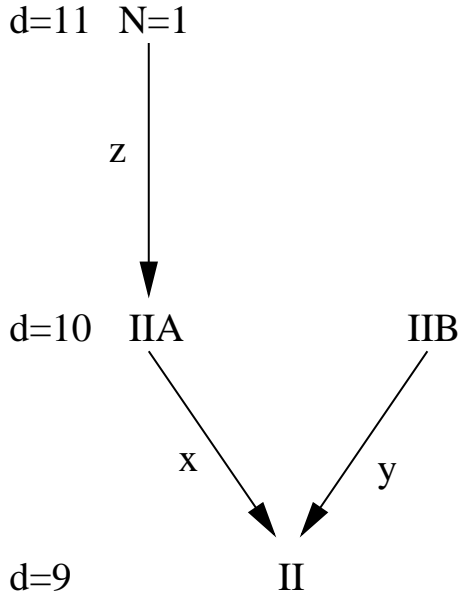}
\caption{Scheme of the different dimensional reductions with the names
of the respective compact coordinates $z,x,y$.}
\label{fig:reductio}
\end{center}
\end{figure}

In Section~\ref{sec-11} we study the M/type~IIA side of the problem.
First, we review the manifestly $SL(2,\mathbb{R})$-covariant
compactification of 11-dimensional supergravity which gives the
standard massless 9-dimensional type~II theory and, then, in
Section~\ref{sec-toro}, we propose a massive generalization which,
upon compactification on a $T^{2}$ gives precisely the massive
9-dimensional type~II theory we obtained in the type~IIB side of our
problem. Figure~\ref{fig:reductio} contains a schematic representation
of the different dimensional reductions involved in this work.

In Section~\ref{sec-pq-7-branes} we study 7-brane solutions and relate
their monodromy properties with the mass matrix of the massive
9-dimensional type~II theory.  This allows us to give physical meaning
to the mass parameters as 7-brane charges.

In Section~\ref{sec-KK789} we study the duality relations between
KK-type branes conjectured in this Introduction and depicted in
Figure~\ref{fig:kkduals}.

Section~\ref{sec-conclusion} contains our conclusions and some
speculations and Figure~\ref{fig:dualbran} which synthesizes our
present knowledge about the duality relations of the different
extended M/string-theory solitons.

Finally in Appendix~\ref{sec-9E-10S-IIA} and
Appendix~\ref{sec-9E-10S-IIB} we relate the fields of the massive
9-dimensional type~II theory we have obtained with the 10-dimensional
fields of the type~IIA and~B theories respectively. These relations
imply the Buscher's T~duality relations between the 10-dimensional
fields themselves which are given in Appendix~\ref{sec-tdual}.

\section{The $Sl(2,\mathbb{R})$-Covariant Generalized Dimensional 
Reduction of Type~IIB Supergravity: An S~Duality Multiplet of
$N=2,d=9$ Massive Supergravities}
\label{sec-generalized}

In this Section we perform the complete generalized dimensional
reduction of type~IIB supergravity in the direction parametrized by
$y$ (see Fig.~\ref{fig:reductio}) using the ideas of
Ref.~\cite{kn:BRGPT} as they were generalized in Ref.~\cite{kn:LLP2}.
As we are going to explain, in the end we will obtain a
three-parameter family (a triplet) of type~II 9-dimensional
supergravities connected by $SL(2,\mathbb{R})$ transformations (in the
adjoint representation).

We are going to perform the generalized dimensional reduction in a
manifestly $SL(2,\mathbb{R})$-covariant way. $SL(2,\mathbb{R})$
symmetry is manifest in the Einstein-frame. However, T~duality, being
a stringy symmetry, is better described in string frame. Thus we will
spend some time relating the fields appearing in both frames.  Since
reducing an action is easier than reducing equations of motion, we are
going to use the {\it non-self-dual} (NSD) action introduced in
Ref.~\cite{kn:BBO}.  We study these two points in the following
subsection and we perform the actual reduction in the next section.


\subsection{An Overview of Type~IIB Supergravity: The NSD Action 
and $Sl(2,R)$ Symmetry}

It is well-known \cite{kn:JHS} that it is not possible to write a
covariant action whose minimization gives the equations of motion
10-dimensional type~IIB supergravity. The problematic equation of
motion is the self-duality of the 5-form field strength. However, we
can use it to find an alternative equation of motion just by replacing
the 5-form field strength by its Hodge dual in the Bianchi identity.
This alternative equation of motion has the conventional form of the
equation of motion of a 4-form potential and it is possible to find an
action from which to derive this and the other equations of motion
{\it but not self-duality}. This NSD action, supplemented by the
self-duality constraint gives all the equations of motion of the
type~IIB theory.

The (bosonic sector of the) string-frame NSD action is\footnote{From
now on we denote with hats and double hats 10- and 11-dimensional
objects respectively. Our conventions are essentially those of
Ref.~\cite{kn:BLO} but we change the symbols denoting NS-NS fields in
the type~IIB theory to distinguish them from those of the type~IIA. In
particular we use the index-free notation of that reference: when
indices are not explicitly shown, they are assumed to be completely
antisymmetrized with weight one, and the definition of field strengths
and gauge transformations are inspired by those of
Refs.~\cite{kn:GHT,kn:BCT}.}

\begin{equation}
\begin{array}{rcl}
S_{\rm NSD} & = &
\int d^{10}\hat{x}\
\sqrt{|\hat{\jmath}|}\ \left\{ e^{-2\hat{\varphi}}
\left[ \hat{R}(\hat{\jmath}) -4\left( \partial\hat{\varphi} \right)^{2}
+{\textstyle\frac{1}{2\cdot 5!}} \hat{\cal H}^{2}\right]\right.\\
& & \\
& & 
\hspace{1cm}+{\textstyle\frac{1}{2}} \left( \hat{G}^{(0)} \right)^{2}
+{\textstyle\frac{1}{2\cdot 3!}}\left(\hat{G}^{(3)}\right)^{2}
+{\textstyle\frac{1}{4\cdot 3!}}\left(\hat{G}^{(5)}\right)^{2} \\
& & \\
& &
\hspace{1cm}\left.-{\textstyle\frac{1}{192}} \frac{1}{\sqrt{|\hat{\jmath}|}}\
\epsilon\  \partial \hat{C}^{(4)}\partial \hat{C}^{(2)} \hat{\cal B}
\right\}\, ,\\
\end{array}
\label{eq:stringIIBaction}
\end{equation}

\noindent where $\{\hat{\jmath}_{\hat{\mu}\hat{\nu}},
\hat{\cal B}_{\hat{\mu}\hat{\nu}},
\hat{\varphi}\}$ are the NS-NS fields: The type~IIB string metric,
the type~IIB NS-NS 2-form  and the type~IIB dilaton respectively.

\begin{equation}
\hat{\cal H}_{\hat{\mu}\hat{\mu}\hat{\rho}}=3\partial_{[\hat{\mu}}
\hat{\cal B}_{\hat{\nu}\hat{\rho}]}\, ,   
\hspace{1cm}
\left(\hat{\cal H}=3\partial\hat{\cal B}\right)\, ,
\end{equation}

\noindent is the NS-NS 2-form field strength. $\{\hat{C^{(0)}},
\hat{C^{(2)}}{}_{\hat{\mu}\hat{\nu}},
\hat{C^{(4)}}{}_{\hat{\mu}\hat{\nu}\hat{\rho}\hat{\sigma}}\}$ are the RR
potentials. Their field strengths and gauge transformations are

\begin{equation}
\left\{
\begin{array}{rcl}
\hat{G}^{(1)} & = & \partial \hat{C}^{(0)}\, ,\\
& & \\
\hat{G}^{(3)} & = & 3\left(\partial \hat{C}^{(2)}
-\partial \hat{\cal B} \hat{C}^{(0)}\right)\, ,\\
& & \\
\hat{G}^{(5)} & = & 5\left(\partial \hat{C}^{(4)}
-6\partial \hat{\cal B} \hat{C}^{(2)}\right)\, .\\
\end{array}
\right.
\end{equation}

\noindent and 

\begin{equation}
\left\{
\begin{array}{rcl}
\delta\hat{C}^{(0)} & = & 0\, ,\\
& & \\
\delta\hat{C}^{(2)} & = & 2\partial\hat{\Lambda}^{(1)}\, ,\\
& & \\
\delta\hat{C}^{(4)} & = & 4\partial\hat{\Lambda}^{(3)}
+6\hat{\cal B}\partial\hat{\Lambda}^{(1)}\, ,\\
\end{array}
\right.
\end{equation}

\noindent respectively.

The equations of motion derived from the above action have to be
supplemented by the self-duality condition

\begin{equation}
\hat{G}^{(5)} = +\ {}^{\star}\hat{G}^{(5)}\, .
\end{equation}

In the original version of the 10-dimensional, chiral $N=2$
supergravity \cite{kn:JHS} the theory has a classical $SU(1,1)$ global
symmetry. The two scalars parametrize the coset $SU(1,1)/U(1)$, $U(1)$
being the maximal compact subgroup of $SU(1,1)$, and transform under a
combination of a global $SU(1,1)$ transformation and a local $U(1)$
transformation which depends on the global $SU(1,1)$ transformation.
They are combinations of the dilaton and the RR scalar.  The group
$SU(1,1)$ is isomorphic to $SL(2,\mathbb{R})$, the conjectured
classical S~duality symmetry group for the type~IIB string theory
\cite{kn:HT}.  A simple field redefinition \cite{kn:BHO} is enough to
rewrite the action in terms of two real scalars parametrizing the
coset $SL(2,\mathbb{R})/SO(2)$ which can now be identified with the
dilaton $\hat{\varphi}$ and the RR scalar $\hat{C}^{(0)}$.

In order to make the S~duality symmetry manifest, we first have to
rescale the metric as to go to the Einstein frame:

\begin{equation}
\hat{\jmath}_{E\ \hat{\mu}\hat{\nu}}
=e^{-\hat{\varphi}/2}\hat{\jmath}_{\hat{\mu}\hat{\nu}}\, . 
\end{equation}

We now have to make some further field redefinitions. For instance,
while the NS-NS and RR 2-forms we are using form an $SL(2,\mathbb{R})$
doublet, their field strengths do not. Furthermore, our self-dual RR
4-form potential $\hat{C}^{(4)}$ is not $SL(2,\mathbb{R})$-invariant.
Thus, for the purpose of exhibiting the $SL(2,\mathbb{R})$ symmetry it
is convenient to perform the following field
redefinitions\footnote{Our conventions are such that all fields are
either invariant or transform {\it covariantly} as opposed to {\it
contravariantly}.}:

\begin{equation}
\left\{
\begin{array}{rcl}
\hat{\vec{\cal B}} & = & 
\left(
\begin{array}{c}
\hat{C}^{(2)}  \\
\hat{\cal B}   \\
\end{array}
\right)\, , \\
& & \\
\hat{D} & = & \hat{C}^{(4)} -3\hat{\cal B}\hat{C}^{(2)}\, ,\\
\end{array}
\right.
\end{equation}

\noindent These new fields undergo the following gauge transformations:

\begin{equation}
\left\{
\begin{array}{rcl}
\delta\hat{\vec{\cal B}} & = & 2\hat{\vec{\Sigma}}\, ,\\
& & \\
\delta\hat{D} & = & 4\partial\hat{\Delta} 
+2\hat{\vec{\Sigma}}{}^{\ T} \eta\ \hat{\vec{\cal H}}\, ,\\
\end{array}
\right.
\end{equation}

\noindent and have field strengths

\begin{equation}
\label{eq:fieldstrengths10B}
\left\{
\begin{array}{rcl}
\hat{\vec{\cal H}} & = & 3\partial \hat{\vec{\cal B}}\, ,\\
& & \\
\hat{F} & = & \hat{G}^{(5)}=+\ {}^{\star} \hat{F} \\
& & \\
& = & 5\left(\partial \hat{D} 
-\hat{\vec{\cal B}}{}^{\ T}\eta\ \hat{\vec{\cal H}}\right)\, ,\\
\end{array}
\right.
\end{equation}

\noindent where $\eta$ is the $2\times 2$ matrix

\begin{equation}
\eta= i\sigma^{2} =
\left(
\begin{array}{rc}
0  & 1 \\
-1 & 0 \\
\end{array}
\right)
=
-\eta^{-1}
=
-\eta^{T}\, ,
\end{equation}

Given the isomorphism $SL(2,\mathbb{R})\sim Sp(2,\mathbb{R})$, it can
be identified with an invariant metric:

\begin{equation}
\label{eq:propertyeta}
\Lambda \eta \Lambda^{T}=\eta\, ,
\,\,
\Rightarrow
\,\,
\eta\Lambda\eta^{T} = (\Lambda^{-1})^{T}\, ,
\hspace{.5cm}
\Lambda \in SL(2,\mathbb{R})\, .
\end{equation}

Finally, it is convenient to define the $2\times 2$ matrix $\hat{\cal
  M}_{ij}$

\begin{equation}
\hat{\cal M}
=e^{\hat{\varphi}}
\left(
\begin{array}{cc}
|\hat{\lambda}|^{2} &    \hat{C}^{(0)}  \\
& \\
\hat{C}^{(0)}       &  1          \\
\end{array}
\right)\, ,
\hspace{1cm}
\hat{\cal M}^{-1}
=e^{\hat{\varphi}}
\left(
\begin{array}{cc}
1                     &  -\hat{C}^{(0)}      \\
& \\
-\hat{C}^{(0)}        &  |\hat{\lambda}|^{2} \\
\end{array}
\right)\, ,
\end{equation}

\noindent where $\hat{\lambda}$ is the complex scalar

\begin{equation}
\hat{\lambda} =\hat{C}^{(0)} +ie^{-\hat{\varphi}}\, .
\end{equation}

Observe that $\hat{\cal M}$ is a symmetric $SL(2,\mathbb{R})$ matrix and
therefore, as a consequence of Eq.~(\ref{eq:propertyeta}) it has the
property

\begin{equation}
\hat{\cal M}^{-1} = \eta \hat{\cal M} \eta^{T}\, ,
\end{equation}

To see that $\hat{\lambda}$ parametrizes the $SL(2,\mathbb{R})/SO(2)$
coset, it is convenient to consider how one arrives at $\hat{\cal
M}$. First one considers the non-symmetric $SL(2,\mathbb{R})$ matrix
$\hat{V}$

\begin{equation}
\hat{V} =
\left( 
\begin{array}{rr}
e^{-\hat{\varphi}/2} & e^{\hat{\varphi}/2}\hat{C}^{(0)} \\
& \\
0                    & e^{\hat{\varphi}/2}              \\
\end{array}
\right)\, .
\end{equation}

This $SL(2,\mathbb{R})$ matrix is generated by only two of the three
$SL(2,\mathbb{R})$ generators and it should cover the
$SL(2,\mathbb{R})/SO(2)$ coset.  The choice for the form of $\hat{V}$
can be understood as a choice of gauge or as a choice of coset
representatives.  However, an arbitrary $SL(2,\mathbb{R})$
transformation $\Lambda$ will transform $\hat{V}$ into a
non-upper-triangular matrix $\Lambda\hat{V}$ (which is not a coset
representative).  A further $\Lambda$-dependent $SO(2)$-transformation
$h$ will, by using the definition of a coset, take us to another coset
representative $\hat{V}^{\prime}=\Lambda\hat{V}h$. The transformation
$h$ will be {\it local} but not arbitrary. It can be thought of as a
compensating gauge transformation. The condition that
$\hat{V}^{\prime}$ is upper-triangular fully determines
$h(\Lambda,\hat{V})$ and the transformations of $\hat{C}^{(0)}$ and
$\hat{\varphi}$:

\begin{equation}
\begin{array}{rcl}
\hat{V}^{\prime} & = &
\left( 
\begin{array}{rr}
e^{-\hat{\varphi}^{\prime}/2} & e^{\hat{\varphi}^{\prime}/2}
\hat{C}^{(0)\ \prime} \\
& \\
0                             & e^{\hat{\varphi}^{\prime}/2} \\
\end{array}
\right) = \Lambda \hat{V} h=\\
& & \\
& & 
\hspace{-1cm}
=
\left( 
\begin{array}{cc}
a & b \\
& \\
c & d \\
\end{array}
\right)
\left( 
\begin{array}{rr}
e^{-\hat{\varphi}/2} & e^{\hat{\varphi}/2}\hat{C}^{(0)} \\
& \\
0                    & e^{\hat{\varphi}/2}              \\
\end{array}
\right)
\left( 
\begin{array}{cc}
\cos{\theta} & \sin{\theta} \\
& \\
-\sin{\theta} & \cos{\theta} \\
\end{array}
\right)\, ,\\
\end{array}
\end{equation}

\noindent where $ad-bc=1$. The result is that the parameter 
$\theta$ of the compensating transformation $h$ is given by

\begin{equation}
\tan{\theta} = \frac{c}{e^{\hat{\varphi}} 
\left(c \hat{C}^{(0)} +d\right)}\, ,
\end{equation}

\noindent and the transformation of the scalars can be 
written in the compact form

\begin{equation}
\hat{\lambda}^{\prime} = \frac{a\hat{\lambda} +b}{c\hat{\lambda} +d}\, .  
\end{equation}

The symmetric matrix $\hat{\cal M}$ is now 
$\hat{\cal M}=\hat{V}\hat{V}^{T}$ and
transforms under $\Lambda\in SL(2,\mathbb{R})$ according to

\begin{equation}
\hat{\cal M}^{\prime} = \Lambda \hat{\cal M} \Lambda^{T}\, ,
\end{equation}

\noindent which is completely equivalent to the above transformation
of $\hat{\lambda}$. Observe that it is not necessary to worry about
the $h$-transformations anymore.

It is also worth stressing that the only $SL(2,\mathbb{R})$
transformations that leave invariant $\hat{\lambda}$ or, equivalently,
$\hat{\cal M}$ or $\hat{V}$ are $\pm\mathbb{I}_{2\times 2}$. This is
an important point: $SO(2)$ is sometimes referred to as the
``stability subgroup''. Had we defined the coset by the equivalence
relation $\hat{V}\sim h\hat{V}\, ,\,\, h\in SO(2)$, then, by
definition, $\hat{V}$ would have been invariant under any $\Lambda \in
SO(2)$. Then, $SO(2)$ would have been the subgroup of
$SL(2,\mathbb{R})$ leaving invariant the coset scalars. This is,
however, {\it not} the way in which this coset is constructed and (as
it can be explicitly checked) there is no stability subgroup of
$SL(2,\mathbb{R})$ in that sense apart from this almost trivial
$\mathbb{Z}_{2}$.

Under this $\Lambda$, the doublet of 2-forms transforms 

\begin{equation}
\hat{\vec{\cal B}}{}^{\prime}
 =  
\Lambda \hat{\vec{\cal B}}\, ,
\end{equation}

\noindent and the 4-form $\hat{D}$ and the Einstein metric are inert.

Now, it is a simple exercise to rewrite the NSD type~IIB action in 
the following manifestly S~duality invariant form

\begin{equation}
\begin{array}{rcl}
\hat{S}_{\rm NSD} & = &
\frac{1}{16\pi G_{N}^{(10)}} \int d^{10}\hat{x}\
\sqrt{|\hat{\jmath}_{E}|} \left\{ \hat{R}(\hat{\jmath}_{E})
+{\textstyle\frac{1}{4}}
{\rm Tr}\left(\partial\hat{\cal M} \hat{\cal M}^{-1}\right)^{2}
\right.\\
& & \\
& & 
\left.
+{\textstyle\frac{1}{2\cdot 3!}} 
\hat{\vec{\cal H}}{}^{\ T}\hat{\cal M}^{-1}\hat{\vec{\cal H}}
+{\textstyle\frac{1}{4\cdot 3!}}\hat{F}^{2}
-{\textstyle\frac{1}{2^{7}\cdot 3^{3}}} 
\frac{1}{\sqrt{|\hat{\jmath}_{E}|}}\
\epsilon\ \hat{D}\ \hat{\vec{\cal H}}{}^{\ T}\eta\ \hat{\vec{\cal H}}
\right\}\, , \\
\end{array}
\label{eq:SinvariantIIBaction}
\end{equation}

It is easy to find how the fields $\hat{\cal H}, \hat{G}^{(3)},
\hat{C}^{(4)}$ in the action Eq.~(\ref{eq:stringIIBaction}) transform
under $SL(2,\mathbb{R})$:

\begin{equation}
\left\{
\begin{array}{rcl}
\hat{\cal H}^{\prime} & = & 
\left(d +c \hat{C}^{(0)}\right) \hat{\cal H} +c \hat{G}^{(3)}\, , \\
& & \\
\hat{G}^{(3)\ \prime} & = & \frac{1}{|c\hat{\lambda} +d|^{2}}
\left[ \left(d +c \hat{C}^{(0)}\right) \hat{G}^{(3)} 
-c e^{-2\hat{\varphi}} \hat{\cal H} \right]\, ,\\
& & \\
\hat{C}^{(4)\ \prime} & = & \hat{C}^{(4)} -3 
\left(
\begin{array}{cc}
\hat{C}^{(2)} & \hat{\cal B} \\
\end{array}
\right)
\left(
\begin{array}{cc}
ac & bc \\
bc & db \\
\end{array}
\right)
\left(
\begin{array}{c}
\hat{C}^{(2)} \\
\hat{\cal B}  \\
\end{array}
\right)\, . \\
\end{array}
\right.
\end{equation}

\noindent $\hat{\lambda}$ transforms as above and we stress 
 that the string metric does transform under $SL(2,\mathbb{R})$:

\begin{equation}
\hat{\jmath}^{\prime} = |c\hat{\lambda} +d| \hat{\jmath}\, .  
\end{equation}


\subsection{Generalized Dimensional Reduction}

Now that we have set up the action we want to reduce, we can proceed.
First, we will explain the generalized KK Ansatz. In this point we
will follow the recipe of Ref.~\cite{kn:LLP2} adapted to our
conventions. Then we will reduce the action and the self-duality
constraint and finally we will eliminate the constraint, obtaining the
action of the 9-dimensional theory.

The fields of the Einstein-frame 9-dimensional theory are the same as
in the massless case:

\begin{equation}
\{g_{E\ \mu\nu}, A_{(3)\ \mu\nu\rho}, \vec{A}_{(2)\ \mu\nu}, 
\vec{A}_{(1)\ \mu}, A_{(1)\ \mu}, K, {\cal M} \}\, ,
\end{equation}

\noindent and only the couplings and symmetries will be different.


\subsubsection{The Kaluza-Klein Ansatz}

As usual in dimensional reductions, we assume the existence of a
Killing vector $\hat{s}^{\hat{\mu}}\partial_{\hat{\mu}}
=\partial_{\underline{y}}$ associated to the coordinate $y$. We choose
adapted coordinates $\hat{x}^{\hat{\mu}}=(x^{\mu},y)$ so that the
metric does not depend on $y$. We normalize the coordinate $y$ such
that it takes values in the interval $[0,1]$ and so $y\sim y+1$. Our
Ansatz for the Einstein-frame Zehnbeins is then that of
Eq.~(\ref{eq:KKbasis}) adapted to ten dimensions and with the scalar
$k$, the length of the (spacelike) Killing vector, relabeled

\begin{equation}
|\hat{s}^{\hat{\mu}}\hat{s}_{\hat{\mu}}|^{1/2}=K^{-3/4}\, ,
 \end{equation}

\noindent for convenience.

Now, instead of assuming that all the other fields in our theory have
vanishing Lie derivatives with respect to $\hat{s}^{\hat{\mu}}$, we
assume that the remaining fields depend on $y$ but in a very specific
way: All the $y$-dependence is introduced by a local
$SL(2,\mathbb{R})$ transformation with parameters linear in $y$,
$\Lambda(y)$:

\begin{equation}
\left\{
\begin{array}{rcl}
\hat{\cal M}(\hat{x})  & \equiv & \Lambda (y) \hat{\cal M}^{\rm b} (x)
 \Lambda^{T} (y)\, ,\\
& & \\
\hat{\vec{\cal B}}(\hat{x}) & \equiv & \Lambda (y) 
\vec{\cal B}^{\rm b} (x)\, ,\\
& & \\
\hat{D}(\hat{x}) & = & \hat{D}^{\rm b} (x)\, , \\
\end{array}
\right.
\label{eq:fieldsAnsatz}
\end{equation}

\noindent where we have denoted by a superscript ${}^{b}$ the {\it
bare} $y$-independent fields.

Obviously, the Ansatz for $\hat{\cal M}$ is equivalent, in terms of
$\hat{\lambda}(\hat{x})$ to

\begin{equation}
\hat{\lambda}   (\hat{x}) =\frac{a(y) \hat{\lambda}^{\rm b} (x) + b (y)}
{c(y) \hat{\lambda}^{\rm b} (x) + d (y)}\, .
\end{equation}

In this scheme $\hat{D}$ cannot depend on $y$ because it is inert
under $SL(2,\mathbb{R})$, but it is worth stressing that the
string-frame metric does depend on $y$. The {\it bare} fields are
$y$-independent and will become the 9-dimensional fields. On the other
hand, they transform under $SL(2,\mathbb{R}$ as the real fields do.

The meaning of this kind of Ansatz is the following: We are
constructing a non-trivial line bundle over the circle parametrized by
$y$ with fiber $\hat{\lambda}$ (or, equivalently $\hat{\cal M}$) and
structure group $SL(2,\mathbb{R})$ (we will later study the
restriction to $SL(2,\mathbb{Z})$). Going once around the circle we go
back to the same $\hat{\cal M}$ up to a global $SL(2,\mathbb{R})$
transformation that we can describe by an $SL(2,\mathbb{R})$ monodromy
matrix $M$. The explicit form of $M$ depends on the explicit form of
$\Lambda (y)$.

Let us now describe more precisely the form of $\Lambda (y)$. If

\begin{equation}
T_{1} = \sigma^{3} =
\left(
\begin{array}{cr}
1 & 0  \\
0 & -1 \\
\end{array}
\right)\, ,
\hspace{.3cm} 
T_{2} = \sigma^{1} =
\left(
\begin{array}{cc}
0 & 1  \\
1 & 0  \\
\end{array}
\right)\, ,
\hspace{.3cm} 
T_{3} = i\sigma^{2} =
\left(
\begin{array}{rc}
0  & 1  \\
-1 & 0  \\
\end{array}
\right)\, ,
\end{equation}

\noindent are the generators of $SL(2,\mathbb{R})$, then the most general
$SL(2,\mathbb{R})$ transformation with local parameters linear in $y$ can be
written in the form

\begin{equation}
\Lambda (y) = \exp{\{{\textstyle\frac{1}{2}}y m^{i}T_{i}\}}\, .
\end{equation}

The three real parameters $m^{i}$ fully determine $\Lambda (y)$ and
therefore the particular compactification.  These parameters are going
to become masses in the lower-dimensional theory.  We define the {\it
mass matrix} $m$

\begin{equation}
\label{eq:massmatrix}  
m \equiv 
\left(\partial_{\underline{y}}\Lambda   \right) \Lambda^{-1}
= {\textstyle\frac{1}{2}}m^{i}T_{i} =
{\textstyle\frac{1}{2}}
\left(
\begin{array}{cc}
m^{1}       & m^{2}+m^{3} \\
& \\
m^{2}-m^{3} &  -m_{1}     \\
\end{array}
\right)\, .
\end{equation}

\noindent This matrix belongs to the Lie algebra $sl(2,\mathbb{R})$ and 
therefore it transforms in the (irreducible) adjoint representation:

\begin{equation}
m^{\prime} = \Lambda m \Lambda^{-1}\, ,
\end{equation}

\noindent and thus the three $m^{i}$ transform as a triplet (a vector of 
$SO(2,1)\sim SL(2,\mathbb{R})$). The expression 

\begin{equation}
\label{eq:alpha}
\alpha^{2}={\rm Tr}\ ( m^{2})={\textstyle\frac{1}{4}}m^{i}m^{j}h_{ij}\, ,
\hspace{1cm}
h_{ij}={\rm diag}(++-)\, ,
\end{equation}

\noindent where $h_{ij}$ is the Killing metric, is thus 
$SL(2,\mathbb{R})$-invariant.  Furthermore, the mass matrix satisfies

\begin{equation}
\eta m\eta^{-1} = -m^{T}\, .  
\end{equation}

\noindent Observe that the parameters $m^{1},m^{2}$ are associated to
non-compact generators of $SL(2,\mathbb{R})$, while $m^{3}$ is associated to
the maximal compact subgroup of $SL(2,\mathbb{R})$ ($SO(2)$). Thus, we are
bound to get mass terms with the wrong sign (for instance in terms
like Eq.~(\ref{eq:alpha})) but we must keep the three mass parameters
in order to have full $SL(2,\mathbb{R})$-covariance and the most general
9-dimensional massive type~II supergravity.

Our Ansatz generalizes that of Ref.~\cite{kn:LLP2}, which only had two
independent parameters: $m^{1}, m^{2}=m^{3}$. The authors argued that
generalized dimensional reduction using $SL(2,\mathbb{R})$ y-dependent
transformations in the stability subgroup $SO(2)$ (i.e.~those
generated by $T_{3}$ and associated to $m^{3}$ in our conventions)
would have no effect. As we discussed in the previous Section, there
is no stability subgroup for the coset scalars. Furthermore, since the
three mass parameters we just defined transform irreducibly, the three
of them are required to obtain $SL(2,\mathbb{R})$-covariant families
of theories. Finally, the $SL(2,\mathbb{R})$ transformation $S=\eta$
is inside the excluded $SO(2)$ and this is one of the generators of
the quantum S~duality group $SL(2,\mathbb{Z})$.

$\Lambda (y)$ will only manifest itself through the mass matrix in the
lower-dimensional theory.  However, in order to reconstruct the
10-dimensional fields we need to know it explicitly.  The explicit
form of $\Lambda (y)$ reads

\begin{equation}
\label{eq:lambday}
\Lambda (y)=
\left(
\begin{array}{cc}
\cosh{\alpha y} +\frac{m^{1}}{2\alpha}\sinh{\alpha y} & 
\frac{m^{2}+m^{3}}{2\alpha} \sinh{\alpha y} \\
& \\
\frac{m^{2}-m^{3}}{2\alpha} \sinh{\alpha y} &
\cosh{\alpha y} -\frac{m^{1}}{2\alpha}\sinh{\alpha y} \\
\end{array}
\right)\, , 
\end{equation}

\noindent where $\alpha$ was defined in Eq.~(\ref{eq:alpha}).

It is easy to see from our definition of $\Lambda (y)$ that the
monodromy matrix will be

\begin{equation}
M(m^{i}) = \exp{\{{\textstyle\frac{1}{2}}m^{i}T_{i}\}}
= \Lambda (y=1)\, ,
\end{equation}

\noindent and 

\begin{equation}
\left\{
\begin{array}{rcl}
\hat{\cal M}(x,y+1)  & = & M\hat{\cal M}(x,y) M^{T}\, ,\\
& & \\
\hat{\vec{\cal B}}(x,y+1) & = & M\vec{\cal B}(x,y)\, .\\
\end{array}
\right.
\label{eq:fieldmonodromy}
\end{equation}

Quantum-mechanically, the monodromy matrices can only be
$SL(2,\mathbb{Z})$ matrices. It is convenient to describe the most
general $SL(2,\mathbb{Z})$ monodromy matrix by for integers $n^{i},n\,
,\,\, i=1,2,3$ subject to the constraint

\begin{equation}
n^{i}n_{i} = n^{2}-1\, .
\end{equation}

Given that this constraint is satisfied, then we simply make the
identifications

\begin{equation}
\alpha = \cosh^{-1}{n}\, ,
\hspace{1cm}
m^{i} = \frac{2\alpha}{\sqrt{n^{2}-1}} n^{i}\, ,
\end{equation}

\noindent and write the monodromy matrix as follows:

\begin{equation}
\label{eq:integermonodromy}
M=  
\left(
\begin{array}{cc}
n+n^{1} & n^{2}+n^{3} \\
& \\
n^{2}-n^{3} & n-n^{1}\\
\end{array}
\right)\, . 
\end{equation}

Thus, in our conventions, the mass parameters $m^{i}$ will be
naturally quantized in terms of the three integers $n^{i}$ which also
transform in the ``adjoint'' of $SL(2,\mathbb{Z})$. $n$ is
$SL(2,\mathbb{Z})$-invariant.

In Section~\ref{sec-pq-7-branes} we will relate the integers $n^{i}$
to the charges of 7-branes.

We can now perform the dimensional reduction.


\subsubsection{Dimensional Reduction}
\label{sec-reduction}

Using the standard techniques \cite{kn:SS} we get with the
just-described Ansatz the NSD 9-dimensional action

\begin{equation}
\label{eq:NSD9}
\begin{array}{rcl}
S_{\rm NSD} & = & \\
& & \\
& & 
\hspace{-2cm}
\int d^{9}x \sqrt{|g|}\ \left\{  K^{-3/4}\ 
\left[\ R (g)\ +{\textstyle\frac{1}{4}} {\rm Tr} 
\left(D {\cal M} {\cal M}^{-1}\right)^{2} 
-{\textstyle\frac{1}{4}}K^{-3/2}F_{(2)}^{2} 

\right.
\right.
\\
& & \\
& &
-{\textstyle\frac{1}{4}}K^{3/2} 
\vec{F}_{(2)}^{\ T}{\cal M} ^{-1}\vec{F}_{(2)}
+{\textstyle\frac{1}{2\cdot 3!}} 
\vec{F}_{(3)}^{\ T}{\cal M}^{-1}\vec{F}_{(3)}
-{\textstyle\frac{1}{4\cdot 4!}}K^{3/2}F_{(4)}^{2}
\\
& & \\
& & 
\left.
+{\textstyle\frac{1}{4\cdot 5!}}F_{(5)}^{2}
-K^{3/2}{\cal V}\left({\cal M}\right)\right]\\
& & \\
& &
+{\textstyle\frac{1}{2^{7}\cdot 3^{2}\cdot 5}} 
\frac{1}{\sqrt{|g|}}\
 \epsilon 
\left\{
\left( F_{(5)} -5A_{(1)}F_{(4)} \right) \times
\right. \\
& & \\
& & 
\times
\left[
2 \left( \vec{F}_{(3)} -3A_{(1)}\vec{F}_{(2)} \right)^{T} \eta\vec{A}_{(1)}
+3\vec{F}_{(2)}^{\ T} \eta \vec{A}_{(2)}
\right] \\
& & \\
& & 
\left.
-5 F_{(4)}\left( \vec{F}_{(3)} -3A_{(1)}\vec{F}_{(2)} \right)^{T} 
\eta \vec{A}_{(2)}
\right\}\, ,\\
\end{array}
\end{equation}

\noindent  and the 9-dimensional duality constraint

\begin{equation}
\label{eq:selfdual9}
F_{(5)} = -K^{3/4}\ {}^{\star}F_{(4)}\, ,
\end{equation}

\noindent where the field strengths are defined as follows:

\begin{equation}
\label{eq:fieldstrengths}
\left\{
\begin{array}{rcl}
{\cal D} {\cal M} & = &\partial {\cal M}
-\left(m {\cal M} +{\cal M}m^{T} \right) A_{(1)}\, ,\\
& & \\
F_{(2)} & = &2\partial A_{(1)}\, ,\\  
& & \\
\vec{F}_{(2)} & = & 2\partial\vec{A}_{(1)} -m\vec{A}_{(2)}\, ,\\
& & \\
\vec{F}_{(3)} & = & 3\partial \vec{A}_{(2)} +3A_{(1)}\vec{F}_{(2)}\, ,\\
& & \\ 
F_{(4)} & = & 4\partial A_{(3)} -3 \vec{A}_{(2)}^{\ T}\ \eta\ \vec{F}_{(2)}
+2\vec{A}_{(1)}^{\ T}\ \eta\ \vec{F}_{(3)}
+6A_{(1)}\vec{A}_{(1)}^{\ T}\ \eta\ \vec{F}_{(2)}\, ,\\
& & \\
F_{(5)} & = & 5\partial A_{(4)} -5 \vec{A}_{(2)}^{\ T}\ \eta\ \vec{F}_{(3)}
+15 A_{(1)}\vec{A}_{(2)}{}^{T}\eta \vec{F}_{(2)}+5A_{(1)} F_{(4)}\, ,\\
\end{array}
\right.
\end{equation}

\noindent and 

\begin{equation}
{\cal V}\left({\cal M}\right) ={\textstyle\frac{1}{2}}
{\rm Tr} \left(m^{2} +m {\cal M}m^{T}{\cal M}^{-1}\right)\, ,
\end{equation}

\noindent is the scalar potential.

The 10- and 9-dimensional fields are related as follows:

\begin{equation}
\begin{array}{rclrcl}
\hat{\cal M}^{\rm b} & = & {\cal M}\, ,
\hspace{2cm} &
\hat{D}_{\mu_{1}\mu_{2}\mu_{3}\underline{y}} & = & 
-A_{(3)\ \mu_{1}\mu_{2}\mu_{3}}\, ,\\
& & & & & \\
\vec{\cal B}^{\rm b}{}_{\mu \underline{y}} & = & -\vec{A}_{(1)\ \mu}\, , &
\hat{D}_{\mu_{1}\cdots\mu_{4}} & = & A_{(4)\ \mu_{1}\cdots\mu_{4}}\, .\\   
& & & & & \\
\vec{\cal B}^{\rm b}{}_{\mu\nu} & = &   \vec{A}_{(2)\ \mu\nu}\, , & & & \\
\end{array}
\end{equation}

\subsubsection{Elimination of the Self-Duality Constraint and Rescaling
of the Metric}

In order to eliminate the self-duality constraint
Eq.~(\ref{eq:selfdual9}) we first Poincar\'e-dualize the NSD action
with respect to the 4-form potential. First, we add the Lagrange
multiplier term

\begin{equation}
\begin{array}{rcl}
{\textstyle\frac{1}{2^{5}\cdot 3^{2}}} \int d^{9}x\ 
\epsilon \partial \tilde{A}_{(3)}\partial A_{(4)} & = & 
\\
& & \\
& & 
\hspace{-5cm}
{\textstyle\frac{1}{2^{5}\cdot 3^{2}}} \int d^{9}x\ 
\epsilon \partial \tilde{A}_{(3)}
\left[
F_{(5)} +5 \vec{A}_{(2)}^{\ T}\ \eta\ \vec{F}_{(3)}
-15 A_{(1)}\vec{A}_{(2)}{}^{T}\eta \vec{F}_{(2)} -5A_{(1)} F_{(4)}
\right]\, ,
\end{array}
\end{equation}

\noindent to the NSD action (\ref{eq:NSD9}). The equation of motion of
the Lagrange multiplier field $\tilde{A}_{(3)}$ enforces the Bianchi
identity of $F_{(5)}$ and we can consider the new action as a
functional of $F_{(5)}$ instead of $A_{(4)}$ which does not occur
explicitly. The equation of motion for $F_{(5)}$ is nothing but

\begin{equation}
F_{(5)} = -K^{3/4}\ {}^{\star}\tilde{F}_{(4)}\, ,
\end{equation}

\noindent where $\tilde{F}_{(4)}$ is like $F_{(4)}$ but with $A_{(4)}$
replaced by $\tilde{A}_{(4)}$. This equation is purely algebraic and
we can use it to eliminate $F_{(5)}$ in the NSD action (\ref{eq:NSD9})
plus the Lagrange multiplier term. The result is an action the depends
both on $A_{(4)}$ and $\tilde{A}_{(4)}$. Now, we simply observe that
the equation of motion for $F_{(5)}$ has the same form as the
self-duality constraint Eq.~(\ref{eq:selfdual9}) and therefore,
eliminating the self-duality constraint amounts to the simple
identification

\begin{equation}
F_{(4)} = \tilde{F}_{(4)}\, .
\end{equation}


The result of these manipulations plus a Weyl rescaling to go to the
Einstein frame (the metric $g$ is neither the string metric nor
Einstein's)

\begin{equation}
g_{\mu\nu} =K^{3/14}g_{E\ \mu\nu}\, .
\end{equation}

\noindent is the action of the type~II massive supergravity:


\begin{equation}
\begin{array}{rcl}
S & = & 
\int d^{9}x\ \sqrt{|g_{E}|}\
\left\{
R_{E} +\frac{9}{14}\left(\partial\log{K} \right)^{2}
+{\textstyle\frac{1}{4}} {\rm Tr} 
\left({\cal D} {\cal M} {\cal M}^{-1}\right)^{2} 
-{\textstyle\frac{1}{4}} K^{-12/7}F_{(2)}^{2}
\right.\\
& & \\
& & 
\hspace{-1.3cm}
-{\textstyle\frac{1}{4}} K^{\frac{9}{7}} 
\vec{F}_{(2)}^{\ T}{\cal M}^{-1}\vec{F}_{(2)}
+{\textstyle\frac{1}{2\cdot 3!}}K^{-3/7} 
\vec{F}_{(3)}^{\ T} {\cal M}^{-1}\vec{F}_{(3)}
-{\textstyle\frac{1}{2\cdot 4!}} K^{6/7} F_{(4)}^{2}
-K^{12/7}{\cal V}\left({\cal M}\right)\\
& & \\
& &
\hspace{-1.3cm}
-{\textstyle\frac{1}{2^{7}\cdot 3^{2}}} 
\frac{1}{\sqrt{|g_{E}|}}\
\epsilon 
\left\{
16(\partial A_{(3)})^{2}A_{(1)} 
\right.
\\
& & \\
& & 
\hspace{-1.3cm}
+24 \partial A_{(3)}
\left[
\partial\vec{A}_{(2)}^{\ T}\eta \vec{A}_{(2)}-
\left(
4\vec{A}_{(2)}^{\ T}\eta \partial\vec{A}_{(1)}
+2\vec{A}_{(1)}^{\ T}\eta \partial\vec{A}_{(2)}
-\vec{A}_{(2)}^{\ T}\eta m\vec{A}_{(2)}
\right)
A_{(1)} 
\right] 
\\
& & \\
& & 
\hspace{-1.3cm}
-36 
\left(
\vec{A}_{(2)}^{\ T}\eta \partial\vec{A}_{(1)}
+\vec{A}_{(1)}^{\ T}\eta \partial\vec{A}_{(2)} 
\right)
\partial\vec{A}_{(2)}^{\ T}\eta \vec{A}_{(2)}
\\
& & \\
& &
\hspace{-1.3cm}
-36 
\left(
\vec{A}_{(2)}^{\ T}\eta \partial\vec{A}_{(1)}
-\vec{A}_{(1)}^{\ T}\eta \partial\vec{A}_{(2)} 
\right)^{2} A_{(1)} 
\\
& & \\
& & 
\hspace{-1.3cm}
+9\vec{A}_{(2)}^{\ T}\eta m\vec{A}_{(2)}
\left[
\partial\vec{A}_{(2)}^{\ T}\eta \vec{A}_{(2)}
-4\left(\vec{A}_{(2)}^{\ T}\eta \partial\vec{A}_{(1)}
-\vec{A}_{(1)}^{\ T}\eta \partial\vec{A}_{(2)} \right)A_{(1)}
\right. 
\\
& & \\
& & 
\hspace{-1.3cm}
\left.
\left.
\left.
+\right(\vec{A}_{(2)}^{\ T}\eta m\vec{A}_{(2)}\left)A_{(1)}\right]
\right\}
\right\}
\, . \\
\end{array}
\label{eq:d9actionsl2rpotentials}  
\end{equation}

\noindent whose topological term, in order to facilitate comparison
with the results of Section~\ref{sec-11}, was rewritten in terms of
potentials only (no field strengths) by integrating several times by
parts and using algebraic properties like

\begin{equation}
\left(\vec{A}_{(1)}^{\ T}\eta \partial\vec{A}_{(2)}  \right)
\left(\partial\vec{A}_{(2)}^{\ T}\eta \vec{A}_{(2)} \right)=
-{\textstyle\frac{1}{2}}
\left(\vec{A}_{(1)}^{\ T}\eta\vec{A}_{(2)}  \right)
\left(\partial\vec{A}_{(2)}^{\ T}\eta\partial \vec{A}_{(2)} \right)\, .
\end{equation}


\subsubsection{Gauge and Global Symmetries of the 9-Dimensional Theory}

The local symmetries of the 9-dimensional theory
(\ref{eq:d9actionsl2rpotentials}) have three different origins: The
gauge transformations of the 2-form fields:

\begin{equation}
\delta \hat{\vec{\cal B}} = 2\partial\hat{\vec{\Sigma}}\, ,
\end{equation}

\noindent the gauge transformations of the 4-form

\begin{equation}
\delta \hat{D} = 4\partial\hat{\Delta}
-{\textstyle\frac{2}{5}}\hat{\vec{\Sigma}}{}^{T}\eta\hat{\vec{\cal H}}\, ,
\end{equation}

\noindent and the $y$-independent reparametrizations of the compact 
coordinate $y$

\begin{equation}
\delta \hat{x}^{\hat{\mu}}=\delta^{\hat{\mu}\underline{y}} \chi (x)\, .  
\end{equation}

The dependence of the 10-dimensional fields on $y$, inexistent in
standard dimensional reduction, induces new terms (the transport
terms) in the $\chi$-transformations.

The 9-dimensional fields have the following {\it infinitesimal} $\chi$
gauge transformations and finite
$\Sigma_{(0)},\vec{\Sigma}_{(0)},\vec{\Sigma}_{(1)},\Sigma_{(3)}$
gauge transformations:

\begin{equation}
\label{eq:infimassivegauge}
\left\{
\begin{array}{rcl}
\delta {\cal M} & = & \chi\left(m {\cal M} +{\cal M}m^{T} \right)\, ,\\
& & \\
\delta A_{(1)} & = & \partial\chi\, ,\\
& & \\
\delta \vec{A}_{(1)} & = & \partial \vec{\Sigma}_{(0)} +m\vec{\Sigma}_{(1)}
+\chi m \vec{A}_{(1)}\, ,\\
& & \\
\delta \vec{A}_{(2)} & = & 2\partial\vec{\Sigma}_{(1)}
+2\partial\chi \vec{A}_{(1)} +\chi m \vec{A}_{(2)}\, ,\\
& & \\
\delta A_{(3)} & = & 3\partial\Sigma_{(2)} 
+\frac{3}{2}\vec{\Sigma}_{(1)}{}^{T}\eta \vec{F}_{(2)}
-\frac{3}{2}\vec{\Sigma}_{(0)}{}^{T}\eta \partial\vec{A}_{(2)}\, ,\\
& & \\
\delta A_{(4)} & = & 4\partial\Sigma_{(3)} 
+6\vec{\Sigma}_{(1)}\eta \partial \vec{A}_{(2)} 
+4\partial\chi A_{(3)}\, .\\
\end{array}
\right.
\end{equation}



The $\chi$-transformations can be exponentiated:

\begin{equation}
\left\{
\begin{array}{rcl}
V^{\prime} & = & e^{\chi m} V\, ,\\
& & \\
{\cal M}^{\prime} & = & e^{\chi m} {\cal M}\ e^{\chi m^{T}}\, ,\\
& & \\
A_{(1)}^{\prime} & = & A_{(1)} +\partial\chi\, ,\\
& & \\
\vec{A}_{(1)}^{\prime} & = & e^{\chi m}\vec{A}_{(1)}\, ,\\
\end{array}
\right.
\hspace{1cm}
\left\{
\begin{array}{rcl}
\vec{A}_{(2)}^{\prime} & = &  e^{\chi m}\left(\vec{A}_{(2)}
+2\partial\chi \vec{A}_{(1)}\right)\, ,\\
& & \\
A_{(3)}^{\prime} & = & A_{(3)}\, ,\\
& & \\
A_{(4)}^{\prime} & = & 4\partial\chi A_{(3)}\, .\\
\end{array}
\right.
\end{equation}

Under the $\chi$-transformations, the field strengths transform covariantly
instead of being invariant:

\begin{equation}
\left\{
\begin{array}{rcl}
\left(D{\cal M}\right)^{\prime} & = & 
e^{\chi m}D {\cal M}\ e^{\chi m^{T}}\, ,\\
& & \\
\vec{F}^{\prime}_{(2,3)} & = & e^{\chi m} 
\vec{F}_{(2,3)} \, ,\\
& & \\
F^{\prime}_{(4,5)} & = & F_{(4,5)} \, .\\
\end{array}
\right.
\end{equation}

We could easily define field strengths invariant under
$\chi$-transformations: For instance

\begin{equation}
\tilde{\vec{F}}_{(2,3)} = V^{-1} \vec{F}_{(2,3)}\, ,
\end{equation}

\noindent as was done in Ref.~\cite{kn:BRGPT}, but we will choose 
not to do so.

It is trivial to check the invariance of the action
(\ref{eq:d9actionsl2rpotentials}) under the above gauge
transformations.

The action Eq.~(\ref{eq:d9actionsl2rpotentials}) enjoys some global
invariances as well, namely rescalings of $K$ and $SL(2,\mathbb{R})$
transformations.  The latter are the most interesting. Their action on
the fields ${\cal M},\vec{A}_{(1)\ \mu},\vec{A}_{(2)\ \mu\nu}$ is

\begin{equation}
{\cal M}^{\prime}=  \Lambda {\cal M}\Lambda^{T}\, ,
\hspace{1cm}
\vec{A}_{(1,2)}^{\prime}= \Lambda\vec{A}_{(1,2)}\, .
\end{equation}

\noindent  As was said before, the mass matrix belongs to
the Lie algebra $sl(2,\mathbb{R})$ and transforms in the adjoint
representation:

\begin{equation}
m^{\prime} = \Lambda m \Lambda^{-1}\, ,
\end{equation}

\noindent and thus the three $m^{i}$ transform as a triplet (a vector of 
$SO(2,1)\sim SL(2,\mathbb{R})$).

Finally, the theory is also invariant under constant rescalings of the
fields:

\begin{equation}
\begin{array}{rclrcl}
K & \rightarrow & e^{14\alpha}K\, , \hspace{2cm}&
m & \rightarrow & e^{-12\alpha}m\; ,\\
& & \\
A_{(1)} & \rightarrow & e^{12\alpha}A_{(1)}\, , & 
\vec{A}_{(1)} & \rightarrow & e^{-9\alpha}\vec{A}_{(1)} \; ,\\
& & \\
A_{(3)} & \rightarrow & e^{-6\alpha}A_{(3)}\, , & 
\vec{A}_{(2)} & \rightarrow & e^{3\alpha}\vec{A}_{(2)} \; .\\
\end{array}
\end{equation}


\section{An Alternative Recipe for Generalized Dimensional Reduction:
Gauging of Global Symmetries}
\label{sec-generalizedgauging}

In this Section we will apply an alternative recipe for generalized
dimensional reduction to type~IIB supergravity.  The general idea is
that gauging the global symmetry and imposing that the gauge field
takes non-vanishing and constant values in the internal direction
only, is equivalent to applying generalized Scherk-Schwarz reduction.
In order to demonstrate this, the algorithm will be applied to the NSD
IIB action, albeit written in terms of forms.  The conventions for
forms are the ones used in Ref.~\cite{kn:Nak} and in particular we
need

\begin{equation}
 \int F_{(p)}{}^{\star}F_{(p)} \;=\; 
 \int d^{d}x\sqrt{|g|}\ {\textstyle\frac{1}{p!}}
       F_{(p)\ \mu_{1}...\mu_{p}}F_{(p)}{}^{\mu_{1}...\mu_{p}}
 \; ,
\end{equation}

The NSD IIB action written in forms reads

\begin{equation}
\begin{array}{rcl}
S_{IIB} & = & \int d^{10}x \sqrt{|\hat{g}|}\,\left[
\hat{R}(\hat{g})-{\textstyle\frac{1}{4}}\, {\rm Tr}
\left( \partial_{\hat{\mu}}\hat{\cal M}\cdot 
\partial^{\hat{\mu}}\hat{\cal M}^{-1}\right)\right] \\
& & \\
& & 
+\int_{10}\left\{ {\textstyle\frac{1}{2}} \hat{\vec{H}}{}^{T}
\hat{\cal M}^{-1}\ {}^{\star}\hat{\vec{H}} 
+{\textstyle\frac{1}{4}} \hat{F}_{(5)} {}^{\star}\hat{F}_{(5)}
+{\textstyle\frac{1}{4}} \hat{F}_{(5)} \hat{\vec{B}}{}^{T}\eta \hat{\vec{H}} 
\right\} \, ,\\
\end{array}
\end{equation}

\noindent where we have defined

\begin{equation}
\left\{
\begin{array}{rcl}
\hat{\vec{H}} & = & d\hat{\vec{B}} \; , \\
& & \\
\hat{F}_{(5)} & = & d\hat{D}-{\textstyle\frac{1}{2}} 
\hat{\vec{B}}{}^{T}\eta \hat{\vec{H}} \, , \\
\end{array}
\right.
\end{equation}

\noindent which are nothing else than the definitions in
Eqs.~(\ref{eq:fieldstrengths10B}), but written in terms of forms.

In order to follow through the above procedure, we start by gauging
the $SL(2,\mathbb{R})$ symmetry. We introduce a covariant derivative
through

\begin{equation}
\left\{
\begin{array}{rcl}
\partial_{\hat{\mu}}\hat{\cal M} & \rightarrow & 
{\cal D}_{\hat{\mu}}\hat{\cal M} \;=\; 
\partial_{\hat{\mu}}\hat{\cal M}+\hat{\cal E}_{\hat{\mu}}\hat{\cal M} 
+\hat{\cal M}\hat{\cal E}_{\hat{\mu}}^{T}\, ,\\
& &  \\
d\hat{\vec{B}} & \rightarrow & {\cal D} \hat{\vec{B}} \;=\; 
d\hat{\vec{B}} \,+\, \hat{\cal E}\wedge \hat{\vec{B}} \, ,\\
\end{array}
\right.
\end{equation}

\noindent and one finds that $\hat{\cal E}$ has to transform
as a gauge field

\begin{equation}
\hat{\cal E} \rightarrow  \Lambda^{-1}\hat{\cal E}\Lambda
\,+\, \Lambda^{-1}d\Lambda \; .
\end{equation}

\noindent Now, applying the same KK Ansatz for the metric as was used
in the preceding section, one sees that the covariant derivatives on
$\hat{\cal M}$ get transformed into, changing notation such that
${\cal E}$ is the constant matrix in the internal direction,

\begin{equation}
\left\{
\begin{array}{rcl}
{\cal D}_{a}\hat{\cal M} & = & 
\partial_{a}{\cal M} -A_{(1)a}\left[ {\cal E}{\cal M} 
+{\cal M}{\cal E}^{T}\right]\, ,  \\
& & \\
{\cal D}_{y}{\cal M} & = & 
K^{3/4}\left( {\cal E}{\cal M} +{\cal M}{\cal E}^{T}\right)\, . \\
\end{array}
\right.
\end{equation}

\noindent Clearly ${\cal E}$ is going to be the mass matrix $m$.
This then means that we can write down

\begin{equation}
\left\{
\begin{array}{rcl}
{\rm Tr}
\left(
\partial\hat{\cal M} \partial\hat{\cal M}^{-1}
\right) 
& \rightarrow & 
{\rm Tr}
\left(
\partial{\cal M} \partial{\cal M}^{-1}
\right)  \\
& & \\
& &
+2A_{(1)\mu} {\rm Tr}
\left[ 
{\cal M}^{-1}\partial^{\mu}{\cal M}
\left({\cal M}^{-1}{\cal E}{\cal M} +{\cal E}^{T}\right) 
\right]\\
& & \\
& & 
+2
\left( 
K^{\frac{3}{2}} - A_{(1)}^{2}
\right)\, 
{\rm Tr}
\left({\cal M}^{-1}{\cal E} {\cal M}{\cal E}^{T} +{\cal E}^{2}\right)
\end{array}
\right.
\end{equation}

\noindent One will readily acknowledge that this is exactly the
result found in Section~\ref{sec-reduction} with ${\cal E}=m$.

Decomposing $\hat{\vec{B}}$ as 

\begin{equation}
\hat{\vec{B}} \;=\; \vec{A}_{(2)}\;-\; \vec{A}_{(1)}d\underline{y}\; ,
\end{equation}

\noindent one finds that the reduction of $\hat{\vec{H}}$ leads to

\begin{equation}
\left\{
\begin{array}{rcl}
\hat{\vec{H}} & = &  \vec{F}_{(3)} \, - \,
K^{\frac{3}{4}}\vec{F}_{(2)}dy \; , \\
& & \\
\vec{F}_{(2)} &=& d\vec{A}_{(1)} \;-\; {\cal E}\vec{A}_{(2)} \; ,  \\
& & \\
\vec{F}_{(3)} &=& d\vec{A}_{(2)} \;+\; A_{(1)}\vec{F}_{(2)} \; .\\
\end{array}
\right.
\end{equation}

\noindent This then allows us to reduce the $\hat{\vec{H}}$ 
term in the action as

\begin{equation}
\int_{10}\hat{\vec{H}}{}^{T}{\cal M}^{-1}\ {}^{\star}\hat{\vec{H}} \;=\; 
\int_{9} \left[ K^{-\frac{3}{4}} \vec{F}_{(3)}^{\ T}{\cal M}^{-1}\ 
{}^{\star} \vec{F}_{(3)} -K^{\frac{3}{4}} \vec{F}_{(2)}^{\ T}{\cal M}^{-1}\ 
{}^{\star}\vec{F}_{(2)} \right]\, .
\end{equation}

\noindent Doing the same thing on the 5-form field strength, we find
that

\begin{equation}
\int_{10 }\hat{F}_{(5)}{}^{\star}\hat{F}_{(5)} \;=\; 
\int_{9}\left[K^{-\frac{3}{4}} F_{(5)}{}^{\star}F_{(5)} 
-K^{\frac{3}{4}}F_{(4)}{}^{\star}F_{(4)}\right] \; ,
\end{equation}

\noindent where we have used

\begin{equation}
\left\{
\begin{array}{rcl}
\hat{D} & =& A_{(4)} \;-\; A_{(3)}d\underline{y} \; ,\\
    & &           \\
F_{(4)} &=& dA_{(3)} 
        \,+\,\frac{1}{2}\vec{A}_{(1)}^{\ T}\eta\vec{F}_{(3)}
        \,-\, \frac{1}{2}\vec{A}_{(2)}^{\ T}\eta\vec{F}_{(2)}
        \,+\, \frac{1}{2}A_{(1)}\vec{A}_{(1)}^{\ T}\eta\vec{F}_{(2)} \; , \\
    & &           \\
F_{(5)} &=& dA_{(4)} 
        \,+\, A_{(1)}F_{(4)} 
        \,+\, \frac{1}{2}A_{(1)}\vec{A}_{(2)}^{\ T}\eta\vec{F}_{(2)}
        \,-\, \frac{1}{2}\vec{A}_{(2)}^{\ T}\eta\vec{F}_{(3)} \, .
\end{array}
\right.
\end{equation}

\noindent Now, reducing the CS-term and dualizing the $d=9$
5-form field strength we end up with the following contribution to the
$d=9$ action

\begin{equation}
\begin{array}{rcl}
S_{(4)} & = & \int_{9}
\left\{
     -\frac{1}{2}K^{\frac{3}{4}}F_{(4)}{}^{\star}F_{(4)} 
     -\frac{1}{2}F_{(4)}F_{(4)}A_{(1)}
     +\frac{1}{2}F_{(4)}\vec{A}_{(2)}^{\ T}\eta
     \left(
         \vec{F}_{(3)}-A_{(1)}\vec{F}_{(2)}
     \right)
\right. \\
& & \\
&+& \frac{1}{8}
     \left. 
       \left[
          \vec{A}_{(2)}^{\ T}\eta\vec{F}_{(2)}
          -A_{(1)}^{\ T}\eta
           \left(
              \vec{F}_{(3)}-A_{(1)}\vec{F}_{(2)}
           \right)
       \right]
       \vec{A}_{(2)}^{\ T}\eta
       \left(
          \vec{F}_{(3)}-A_{(1)}\vec{F}_{(2)}
       \right)
     \right\} \, .
\end{array}
\end{equation}

\noindent Comparing the above results with the results in
Eq.~(\ref{eq:d9actionsl2rpotentials}) one can see that both ways of
reducing lead to the same thing.


\subsection{Derivation of the massive transformations}

Before the gauging, in $d=10$, we have the invariance

\begin{equation}
\delta \hat{\vec{B}} \;=\; d\hat{\vec{\cal N}} \: , 
\end{equation}

\noindent and we want to find the effect of these transformations after
the gauging and the reduction: These will turn out to be related to
{\it some of} the massive transformations.

When gauging the action, we have to covariantize the corresponding
transformations.
Since the $SL(2,\mathbb{R})$ acts on the $\hat{\vec{B}}$ fields,
it is only natural to introduce the covariantized transformation rules

\begin{equation}
\delta \hat{\vec{B}} \;=\; d\hat{\vec{\cal N}} 
                     \;\rightarrow\; \delta \hat{\vec{B}} \,=\, 
{\cal D}\hat{\vec{\cal N}} \,=\, d\hat{\vec{\cal N}} 
                           \,+\, \hat{\cal E}\wedge\hat{\vec{\cal N}} \; ,
\end{equation}

\noindent under which the field strength for the $\hat{\vec{B}}$ 
field transforms as

\begin{equation}
 \delta \hat{\vec{H}} \;=\; F(\hat{\cal E})\wedge \hat{\vec{\cal N}} \; ,
\end{equation}

\noindent where we have defined $F(\hat{\cal E}) =d\hat{\cal E}+
\hat{\cal E}\wedge\hat{\cal E}$.  This looks worse than it actually
is: Since we take the gauge field to be constant and in one direction
only, the field strength for the gauge field $\hat{\cal E}$ is
identically zero, rendering the variation for $\hat{\vec{H}}$ nil.

Splitting the $\hat{\vec{B}}$ fields then as before, and defining

\begin{equation}
\hat{\vec{\cal N}} \;=\; \vec{\Sigma}_{(1)} 
             \,-\, \vec{\Sigma}_{(0)}d\underline{y}\; , 
\end{equation}

\noindent one finds the following massive transformations 

\begin{equation}
\left\{
\begin{array}{rcl}
\delta\vec{A}_{(2)} &=& d\vec{\Sigma}_{(1)} \; , \\
                    & &    \\
\delta\vec{A}_{(1)} &=& d\vec{\Sigma}_{(0)} \;+\;
                        {\cal E}\vec{\Sigma}_{(1)} \, .
\end{array}
\right.
\end{equation}

\noindent One can then see that the field strengths for the $d=9$
fields $\vec{A}_{(2)}$ and $\vec{A}_{(1)}$ are indeed invariant under
these transformations, and are $SL(2,\mathbb{R})$ invariant.

Under the $d=10$ transformation $\delta\hat{\vec{B}}= d\hat{\vec{\cal
    N}}$ one finds that

\begin{equation}
\delta \hat{F}_{(5)} \;=\; d \delta \hat{D}
-{\textstyle\frac{1}{2}}(\delta \hat{\vec{B}})^{T}\eta \hat{\vec{H}} \; ,
\end{equation}

\noindent because $\hat{\vec{H}}$ is invariant. Now, using the facts

\begin{equation}
d\hat{\cal E}=0 \; ,\; 
\hat{\cal E}\wedge \hat{\cal E}=0 \; ,\; 
(\hat{\cal E}\wedge \hat{\vec{\cal N}}){}^{T}
       = - \hat{\vec{\cal N}}{}^{T}\wedge \hat{\cal E}^{\ T} \; ,\;
\hat{\cal E}^{\ T}\eta \,=\, -\eta \hat{\cal E} \; ,
\end{equation}

\noindent one finds that the variation reads

\begin{equation}
\delta \hat{F}_{(5)} \;=\; d \delta \hat{D}
-{\textstyle\frac{1}{2}}d
\left( 
\hat{\vec{\cal N}}{}^{\ T}\eta \hat{\vec{H}}
\right) \; .
\end{equation}

\noindent This then means that iff

\begin{equation}
\delta \hat{D} \;=\; d\hat{\Delta}^{(3)} +{\textstyle\frac{1}{2}}
\hat{\vec{\cal N}}{}^{\ T}\eta \hat{\vec{H}} \; ,
\end{equation}

\noindent the 5-form field strength is invariant.

Dimensional reduction of the above transformation rule, leads to the
variation rule for the 3-form, i.e.

\begin{equation}
\delta A_{(3)} \;=\; d\Delta^{(2)}
+{\textstyle\frac{1}{2}}\vec{\Sigma}_{(1)}^{\ T}\eta\vec{F}_{(2)}
-{\textstyle\frac{1}{2}}\vec{\Sigma}_{(0)}{}^{T}\eta
       \left[
          \vec{F}_{(3)}-A_{(1)}\vec{F}_{(2)}
       \right] \; .
\end{equation}

\noindent Clearly, these transformations correspond to the non-$\chi$
transformations found in the preceding subsection.


\section{11-Dimensional Origin of $N=2,d=9$ Massive Supergravities}
\label{sec-11}

In this Section we construct an 11-dimensional action which, upon
dimensional reduction (zero-mode compactification) over a 2-torus
gives the massive 9-dimensional type~II supergravity action
Eq.~(\ref{eq:d9actionsl2rpotentials}). In
Section~\ref{sec-generalized} it was important for us to keep
$SL(2,\mathbb{R})$-covariance throughout the dimensional reduction and
as a result we got a general action which describes a 3-parameter
family of massive 9-dimensional type~II supergravities. The three mass
parameters transform in the adjoint representation of
$SL(2,\mathbb{R})$ and thus, an $SL(2,\mathbb{R})$ transformation
takes us from one member of the family (a supergravity theory) to
another one.

Thus, in order to make contact with that result from an 11-dimensional
(that is, from a type~IIA/M-theoretical) starting point, it is
important to have full control over the $SL(2,\mathbb{R})\subset
GL(2,\mathbb{R})$ symmetry that arises in the dimensional reduction in
two dimensions.
This symmetry in the type~IIA side exactly corresponds to the
S~duality of the type~IIB side \cite{kn:BHO,kn:BJO,kn:Be,kn:As}. Thus,
we will first reduce standard 11-dimensional supergravity making this
symmetry manifest.


\subsection{Compactification of 11-Dimensional Supergravity on $T^{2}$
 and $Sl(2,\mathbb{R})$ Symmetry}
\label{sec-toro}

The bosonic fields of $N=1,d=11$ supergravity \cite{kn:CJS} are the
Elfbein and a 3-form potential

\begin{equation}
\left\{\hat{\hat{e}}_{\hat{\hat{\mu}}}{}^{\hat{\hat{a}}},
\hat{\hat{C}}_{\hat{\hat{\mu}}\hat{\hat{\nu}}\hat{\hat{\rho}}}
\right\}\, .
\end{equation}

The field strength of the 3-form is

\begin{equation}
\hat{\hat{G}} = 4\partial \hat{\hat{C}}\, ,
\end{equation}

\noindent and is obviously invariant under the gauge transformations

\begin{equation}
\label{eq:3formgauge}
\delta \hat{\hat{C}}= 3\partial\hat{\hat{\chi}}\, ,
\end{equation}

\noindent where $\hat{\hat{\chi}}$ is a 2-form. 
The action for these bosonic fields is

\begin{equation}
\hat{\hat{S}}= \int d^{11}x\
\sqrt{|\hat{\hat{g}}|}\ \left[\hat{\hat{R}}
-{\textstyle\frac{1}{2\cdot 4!}} \hat{\hat{G}}{}^{2}
-{\textstyle\frac{1}{6^{4}}} \frac{1}{\sqrt{|\hat{\hat{g}}}|}
\hat{\hat{\epsilon}} \partial\hat{\hat{C}}
\partial \hat{\hat{C}} \hat{\hat{C}}
\right]\, .
\label{eq:11daction}
\end{equation}

We have $2$ mutually commuting Killing vectors
$\{\hat{\hat{k}}_{(m)}{}^{\hat{\hat{\mu}}}\}$ and use coordinates
adapted to both of them: $\{\hat{\hat{x}}{}^{\hat{\hat{\mu}}}\}
=\{x^{\mu},x^{m}\}$ with $m=9,10$ and $x^{9}=x, x^{10}=z$ and

\begin{equation}
\hat{\hat{k}}_{(m)}{}^{\hat{\hat{\mu}}}
\frac{\partial}{\partial \hat{\hat{x}}{}^{\hat{\hat{\mu}}}}
=\frac{\partial}{\partial x^{m}}\, .
\end{equation}

In these coordinates 

\begin{equation}
\hat{\hat{k}}_{(m)}{}^{\hat{\hat{\mu}}}
\hat{\hat{k}}_{(n)}{}^{\hat{\hat{\nu}}}  
\hat{\hat{g}}_{\hat{\hat{\mu}}\hat{\hat{\nu}}}
= \hat{\hat{g}}_{mn}\, .
\end{equation}

This is the internal space metric and it is in general non-diagonal,
so the Killing vectors are not mutually orthogonal in general.

The standard KK Ansatz is\footnote{This is not exactly the standard KK
  Ansatz, which includes a rescaling of the lower-dimensional metric
  to end up in the Einstein conformal frame. We will perform the
  rescaling as a second step for pedagogical reasons.}

\begin{equation}
\left( \hat{\hat{e}}_{\hat{\hat{\mu}}}{}^{\hat{\hat{a}}} \right) = 
\left(
\begin{array}{cr}
e_{\mu}{}^{a} & e_{m}{}^{i}A^{(m)}{}_{\mu} \\
&\\
0             & e_{m}{}^{i}                \\
\end{array}
\right)
\, , 
\hspace{1cm}
\left(\hat{\hat{e}}_{\hat{\hat{a}}}{}^{\hat{\hat{\mu}}} \right) =
\left(
\begin{array}{cr}
e_{a}{}^{\mu} & -A^{(m)}{}_{a}   \\
& \\
0             &  e_{i}{}^{m}  \\
\end{array}
\right)\, , 
\label{eq:elfbein}
\end{equation}

\noindent where $A^{(m)}{}_{a}=e_{a}{}^{\mu}A^{(m)}{}_{\mu}$. 
For the metric, this means the following decomposition in
9-dimensional fields:

\begin{equation}
\left\{
\begin{array}{rcl}
\hat{\hat{g}}_{\mu\nu} & = & g_{\mu\nu} 
+G_{mn} A^{(m)}{}_{\mu}A^{(n)}{}_{\nu}\, ,\\
& & \\
\hat{\hat{g}}_{\mu m} & = & G_{mn}A^{(n)}{}_{\mu} 
=\hat{\hat{k}}_{(m)\ \mu}\, ,\\
& & \\
\hat{\hat{g}}_{mn} & = & G_{mn} 
=\hat{\hat{k}}_{(m)}{}^{\hat{\hat{\mu}}}
\hat{\hat{k}}_{(n)\ \hat{\hat{\mu}}}\, .\\
\end{array}
\right.
\end{equation}

The inverse relations are given in Appendix~\ref{sec-9E-10S-IIA}.

From now on we will write the internal metric in matrix form and the
two KK vectors in a column vector form:

\begin{equation}
G \equiv 
\left(  
\begin{array}{cc}
G_{\underline{xx}} & G_{\underline{xz}} \\
G_{\underline{zx}} & G_{\underline{zz}} \\
\end{array}
\right)\, ,
\hspace{1cm}
\vec{A}_{\mu} \equiv 
\left( 
\begin{array}{c}
A^{(\underline{x})}{}_{\mu}\\
\\
A^{(\underline{z})}{}_{\mu}\\
 \end{array}
\right)\, .  
\end{equation}

Under global transformations in the internal space

\begin{equation}
x^{m\ \prime} = \left(R^{-1\ T}\right)^{m}{}_{n}\ x^{n} +a^{m}\, ,
\hspace{1cm}
R\in GL(2,\mathbb{R})\, ,
\end{equation}

\noindent objects with internal space indices transform as follows:

\begin{equation}
G^{\prime}= RGR^{T}\, ,  
\hspace{1cm}
\vec{A}^{\prime}_{\mu} = (R^{-1})^{T} \vec{A}_{\mu}\, .
\end{equation}

We know that $GL(2,\mathbb{R})$ can be decomposed in
$SL(2,\mathbb{R})\times \mathbb{R}^{+}\times\mathbb{Z}_{2}$ and any
matrix $R$ can therefore be decomposed into

\begin{equation}
R= a\Lambda (\sigma^{1})^{\alpha}\, ,
\hspace{.5cm}
\Lambda \in SL(2,\mathbb{R})\, ,
\hspace{.5cm}
\sigma^{1} =
\left(
\begin{array}{cc}
0 & 1  \\
1 & 0  \\
\end{array}
\right)\, ,
\hspace{.5cm}
\alpha=0,1\, ,
\hspace{.5cm}
a\in \mathbb{R}^{+}\, .
\end{equation}

The effect of a $\mathbb{Z}_{2}$ transformation $\sigma^{1}$ is the
relabeling of the two internal coordinates and we will ignore it.
Thus, we will focus on $GL(2,\mathbb{R})/\mathbb{Z}_{2}\sim
SL(2,\mathbb{R})\times \mathbb{R}^{+}$.  We want to separate fields
that transform under the different factors. First we define the
symmetric $SL(2,\mathbb{R})$ matrix\footnote{The minus sign is due to
our mostly minus signature which makes the internal metric negative
definite. We want ${\cal M}$ to be positive definite.}

\begin{equation}
{\cal M} = -G/| {\rm det}\ G |^{1/2}\, ,
\label{eq:defM}\end{equation}

\noindent and the scalar

\begin{equation}
K= |{\rm det}\ G|^{1/2}\, .
\end{equation}

Now, under $SL(2,\mathbb{R})$ only ${\cal M}$ and $\vec{A}_{\mu}$
transform:

\begin{equation}
{\cal M}^{\prime} = \Lambda {\cal M} \Lambda^{T}\, ,  
\hspace{.5cm}
\vec{A}_{\mu}^{\prime} = (\Lambda^{-1})^{T}\vec{A}_{\mu}\, ,
\end{equation}

\noindent that is, $\vec{A}_{\mu}$ transforms contravariantly,
while under $\mathbb{R}^{+}$ rescalings only $K$ and $\vec{A}_{\mu}$
transform:

\begin{equation}
K^{\prime} = a K\, ,
\hspace{.5cm}
\vec{A}_{\mu}^{\prime} = a \vec{A}_{\mu}\, .
\end{equation}

It is convenient for our purposes to use a slightly different set of
vector fields $\vec{A}_{(1)\ \mu}$ transforming covariantly under
$SL(2,\mathbb{R})$, defined as follows:

\begin{equation}
\vec{A}_{(1)\ \mu} = \eta \vec{A}_{\mu}\, ,
\hspace{.5cm}
\vec{F}_{(2)\ \mu\nu} = 2\partial_{[\mu}\vec{A}_{(1)\ \nu]}\, ,
\hspace{.5cm}
\vec{A}_{(1)\ \mu}^{\prime} = a\Lambda \vec{A}_{(1)\ \mu}\, .
\label{eq:defA1}\end{equation}

Using the standard techniques, the above Elfbein Ansatz and rescaling
the resulting 9-dimensional metric to the Einstein frame

\begin{equation}
g_{\mu\nu} =K^{-2/7}g_{E\ \mu\nu}\, ,  
\label{eq:masscale}\end{equation}

\noindent one finds

\begin{equation}
\label{eq:curvature}
\begin{array}{rcl}
\int d^{11}\hat{\hat{x}}\ \sqrt{|\hat{\hat{g}}|}\ 
\left[\ \hat{\hat{R}}\ \right] & = & 
\int d^{9}x\ \sqrt{|g_{E}|}\
\left[
R_{E} +\frac{9}{14}\left(\partial\log{K} \right)^{2}
\right.\\
& & \\
& & 
\left.
+{\textstyle\frac{1}{4}} {\rm Tr} 
\left(\partial {\cal M} {\cal M}^{-1}\right)^{2} 
-{\textstyle\frac{1}{4}} K^{\frac{9}{7}} 
\vec{F}_{(2)}^{\ T}{\cal M}^{-1}\vec{F}_{(2)}
\right]\, . 
\end{array}
\end{equation}

The 3-form term can be reduced along the same lines and we decompose the
11-dimensional 3-form potential into the 9-dimensional fields $A_{(3)\ 
  \mu\nu\rho}, \vec{A}_{(2)\ \mu\nu}$ and $A_{(1)\ \mu}$ as follows:

\begin{equation}
\left\{
\begin{array}{rcl}
\hat{\hat{C}}_{\mu \nu \rho} & = &  A_{(3)\ \mu\nu\rho}
+\frac{3}{2}\vec{A}_{(1)\ [\mu}^{\ T}\eta \vec{A}_{(2)\ \nu\rho]}
+3A_{(1)\ [\mu}\vec{A}_{(1)\ \nu}^{\ T}\eta\vec{A}_{(1) \rho]}\, ,\\
& & \\
\left( 
\begin{array}{c}
\hat{\hat{C}}_{\mu\nu \underline{x}} \\
\hat{\hat{C}}_{\mu\nu \underline{z}} \\
\end{array}
\right) 
& = & \vec{A}_{(2)\ \mu\nu}
-2A_{(1)\ [\mu}\vec{A}_{(1)\ \nu]}\, ,\\
& & \\
\left(
\begin{array}{cc}
0 & 
\hat{\hat{C}}_{\mu \underline{x}\underline{z}}  \\
& \\
\hat{\hat{C}}_{\mu \underline{z}\underline{x} } & 0     \\
\end{array}
\right) 
& = & +\eta A_{(1)\ \mu}\, ,\\
\end{array}
\right.
\end{equation}
%

The corresponding 9-dimensional field strengths $F_{(4)},
\vec{F}_{(3)}$ and $F_{(2)}$ are defined exactly by the massless limit
of Eq.~(\ref{eq:fieldstrengths}). The relation with the 11-dimensional
field strength $\hat{\hat{G}}$ is



\begin{equation}
\left\{
\begin{array}{rcl}
\hat{\hat{G}}_{\mu\nu \rho \sigma} & = & 
F_{(4)\ \mu\nu\rho\sigma} 
-4\vec{A}_{(1)\ [\mu}^{\ T}\eta \vec{F}_{(3)\ \nu\rho\sigma]} \\
& & \\
& & 
+5 \vec{A}_{(1)\ [\mu}^{\ T}\eta \vec{A}_{(1)\ \nu} F_{(2)\ \rho\sigma]}\, ,\\
& & \\
\left(
\begin{array}{c}
\hat{\hat{G}}_{\mu\nu \rho \underline{x}} \\
\hat{\hat{G}}_{\mu\nu \rho \underline{z}} \\
\end{array}
\right) 
& = & 
\vec{F}_{(3)\ \mu\nu\rho}
-3\vec{A}_{(1)\ [\mu}F_{(2)\ \nu\rho]}\, ,\\
& & \\
\left( 
  \begin{array}{cc}
0  & \hat{\hat{G}}_{\mu\nu \underline{x}\underline{z}}  \\
\hat{\hat{G}}_{\mu\nu\underline{z}\underline{x}} & 0   \\
\end{array}
\right) 
& = & \eta F_{(2)\ \mu\nu} \, .\\
& & \\
\end{array}
\right.
\end{equation}

This allows us to decompose the kinetic term as follows:

\begin{equation}
\label{eq:kinetic}
\sqrt{|\hat{\hat{g}}|}
{\textstyle\frac{-1}{2\cdot 4!}} \hat{\hat{G}}{}^{2} =
\sqrt{|g_{E}|} 
\left\{ 
{\textstyle\frac{-1}{2\cdot 4!}} K^{6/7} F_{(4)}^{2}
+{\textstyle\frac{1}{2\cdot 3!}}K^{-3/7} 
\vec{F}_{(3)}^{\ T} {\cal M}^{-1}\vec{F}_{(3)}
-{\textstyle\frac{1}{4}} K^{-12/7}F_{(2)}^{2}
\right\}\, .
\end{equation}

\noindent and the topological term as follows:

\begin{equation}
\label{eq:topolo}
\begin{array}{rcl}
{\textstyle\frac{1}{(144)^{2}}} \
\hat{\hat{\epsilon}} \hat{\hat{G}} \hat{\hat{G}} \hat{\hat{C}} 
& = & 
\frac{1}{3^{2}\cdot 2^{8}}\epsilon \epsilon^{mn}
\left\{
\hat{\hat{G}} \hat{\hat{G}} \hat{\hat{C}}_{mn} 
+4\hat{\hat{G}} \hat{\hat{G}}_{m} \hat{\hat{C}}_{n}
\right\}
\\
& & \\
& = & 
\frac{1}{3^{2}\cdot 2^{7}}\epsilon 
\left[F_{(4)} -4\vec{A}_{(1)}{}^{T}\eta \vec{F}_{(3)} 
+6\vec{A}_{(1)}{}^{T}\eta \vec{A}_{(1)} F_{(2)} \right]\times \\
& & \\
& & 
\times
\left\{ 
\left[ F_{(4)} -4\vec{A}_{(1)}{}^{T}\eta \vec{F}_{(3)} 
+6\vec{A}_{(1)}{}^{T}\eta \vec{A}_{(1)} F_{(2)} \right] A_{(1)}
\right.
\\
& & \\
& & 
\left.
+2 \left[\vec{F}_{(3)} -3\vec{A}_{(1)}F_{(2)}\right]^{T}
\eta \left[\vec{A}_{(2)} +2\vec{A}_{(1)}A_{(1)}\right]
\right\}\, . \\
\end{array}
\end{equation}

Putting all our partial results together,
Eqs.~(\ref{eq:curvature},\ref{eq:kinetic},\ref{eq:topolo}), we arrive
at the action of type~II 9-dimensional supergravity in Einstein frame,
Eq.~(\ref{eq:d9actionsl2rpotentials}), which we obtained through
generalized dimensional reduction of the 10-dimensional type~IIB
theory with the mass matrix set to zero \cite{kn:BHO}.

The fact that upon dimensional reduction the type~IIA and type~IIB
{\it supergravity theories} are identical in nine dimensions is
nothing but the manifestation at the level of the massless modes of
the T~duality existing between the type~IIA and type~IIB {\it
superstring theories} when they are compactified in circles of dual
radii \cite{kn:DLP,kn:DHS}.

There are four important points we would like to stress:

\begin{enumerate}

\item There is no ``hidden symmetry'' of the 9-dimensional type~II
theory corresponding to this T~duality.
 
\item To obtain two {\it identical} actions it is crucial that the two
topological terms come with the same global sign. In the M/type~IIA
side the sign can be changed by the 11-dimensional transformation
$\hat{\hat{C}}\rightarrow -\hat{\hat{C}}$ which is not a symmetry.  In
the type~IIB side, flipping the sign of the 4-form $\hat{D}$ does not
work because it changes the definition of its field strength.
Changing the signs of $\hat{D}$ and, say, $\hat{B}^{(1)}$ leaves
$\hat{F}$ invariant but also leaves invariant the topological term.
Thus, at first sight, there seems to be no~IIB-side version of this
rather trivial M/IIA-side transformation.
  
  It is, however, easy to see that the sign of the topological term in
  the NSD 10-dimensional type~IIB action is directly related to the
  self-duality of the 5-form field strength. Had we considered an
  anti-self-dual 5-form the sign would have been exactly the opposite
  in ten and nine dimensions. The (anti-) self-duality of the 5-form
  is related to the chirality of the theory.
  
  The picture that emerges is therefore the following: There are two
  (otherwise equivalent) 11-dimensional supergravity theories and two
  10-dimensional type~IIA theories that differ only in the sign of the
  action's topological term. Upon dimensional reduction to nine
  dimensions they are related to the two type~IIB theories of opposite
  chiralities.
  
  In the decompactification limit, each of these two 9-dimensional
  (and, thus, non-chiral) theories knows to which chiral
  10-dimensional type~IIB theory it should decompactify.

\item The above observation solves in part the puzzle found in
Ref.~\cite{kn:O2} where it was argued that approximately half of all
extreme black holes are not supersymmetric in type~II theories.
Clearly, those which are not supersymmetric in one of the
11-dimensional supergravities are supersymmetric in the 11-dimensional
supergravity with the sign of the 3-form $\hat{\hat{C}}$ reversed. As
suggested also in Ref.~\cite{kn:KhO1}, the whole picture begs for both
11-dimensional supergravities to be integrated into a
higher-dimensional supergravity from which also the type~IIB would be
derivable, perhaps one of those with the algebras studied in
Ref.~\cite{kn:Ba}. (This argument is completely different from the one
in Ref.~\cite{kn:Ba2} and, in fact, it is in disagreement with it).
  
\item The fact that the two theories (A and B) are identical allows us
to relate the 10-dimensional fields of the two type~II theories.  This
relation provides a generalization of Buscher's T~duality rules
\cite{kn:Bu}.  These type~II Buscher rules were found in
Ref.~\cite{kn:BHO} and they are determined again in
Appendices~\ref{sec-9E-10S-IIA}, \ref{sec-9E-10S-IIB} and
\ref{sec-tdual} in our (more systematic) conventions and extended to
the massive case at hands.

\end{enumerate}


\subsection{$Sl(2,\mathbb{R})$-Covariant Massive 11-Dimensional Supergravity}

So much for the massless case. Now, it is clear that the picture seems
to break down whenever the mass matrix does not vanish. In
Ref.~\cite{kn:BRGPT} the particular case with mass matrix with
$m^{1}=0, m^{2}=m^{3}=m$

\begin{equation}
m_{\rm BRGPT} =
\left(
\begin{array}{cc}
0 & m \\
0 & 0 \\
\end{array}
\right)\, ,
\end{equation}

\noindent was considered. As will be discussed in
Section~\ref{sec-pq-7-branes} this particular choice of mass matrix
corresponds to compactification of the type~IIB on a background with
different species of 7-branes. Since the T~dual of a D-7-brane in a
direction orthogonal to its worldvolume is a type~IIA  D-8-brane, one
expects the theory with mass matrix $m_{\rm BRGPT}$ to correspond to the
type~IIA theory on a background with D-8-branes.

While it is not possible to write the 10-dimensional type~IIB theory
in presence of D-7-branes in a covariant fashion (there is dependence
on the compactifying coordinate $y$) it is possible to write in a
covariant fashion the action for the type~IIA theory in presence of
D-8-branes. As was first first realized in Ref.~\cite{kn:PW}, this
theory has long been known as Romans' massive type~IIA supergravity
\cite{kn:Ro2}. The precise identification, leading to a further
generalization of Buscher's rules was carried out in
Ref.~\cite{kn:BRGPT}. We stress that these T~duality rules are
essentially identical to the original type~II T~duality rules of
Ref.~\cite{kn:BHO} but are deformed in a $y$-dependent fashion in the
type~IIB side of the equations.

Our task in the remainder of this Section will be to generalize the
results of Ref.~\cite{kn:BRGPT}. It is clear from the setting that
this generalization amounts to its
$SL(2,\mathbb{R})$-covariantization: We start from the
compactification of the type~IIB theory on a background containing
D-7-branes and their S~duals and, after T~duality, we expect to find a
type~IIA theory on a background of the T~duals of D-7-branes and their
S~duals.  We will not repeat here the discussion of the Introduction
where we concluded that we must look for a non-covariant
generalization of Romans' type~IIA supergravity.

As a matter of fact, it is easier to generalize the 11-dimensional
theory that gives Romans', given in Ref.~\cite{kn:BLO}. From our point
of view this theory would correspond to 11-dimensional supergravity
with a KK-9M-brane in the background. To find in 9-dimensions an
$SL(2,\mathbb{R})$-covariant result we must consider a theory
describing 11-dimensional supergravity with two KK-9M-branes in the
background.

In what follows we will construct such a theory along the same lines
as Ref.~\cite{kn:BLO} and show that it gives the massive 9-dimensional
type~II theory constructed in Section~\ref{sec-generalized}.

Since each KK-9M-brane is associated to a Killing vector we assume the
presence of the two mutually commuting Killing vectors of the previous
Section and also assume that the Lie derivatives of all fields with
respect to both of them vanishes.

Next, we define the 11-dimensional massive transformations. For a
general tensor, except for $\hat{\hat{C}}$ whose transformation law
will be defined below, they are

\begin{equation}
\delta_{\hat{\hat{\chi}}}L_{\hat{\hat{\mu}}_{1}
\ldots\hat{\hat{\mu}}_{r}}\;=\;
\hat{\hat{\lambda}}{}^{(n)}{}_{\hat{\hat{\mu}}_{1}} 
\hat{\hat{k}}_{(n)}{}^{\hat{\hat{\nu}}}
\hat{\hat{L}}_{\hat{\hat{\nu}}\hat{\hat{\mu}}_{2}
\ldots\hat{\hat{\mu}}_{r}} 
 \, +\ldots\, +\,
\hat{\hat{\lambda}}{}^{(n)}{}_{\hat{\hat{\mu}}_{r}} 
\hat{\hat{k}}_{(n)}{}^{\hat{\hat{\nu}}}
\hat{\hat{L}}_{\hat{\hat{\mu}}_{1}\ldots
\hat{\hat{\mu}}_{\scriptscriptstyle r-1}\hat{\hat{\nu}}} \; ,
\label{eq:masvar.1}
\end{equation}

\noindent where we have defined

\begin{equation}
\hat{\hat{\lambda}}{}^{(n)} 
\equiv
-i_{\hat{\hat{k}}_{\scriptscriptstyle (m)}}\hat{\hat{\chi}} 
Q^{nm} \, ,
\hspace{.5cm}
Q^{nm}=\left( m^{T}\eta\right)^{mn}=
{\textstyle\frac{1}{2}}
\left(
\begin{array}{cc}
-(m^{2}+m^{3}) & m^{1}        \\
& \\
m^{1}          & m^{2} -m^{3} \\
\end{array}
\right)\, .
\label{eq:masvar.2}
\end{equation}

The contraction of a space tensor with the Killing vectors will bear
an $SL(2, \mathbb{R})$ index: The extension of the above rule for
incorporating $SL(2,\mathbb{R})$ indices is found by defining the
inclusion to commute with the massive transformations.

In particular we find that the 11-dimensional metric and $r$-forms
$\hat{\hat{S}}$ transform as

\begin{equation}
\left\{\begin{array}{rcl}
\delta_{\hat{\hat{\chi}}}
\hat{\hat{g}}_{\hat{\hat{\mu}}\hat{\hat{\nu}}} & = &
2\hat{\hat{\lambda}}{}^{(n)}{}_{(\hat{\hat{\mu}}}\,  
\hat{\hat{k}}_{(n)}{}^{\hat{\hat{\rho}}}
\hat{\hat{g}}_{\hat{\hat{\nu}})\hat{\hat{\rho}}} \; ,\\
& & \\
\delta_{\hat{\hat{\chi}}}
\hat{\hat{S}}_{\hat{\hat{\mu}}_{1}\ldots\hat{\hat{\mu}}_{r}} &=&
(-)^{r-1}r\hat{\hat{\lambda}}{}^{(n)}{}_{[\hat{\hat{\mu}}_{1}}\,
\hat{\hat{k}}_{(n)}{}^{\hat{\hat{\rho}}}
\hat{\hat{S}}_{\hat{\hat{\mu}}_{2}\ldots\hat{\hat{\mu}}_{r}]
\hat{\hat{\rho}}}\; .
\end{array}
\right.
\label{eq:masvar.3}
\end{equation}
        
\noindent Observe that these rules imply that 

\begin{equation}
\left\{
\begin{array}{lcl}
\delta_{\hat{\hat{\chi}}} \sqrt{|\hat{\hat{g}}|} 
& = & 0 \; , \\
& &        \\
\delta_{\hat{\hat{\chi}}} \hat{\hat{S}}{}^{2}     
& = & 0 \; ,
\end{array}
\right.
\label{eq:masvar.4}
\end{equation}

\noindent where the latter holds due to the fact that also the metric
varies under the massive transformations, and the former holds due to
the fact that the matrix $Q=m^{T}\eta$ is symmetric.

The 3-form field $\hat{\hat{C}}$ is going to play the role of a
connection-field with respect to the massive transformations and, as
such, does not transform covariantly

\begin{equation}
\delta_{\hat{\hat{\chi}}} \hat{\hat{C}} \;=\; d\hat{\hat{\chi}} \;+\,
\hat{\hat{\lambda}}{}^{(n)}\wedge
\left(i_{\hat{\hat{k}}_{\scriptscriptstyle (n)}}\hat{\hat{C}}\right) \; .
\label{eq:masvar.5}
\end{equation}

\noindent The generalization of the field strength for
$\hat{\hat{C}}$, denoted as before by $\hat{\hat{G}}$, is then found
by requiring that the field strength does transform covariantly.  One
can see that this implies that

\begin{equation}
\hat{\hat{G}} \;=\; d\hat{\hat{C}} \;-\; 
{\textstyle\frac{1}{2}}
\left(i_{\hat{\hat{k}}_{\scriptscriptstyle (n)}}\hat{\hat{C}}\right) \, 
Q^{nm} \,
\left(i_{\hat{\hat{k}}_{\scriptscriptstyle (n)}}\hat{\hat{C}}\right) \; .
\label{eq:masvar.6}
\end{equation}

\noindent Comparing this with a torsionful covariant derivative acting
on a 3-form, one sees that the above equation states that the massive
transformations induce a torsion term in our spacetime
connection. This then means that if we want our $d=11$ theory to be
invariant under the massive transformations, we have to define our
theory in terms of the torsionful connection.

The torsion we need is given by

\begin{equation}
\hat{\hat{T}}_{\hat{\hat{\mu}}\hat{\hat{\nu}}}{}^{\hat{\hat{\rho}}}
= -\left(i_{\hat{\hat{k}}_{(n)}}\hat{\hat{C}}
\right)_{\hat{\hat{\mu}}\hat{\hat{\nu}}} Q^{nm}
\hat{\hat{k}}_{(m)}{}^{\hat{\hat{\rho}}}\, .  
\end{equation}

The torsionful connection $\hat{\hat{\Omega}}$ is then defined
in the standard way, by adding the so-called contorsion-torsion tensor,

\begin{equation}
\hat{\hat{K}}_{\hat{\hat{a}}\hat{\hat{b}}\hat{\hat{c}}} 
\;=\; {\textstyle\frac{1}{2}}
\left(
\hat{\hat{T}}_{\hat{\hat{a}}\hat{\hat{c}}\hat{\hat{b}}} 
\; +\; \hat{\hat{T}}_{\hat{\hat{b}}\hat{\hat{c}}\hat{\hat{a}}} 
\; -\; \hat{\hat{T}}_{\hat{\hat{a}}\hat{\hat{b}}\hat{\hat{c}}} 
\right) \; , 
\label{eq:torsion.2}
\end{equation}

\noindent to the Levi-Civit\`{a} connection $\hat{\hat{\omega}}$, i.e.

\begin{equation}
\hat{\hat{\Omega}}_{\hat{\hat{a}}}{}^{\hat{\hat{b}}\hat{\hat{c}}} 
\;=\; \hat{\hat{\omega}}_{\hat{\hat{a}}}{}^{\hat{\hat{b}}\hat{\hat{c}}} 
\;+\; \hat{\hat{K}}_{\hat{\hat{a}}}{}^{\hat{\hat{b}}\hat{\hat{c}}} \; .
\label{eq:torsion.3}
\end{equation}

{}From the above equation we can obtain the non-vanishing components of
the torsion written directly in 9-dimensional Lorentz coordinates for
future use

\begin{equation}
\left\{
\begin{array}{rcl}
\hat{\hat{T}}_{abi} & = & 
-A_{(2)(n)ab}\eta^{np}{m_{p}}^{q}e_{qi} \; , \\
& & \\
\hat{\hat{T}}_{aij} &=& A_{(1)a} 
{e_{i}}^{p}{m_{p}}^{q}e_{qi} \; .
\end{array}
\right.
\label{eq:torsion.1}
\end{equation}

\noindent Having all this, one can see that the 11-dimensional theory
invariant under the massive transformation reads\footnote{Note that
the cosmological constant part is, apart from correspondence with the
massive $d=9$ theory, arbitrary. However, supersymmetry should
completely determine it.}

\begin{equation}
\begin{array}{rcl}
\hat{\hat{S}} 
& = &
\int d^{11}\hat{\hat{x}}\sqrt{|\hat{\hat{g}}|}\, 
\left\{
\hat{\hat{R}}\left(\hat{\hat{\Omega}}\right) 
+\left(d\hat{\hat{k}}_{(n)}\right)
   {}_{\hat{\hat{\mu}}\hat{\hat{\nu}}}Q^{nm}
   \left( i_{\hat{\hat{k}}_{(m)}}\hat{\hat{C}}\right)
   {}^{\hat{\hat{\mu}}\hat{\hat{\nu}}}
-\textstyle{\frac{1}{2\cdot 4!}} \hat{\hat{G}}{}^{2} 
\right.\\
& & \\
& & 
-2\hat{\hat{K}}_{\hat{\hat{\mu}}\hat{\hat{\nu}}\hat{\hat{\rho}}}
\hat{\hat{K}}{}^{\hat{\hat{\nu}}\hat{\hat{\rho}}\hat{\hat{\mu}}}
+\textstyle{\frac{1}{2}} 
\left( \hat{\hat{k}}_{(n)\ \hat{\hat{\mu}}}Q^{nm}
\hat{\hat{k}}_{(m)}{}^{\hat{\hat{\mu}}}\right)^{2} 
-\left(  \hat{\hat{k}}_{(n)\ \hat{\hat{\mu}}}Q^{nm} 
\hat{\hat{k}}_{(m)\ \hat{\hat{\nu}}}\right)^{2}   
\\
& & \\
& & 
-\frac{1}{6^{4}}
\frac{\hat{\hat{\epsilon}}}{\sqrt{|\hat{\hat{g}}|}}
\left\{
\partial\hat{\hat{C}}\partial\hat{\hat{C}}\hat{\hat{C}}
- \frac{9}{8}\partial\hat{\hat{C}}\hat{\hat{C}}
\left(i_{\hat{\hat{k}}_{(n)}}\hat{\hat{C}}\right)
Q^{nm} 
\left(i_{\hat{\hat{k}}_{(m)}}\hat{\hat{C}}\right)
\right. \\
& & \\
& &
\left.
\left.
+ \frac{27}{80}
\hat{\hat{C}}
\left[
\left(i_{\hat{\hat{k}}_{(n)}}\hat{\hat{C}}\right)
 Q^{nm}
\left(i_{\hat{\hat{k}}_{(m)}}\hat{\hat{C}}\right)
\right]^{2} 
\right\}
\right\}\; ,
\end{array}
\label{eq:masact.11}
\end{equation}

\noindent For the dimensional reduction of the above theory, the
fields will be split in the same way as in the preceding subsection;
The only thing that changes, is the torsion part of the connection and
some terms in the 11-dimensional Chern-Simons term.

Let us first consider the reduction of the curvature term, evaluated
using the connection in Eq.~(\ref{eq:torsion.3}). Using Palatini's
identity for torsionful connections

\begin{eqnarray}
\int_{d}\sqrt{|g|}\, e^{-2\phi}\,R(\Omega ) &=& 
 -\int_{d}\sqrt{|g|}\, e^{-2\phi}\left\{
 {\Omega_{b}}^{ba} {{\Omega_{c}}^{c}}_{a} \,+\,
 {\Omega_{a}}^{bc} {\Omega_{bc}}^{a} \,+\,
 4{\Omega_{b}}^{ba}\partial_{a}\phi
 \right. \nonumber \\
 & & \left.
 -2 {\Omega_{b}}^{ba}{{K_{c}}^{c}}_{a} \,-\,
 2{\Omega_{a}}^{bc} {K_{bc}}^{a}
 \right\} \; ,
\end{eqnarray}

\noindent the facts

\begin{equation}
\left\{
\begin{array}{rcl}
\hat{\hat{K}}_{\hat{\hat{a}}}{}^{\hat{\hat{a}}b} 
& = & A_{(1)}^{b}\hat{\hat{\eta}}^{ij}{e_{i}}^{m}e_{nj}{m_{m}}^{n} 
         \,=\, A_{(1)}^{b} {\rm Tr}\ (m) \,=\, 0 \; , \\
& & \\
\hat{\hat{K}}_{\hat{\hat{a}}}{}^{\hat{\hat{a}}i} & = & 0 \; ,
\end{array}
\right.
\end{equation}

\noindent and the fact that the second term in Eq.~(\ref{eq:masact.11})
annihilates the $\Omega K$-terms whilst applying Palatini's identity
to the case at hand, one can write

\begin{equation}
\begin{array}{rcl}
\int d^{11}\hat{\hat{x}} \sqrt{|\hat{\hat{g}}|}\left\{ 
\hat{\hat{R}}\left(\hat{\hat{\Omega}}\right) 
+ \left(d\hat{\hat{k}}_{(n)}\right)
   {}_{\hat{\hat{\mu}}\hat{\hat{\nu}}}Q^{nm}
   \left( i_{\hat{\hat{k}}_{(m)}}\hat{\hat{C}}\right)
   {}^{\hat{\hat{\mu}}\hat{\hat{\nu}}}
-2\hat{\hat{K}}_{\hat{\hat{\mu}}\hat{\hat{\nu}}\hat{\hat{\kappa}}}
  \hat{\hat{K}}_{\hat{\hat{\nu}}\hat{\hat{\kappa}}\hat{\hat{\mu}}}
\right\}
& = & \\
& & \\
& & 
\hspace{-10cm}
=\int d^{9}x\sqrt{|g|}\ K\,
\left[ 
R(g) -\left(\partial \log{K}\right)^{2}
\,+\,\textstyle{\frac{1}{4}}\left( F_{iab}+\hat{T}_{abi}\right)^{2}
\right.\\
& & \\
& & \hspace{-10cm}
\left.+\textstyle{\frac{1}{4}}\left( 
             {e_{i}}^{n}{e_{j}}^{m}\partial_{a}G_{nm}\;+\;
             2\hat{T}_{a(ij)} \right)^{2} \right)\; .
\end{array}
\label{mascurvred.1}
\end{equation}

\noindent Using now our previous partial results
Eqs.~(\ref{eq:defM},\ref{eq:defA1},\ref{eq:torsion.1}) and rescaling
to the Einstein frame, Eq.~(\ref{eq:masscale}), this can be written as
%

\begin{equation}
=\; 
\int d^{9}x\sqrt{|g_{\scriptstyle E}|}\
\left[
R(g_{\scriptstyle E}) +\textstyle{\frac{9}{14}}(\partial \log K)^{2}
\,+\,
\textstyle{\frac{1}{4}}{\rm T}r\left( {\cal DM} {\cal M}^{-1}\right)^{2}
-\textstyle{\frac{1}{4}}K^{9/7}\vec{F}_{(2)}^{\ T}
{\cal M}^{-1}\vec{F}_{(2)}
\right] \; ,
\end{equation}

\noindent where the field strengths and covariant derivative are
the same as the ones used in Section~\ref{sec-generalized}.

The cosmological constant part is readily reduced by using the 
well-known identity

\begin{equation}
\eta^{mn}\eta^{pq} \;=\; -\eta^{np}\eta^{mq} \,-\, \eta^{pm}\eta^{nq} \, ,
\end{equation}

\noindent and it follows that

\begin{equation}
\begin{array}{rcl}
\textstyle{\frac{1}{2}}\int d^{11}\hat{\hat{x}}\sqrt{|\hat{\hat{g}}|}\ 
\left[
\left(\hat{\hat{k}}_{(n)\ \hat{\hat{\mu}}}
Q^{nm}
\hat{\hat{k}}_{(m)}{}^{\hat{\hat{\mu}}}\right)^{2} 
-\left( \hat{\hat{k}}_{(n)\ \hat{\mu}}
Q^{nm} 
\hat{\hat{k}}_{(m)\ \hat{\hat{\nu}}}\right)^{2}
\right]
& = &  \\
& & \\
& & 
\hspace{-7.5cm}
=-\textstyle{\frac{1}{2}}\int d^{9}x\sqrt{|g_{\scriptstyle E}|}\
K^{12/7}{\rm Tr}\left( m^{2}
\,+\, {\cal M}m{\cal M}^{-1}m^{T}
\right) \; ,
\end{array}
\end{equation}

\noindent which is just the result obtained in the $d=9$ theory.

The effect of the torsion included in definition (\ref{eq:masvar.6}),
can readily be seen to promote the field strengths to their massive
equivalents Eq.~(\ref{eq:fieldstrengths}). As such, it will be no
surprise at all to see that

\begin{equation}
  \begin{array}{rcl}
\int_{11} -\textstyle{\frac{1}{2}} \hat{\hat{G}}{}^{\star}\hat{\hat{G}}
& = &
\int_{9} \left\{
-\textstyle{\frac{1}{2}}K^{6/7}F_{(4)}{}^{\star}F_{(4)} \, +\,
\textstyle{\frac{1}{2}}K^{-3/7} 
\vec{F}_{(3)}^{\ T}{\cal M}^{-1}{}^{\star}\vec{F}_{(3)} \right. \\
& & \\
& & \left.
-\textstyle{\frac{1}{2}} K^{-12/7} F_{(2)}{}^{\star}F_{(2)} 
\right\} \; . \\
\end{array}
\end{equation}

\noindent From the fact that we do not change the decomposition of the
fields while doing the reduction, it is clear that the
$d\hat{C}d\hat{C}\hat{C}$ will lead to the same result as in Eq.
(\ref{eq:topolo}). The other terms can easily be seen to result in

\begin{equation}
\begin{array}{rcl}
\textstyle{\frac{1}{6^{4}}}
\int_{11}\hat{\hat{\epsilon}}\,
\textstyle{\frac{9}{8}}\partial\hat{\hat{C}}\hat{\hat{C}}
\left[
\left(i_{\hat{\hat{k}}_{(n)}}\hat{\hat{C}}\right)
\, Q^{nm}\, 
\left(i_{\hat{\hat{k}}_{(m)}}\hat{\hat{C}}\right)
\right] 
& = & 
\textstyle{\frac{1}{3^{2}2^{7}}}\int_{9}\epsilon\,\left[
 6 F_{(4)}\, \vec{A}_{(2)}^{T}Q\vec{A}_{(2)}\, A_{(1)} 
\right.\\
& & \\
& & 
\left.
-9\left(\vec{A}_{(2)}^{T}Q\vec{A}_{(2)}\right)
\left( \vec{A}_{(2)}^{T}\eta \partial\vec{A}_{(2)}\right)
\right]\, , \\
 & & \\
\textstyle{\frac{1}{6^{4}}}\int_{11}\hat{\hat{\epsilon}}\,
\textstyle{\frac{27}{80}}\hat{\hat{C}}
\left[
\left(i_{\hat{\hat{k}}_{(n)}}\hat{\hat{C}}\right)
 \, Q^{nm}\, 
\left(i_{k_{(m)}}\hat{\hat{C}}\right)
\right]^{2} 
& = & 
\textstyle{\frac{1}{3^{2}2^{7}}}
\int_{9}\epsilon\, 9\left(\vec{A}_{(2)}^{T}Q\vec{A}_{(2)}\right)^{2} A_{(1)} 
\; .\\
\end{array}
\end{equation}

\noindent Adding the above equations to Eq.~(\ref{eq:topolo}) we find
that the effect of the torsion is, once again, precisely to turn the
massless CS term, into the massive CS term of the massive
9-dimensional type~II theory we got by generalized dimensional
reduction of the type~IIB theory.  Thus, we have achieved our second
goal.

The T~duality rules that one can immediately deduce from this relation
between 10-dimensional theories will be worked out in the Appendices.


\section{7-Branes}
\label{sec-pq-7-branes}

In this Section we want to identify the 10-dimensional background of
the type~IIB theory that produces the masses of the 9-dimensional
theory. The T~dual background will be dealt with in
Section~\ref{sec-KK789}.

S~duality is (believed to be) a fundamental non-perturbative symmetry
of type~IIB string theory. This implies that the full spectrum of the
theory has to be S~duality-invariant and thus all the states can be
organized in $SL(2,\mathbb{Z})$ multiplets. Thus, bound states of $q$
fundamental strings and $p$ D-strings, known as $pq$-strings,
transform as doublets under $SL(2,\mathbb{Z})$. A general solution
describing all possible $pq$-strings was constructed in
Ref.~\cite{kn:JHS2} and a dual general solution describing all
possible $pq$-5-branes was recently constructed in Ref.~\cite{kn:LR}.
The D-3-brane, being self-dual, is an $SL(2,\mathbb{Z})$ singlet. The
situation for D-9-branes and D-instantons is unclear, although one
expects to have D-9-brane solutions which only differ in the constant
value of the dilaton.

It is commonly accepted that there are bound states of $p$ D-7-branes
and $q$ NS-NS 7-branes (that we will call Q-7-branes) which transform
as doublets. As we are going to see, this is not so clear and we will
argue that 7-brane states transform as triplets. We will relate the
monodromy matrices of massive 9-dimensional type~II supergravity and
these 7-brane triplets, showing again in this way that the presence of
a background of 7-branes is the origin of the masses.

\subsection{Point-Like (in Transverse Space) 7-Branes}

The extreme D-7-brane solution in the string frame is

\begin{equation}
\left\{
\begin{array}{rcl}
ds^{2} & = & H_{D7}^{-1/2}\left[dt^{2} -d\vec{y}_{7}^{\ 2} \right]
-H_{D7}^{1/2}d\vec{x}_{2}^{\ 2}\, , \\
& & \\
e^{-2(\hat{\varphi}-\varphi_{0})} & = & H_{D7}^{2}\, ,\\
& & \\
\hat{C}^{(8)}{}_{ty^{1}\cdots y^{7}} & = & \pm e^{-\hat{\varphi}_{0}}
H_{D7}^{-1}\, ,\\
\end{array}
\right.
\end{equation}

\noindent where $\vec{y}_{7}
=\left(y^{1}_{7},y^{2}_{7},\ldots,y^{7}_{7}\right)$ are the
worldvolume coordinates and $\vec{x}_{2} =
\left(x^{1}_{2},x^{2}_{2}\right)$ are the coordinates of the
2-dimensional transverse space. Any function $H_{D7}$ harmonic in the
transverse space provides a D-7-brane-type solution. A harmonic
function $H_{D7}$ with a single point-like singularity

\begin{equation}
\label{eq:harmonic}
\partial_{x_{2}^{i}}\partial_{x_{2}^{i}} H_{D7} = 
2\pi h_{D7} \delta^{(2)} (\vec{x}_{2})\, ,
\end{equation}

\noindent describes a single D-7-brane placed at $\vec{x}_{2}=0$.  The
positive constant $h_{D7}$ is proportional to the D-7-brane charge and
mass and later on we will determine the precise relation between
them. The two possible signs of the charge are taken care of by the
$\pm$ in $\hat{C}^{(8)}$. The standard solution in $\mathbb{R}^{2}$ to
the above equation is (the additive constant is arbitrary and
momentarily we set to zero)

\begin{equation}
H_{D7} = h_{D7} \log{|\vec{x}_{2}|}\, . 
\end{equation}

The 8-form potential $\hat{C}^{(8)}$ is nothing but the dual of the RR
scalar $\hat{C}^{(0)}$ that occurs in the type~IIB theory (i.e.~their
field strengths are each other's Hodge dual
$\hat{G}^{(1)}={}^{\star}\hat{G}^{(9)}$). This dualization can only be
done ``on shell'', i.e.~using at the same time $\hat{C}^{(0)}$ and
$\hat{C}^{(8)}$ because $\hat{C}^{(0)}$ occurs explicitly in the
type~IIB action. This gives the standard form of $\hat{G}^{(9)}$
suggested in Refs.~\cite{kn:GHT,kn:BCT}.
If we ignore all other fields apart from $\hat{\lambda}$ both
dualizations are equivalent. Using this relation we find

\begin{equation}
\partial_{i} \hat{C}^{(0)} = \pm 
e^{-\hat{\varphi}_{0}}\epsilon_{ij}\partial_{j} H_{D7}\, ,
\end{equation}

\noindent and we can rewrite the solution in terms of just the 
metric and the two real scalars $\hat{C}^{(0)},e^{-\hat{\varphi}}$
that we combine into the single complex scalar $\hat{\lambda}
=\hat{C}^{(0)} +ie^{-\hat{\varphi}}$. For the single D-7-brane we find

\begin{equation}
\hat{\lambda} =
\left\{
\begin{array}{c}
ie^{-\hat{\varphi}_{0}} h_{D7}\log{\omega}\, , \\
\\
ie^{-\hat{\varphi}_{0}} h_{D7}\log{\overline{\omega}}\, , \\
\end{array}
\right.
\hspace{1cm}
\omega=x^{1}_{2}+ix^{2}_{2}\, ,
\end{equation}

\noindent for the upper and lower signs respectively. 

The charge of a D-7-brane is just, with our normalizations (in the
string frame)

\begin{equation}
p = \oint_{\gamma} {}^{\star}\hat{G}^{(9)} = \oint_{\gamma} \hat{G}^{(1)} = 
\oint_{\gamma} d\hat{C}^{(0)} = \Re{\rm e} \oint_{\gamma}d\hat{\lambda}\, . 
\end{equation}

The contour $\gamma$ is any circle around the point in the transverse
space. Using the residue theorem we find for our case that the imaginary
part of the integral is zero and

\begin{equation}
p = \mp 2\pi e^{-\hat{\varphi}_{0}}h_{D7}\, ,
\end{equation}

\noindent so the solution indeed describes an anti-D-7-brane (upper sign, 
$\hat{\lambda}=\hat{\lambda}(\omega)$ a holomorphic function of
$\omega$) or D-7-brane (lower sign,
$\hat{\lambda}=\hat{\lambda}(\overline{\omega})$ a holomorphic
function of $\overline{\omega}$) for

\begin{equation}
h_{D7}=\frac{e^{\hat{\varphi}_{0}}}{2\pi}\, .
\end{equation}

We stress that the transformation that takes us from the D-7-brane to
the anti-D-7-brane with opposite RR charge is

\begin{equation}
\label{eq:reverse}
\hat{\lambda}_{(p)} \rightarrow \hat{\lambda}_{(-p)}
=-\overline{\hat{\lambda}_{(p)}}\, ,
\end{equation}

\noindent and it is not an $SL(2,\mathbb{R})$ transformation.

We have just associated the charge of the D-7-brane to the monodromy
properties of the anti-holomorphic function
$\hat{\lambda}(\overline{\omega})$: If we place at the origin a
D-7-brane of unit charge, described by

\begin{equation}
\label{eq:p=17brane}
\hat{\lambda}_{(p=1)} = -\frac{1}{2\pi i}\log{\overline{\omega}}\, ,
\end{equation}

\noindent and travel once along the path 
$\gamma(\xi)\, ,\,\, \xi\in [0,1]$, around the origin

\begin{equation}
  \begin{array}{rcl}
\hat{\lambda}_{(p=1)}[\gamma(1)] & =  &
\hat{\lambda}_{(p=1)}[\gamma(0)] +1= 
\left(M_{(p=1)} \hat{\lambda}_{(p=1)}\right) [\gamma(0)]\, ,  
\\
& & \\
M_{(p=1)} &  =  &
\left(
\begin{array}{cc}
1 & 1 \\
0 & 1 \\
\end{array}
\right)
=T\, , \\
\end{array}
\end{equation}

\noindent where $M_{(p=1)}$ is the $SL(2,\mathbb{Z})$ monodromy matrix
characterizing the 7-brane with charge $p=1$.  One can then apply
$SL(2,\mathbb{Z})$ transformations $\Lambda$ to generate other
solutions as done in Ref.~\cite{kn:GHZ}. Clearly, the monodromy matrix
transforms in the adjoint representation

\begin{equation}
M^{\prime} = \Lambda M \Lambda^{-1}\, .
\end{equation}

Now, it is usually assumed that there are bound states of two kinds of
7-branes ($pq$-branes) transforming as doublets under
$SL(2,\mathbb{Z})$. In particular, the charge vector of $pq$-7-branes
transforms {\it covariantly} under $SL(2,\mathbb{Z})$, that is

\begin{equation}
\label{eq:rule}
\left(
\begin{array}{c}
p^{\prime} \\
q^{\prime} \\
\end{array}
\right)  
=
\Lambda
\left(
\begin{array}{c}
p \\
q \\
\end{array}
\right)\, .  
\end{equation}

The charge vector of $pq$-strings transforms contravariantly
\cite{kn:JHS2}, that is

\begin{equation}
(p^{\prime}\,\, q^{\prime}) =  (p\,\, q) \Lambda^{-1}\, ,
\end{equation}

\noindent and so does the charge vector of $pq$-5-branes \cite{kn:LR}.
Using the above transformation law, one can generate, starting from
the $(p=1)\equiv (1,0)$ other charge vectors using the
$SL(2,\mathbb{Z})$ matrix $\Lambda_{(p,q)}$

\begin{equation}
\Lambda_{(p,q)} =
\left(
\begin{array}{cc}
p & b \\
q & d \\
\end{array}
\right)\, ,
\hspace{1cm}
\Lambda_{(p,q)}
\left(
\begin{array}{c}
1 \\
0 \\
\end{array}
\right)
=
\left(
\begin{array}{c}
p \\
q \\
\end{array}
\right)\, .
\end{equation}

With the same transformation we generate the supergravity solution
describing the $pq$-7-brane with those charges. The monodromy matrix
that characterizes this solution is

\begin{equation}
M_{(p,q)} =  \Lambda_{(p,q)} M_{(1,0)} \Lambda_{(p,q)}^{-1} =
\left(
\begin{array}{cc}
1-pq   & p^{2} \\
& \\
-q^{2} & 1+pq  \\
\end{array}
\right)\, .
\label{eq:mpq}
\end{equation}

Clearly not any pair $(p,q)$ can be generated in this way from
$(1,0)$. $p$ and $q$ cannot be even at the same time, to start with.
According to the standard lore of S~duality $p$ and $q$ have to be
coprime in order to correspond to stable bound states, and thus this
first objection does not seem serious. Still, there is no proof that
all pairs corresponding to stable states can be generated in this way.

A second problem is that this is not (by far) the most general
$SL(2,\mathbb{Z})$ matrix. Thus, given a certain monodromy matrix we
cannot in general determine to which $(p,q)$ state it corresponds.

But there is a more serious problem: We saw in Eq.~(\ref{eq:reverse})
that the transformation that takes us from the $(1,0)$ state to the
$(-1,0)$ state is not an $SL(2,\mathbb{Z})$ transformation. However,
if the rule Eq.~(\ref{eq:rule}) is true the transformation
$-\mathbb{I}_{2\times 2}$ does the same job. But this transformation
leaves $\hat{\lambda}$ exactly invariant!\footnote{As we said before,
the group acting on $\hat{\lambda}$ is $PSL(2,\mathbb{Z})\equiv
SL(2,\mathbb{Z})/\{ \pm\mathbb{I}_{2\times 2}\}$.}

We conclude that bound states of $p$- and $q$-7-branes cannot
transform according to Eq.~(\ref{eq:rule}), and it is easy to see that
they do not transform contravariantly either. Thus, they cannot
transform as doublets.

It is evident that D-7-branes are not singlets. Thus, the next
possibility to be tested is that 7-branes are triplets, i.e.~they
transform in the adjoint representation. This possibility looks
particularly promising if we stick to the characterization of 7-brane
bound states through monodromy matrices, which transform in the
adjoint representation. Furthermore, there is no $SL(2,\mathbb{Z})$
transformation taking us from the monodromy matrix of the $(p=1)$
state, $T$, to the monodromy matrix of the $(p=-1)$ state, $T^{-1}$.

To clarify completely this issue we are going to make a precise
definition of the charges involved and their relation with the
monodromy matrix. First, we observe that the equations of motion for
the scalars can be written as (we suppress hats here):

\begin{equation}
\label{eq:scalareqs}
\nabla_{\mu}{\cal J}^{\mu}=0\, ,
\hspace{1cm}
{\cal J}_{\mu}= 2\partial_{\mu}{\cal M} {\cal M}^{-1}
=
2
\left(
\begin{array}{cc}
\frac{1}{2}j_{\mu}^{(\varphi)} & j_{\mu} \\
& \\
j_{\mu}^{(0)} & -\frac{1}{2}j_{\mu}^{(\varphi)} \\
\end{array}
\right)\, ,
\end{equation}

\noindent where

\begin{equation}
\left\{
\begin{array}{rcl}
j_{\mu}^{(\varphi)} & = & e^{2\varphi} \partial_{\mu}|\lambda|^{2}\, ,\\
& & \\
j_{\mu}^{(0)} & = & e^{2\varphi} \partial_{\mu} C^{(0)}\, ,\\
& & \\
j_{\mu} & = & -C^{(0)}j_{\mu}^{(\varphi)} +|\lambda|^{2}j_{\mu}^{(0)}\, .
\end{array}
\right.
\end{equation}

The divergences of the first two currents are the dilaton and RR
scalar equations of motion. The divergence of the third current is
zero on shell but it is not an equation of motion. These three
conserved currents can be associated to the three parameters of
$SL(2,\mathbb{R})$. In fact, the Noether current associated to the
global $SL(2,\mathbb{R})$ transformation $\Lambda=e^{m}$ where $m$ is
the mass matrix defined in Eq.~(\ref{eq:massmatrix}) is given by

\begin{equation}
j^{(m)}_{\mu}={\rm Tr}\left({\cal J}_{\mu}m \right)\, .
\end{equation}

Using the current matrix we can define a conserved charge
matrix

\begin{equation}
{\cal Q} \equiv
\left(
\begin{array}{cc}
\frac{\delta}{2}r & \beta q             \\
\gamma p          & -\frac{\delta}{2} r \\
\end{array}
\right)
\equiv
{\textstyle\frac{1}{2}} \oint_{S^{1}} {\cal J} =
\oint_{S^{1}} d{\cal M}{\cal M}^{-1}\, ,
\end{equation}

\noindent where $p,q,r$ are integer charges and $\delta,\beta,\gamma$
are the adequate normalization constants. $r$ is the charge associated
to the dilatation current:

\begin{equation}
2\alpha r= \oint j^{T_{1}}=2\oint j^{(\varphi)}\, ,
\end{equation}

\noindent $p$ is the charge associated to shifts of the RR scalar 

\begin{equation}
2\gamma p= \oint j^{\frac{1}{2}(T_{2}+T_{3})}=2\oint j^{(0)}\, ,
\end{equation}

\noindent and therefore the D-7-brane charge, and $q$ is the charge
associated to the remaining independent transformation

\begin{equation}
2\beta q= \oint j^{\frac{1}{2}(T_{2}-T_{3})}=2\oint j\, .
\end{equation}

Observe that both the current matrix and charge matrix transform in
the adjoint representation under $SL(2,\mathbb{R})$.  Using the scalar
equations of motion as we have written them in Eq.~(\ref{eq:scalareqs}) it
is possible to dualize the scalars on-shell and substitute the current
matrix ${\cal J}^{\mu}$ by the Hodge dual of a 9-form field-strength
matrix. This matrix will also transform in the adjoint
representation\footnote{See {\it Note Added in Proof}.}.

Let the $S^{1}$ be parametrized by $\xi\in[0,1]$: We define

\begin{equation}
{\cal Q} (\xi)
\equiv
\int_{0}^{\xi} d{\cal M}{\cal M}^{-1}\, ,
\hspace{.5cm}
\Rightarrow 
\hspace{.5cm}
d{\cal Q} (\xi) = d{\cal M} {\cal M}^{-1}\, .
\end{equation}

\noindent If ${\cal Q} (\xi)={\cal Q} \xi$, the differential equation
can be integrated giving

\begin{equation}
{\cal M} (\xi)  = e^{\frac{1}{2}{\cal Q}\xi}
{\cal M}_{0}\  e^{\frac{1}{2}{\cal Q}^{T}\xi}\, ,
\end{equation}

\noindent so that the corresponding monodromy matrix reads

\begin{equation}
M_{(p,q,r)} =e^{\frac{1}{2}{\cal Q}}\, .
\end{equation}

The restriction to $M\in SL(2,\mathbb{Z})$ implies the quantization of
the charges $(p,q,r)$. In particular it implies that there are no
allowed {\it quantum} states with $p=q=0, r\neq 0$. This seems to
restrict the number of independent charge to just two: $p$ and
$q$. But it is not easy to talk about the number of independent
integers related by a Diophantic equation: Not any pair $p,q$ is
allowed.

The general form Eq.~(\ref{eq:integermonodromy}) for an
$SL(2,\mathbb{Z})$ matrix is useful to illustrate our result. Let us
take the case $n=1$.  The other three integers $n^{i}$ are a
Pythagorean triplet and can be parametrized by three integers $t,s,l$
with the only restriction that $s$ and $l$ are coprime and one of them
is an even number:

\begin{equation}
n^{1} = \pm t (s^{2} -l^{2})\, ,
\hspace{.5cm}
n^{2} = \pm 2tsl\, ,
\hspace{.5cm}  
n^{3} = \pm t (s^{2} +l^{2})\, .
\end{equation}

This restricted case already produces a monodromy matrix much more
general than the $M_{pq}$ in Eq.~(\ref{eq:mpq}). Only two of the
integers are independent and the three of them can be put in
one-to-one correspondence with the charges $p$ and $q$.

In any case, the important lesson at this stage is that given the
monodromy matrix of a certain 7-brane configuration, the above
relation immediately allows us to find the 7-brane charges.

To finish this Section, let us stress that these solutions are just
examples of the general class of negative-charge 7-brane-type
solutions that we write below in the Einstein frame:

\begin{equation}
\left\{
\begin{array}{rcl}
ds^{2}_{E} & = & dt^{2} - d\vec{y}_{7}^{\ 2} 
-H_{7}d\omega d\overline{\omega}\, ,\\
& & \\
H_{7} & = & |h|^{2}\Im{\rm m}\hat{\lambda}\, ,\\
& & \\
\partial_{\overline{\omega}}\hat{\lambda} & = & 
\partial_{\overline{\omega}}h=0\, .\\
\end{array}
\right.
\end{equation}

The holomorphic function $h$ is nothing but a holomorphic coordinate
change. The solutions with positive charge can be obtained by the
transformation in Eq.~(\ref{eq:reverse}).


\subsubsection{Q-7-Branes}

We can now generate the S~duals of the D-7-brane. The rules found
above allow us to identify their charges. However, we need a
formulation in terms of 8-form potentials to understand physically
whether $r$ represents an independent 7-brane charge or not. We will
present such a formulation elsewhere.

First, we will construct the Q-7-brane.

Any $SL(2,\mathbb{Z})$ transformation can be written as a product of
$S$ and $T$ transformations

\begin{equation}
S=\eta=
\left( 
\begin{array}{rc}
0  & 1 \\
-1 & 0 \\
\end{array}
\right)\, ,
\hspace{1cm}
T=
\left( 
\begin{array}{cc}
1 & 1 \\
0 & 1 \\
\end{array}
\right)\, ,
\end{equation}

\noindent raised to positive or negative powers. Under these 
transformations, the charges $p,q,r$ transform as follows:

\begin{equation}
\left\{
\begin{array}{rcl}
r & \stackrel{S}{\rightarrow} & -r\, , \\
& & \\
q & \stackrel{S}{\rightarrow} & -\gamma/\beta p\, ,\\
& & \\
p & \stackrel{S}{\rightarrow} & -\beta/\gamma q\, ,\\
\end{array}
\right.
\hspace{1cm}
\left\{
\begin{array}{rcl}
r & \stackrel{T}{\rightarrow} & r +2\gamma/\delta p\, , \\
& & \\
q & \stackrel{T}{\rightarrow} & q -\delta/\beta r -\gamma/\beta p\, ,\\
& & \\
p & \stackrel{T}{\rightarrow} & p\, .\\
\end{array}
\right.
\end{equation}

We see that, as expected, from a configuration with only $p$ charge (a
D-7-brane) an $S$ transformation generates a configuration with only
$q$ charge. We call the object described by this kind of solution a
``Q-7-brane'' and, taking the D-7-brane solution in
Eq.~(\ref{eq:p=17brane}) we can immediately find its form:

\begin{equation}
\label{eq:q=1Q7}
\begin{array}{c}
Q7\\
(7,0,2)\\
\end{array}
\,\,\,
\left\{
\begin{array}{rcl}
d\hat{s}^{2}_{IIB} & = & 
\left(H_{D7}^{2}+A^{2}\right)^{1/2}
\left[H_{D7}^{-1/2}\left( \eta_{ij}dy^{i}dy^{j} -dy^{2}\right)
-H_{D7}^{1/2}d\omega d\overline{\omega}\right]\, ,\\
& & \\
\hat{\lambda} & = & -1/(-A +iH_{D7})\, ,\\
\end{array}
\right.
\end{equation}

\noindent where

\begin{equation}
H_{D7} = {\textstyle\frac{1}{4\pi}} \log{\omega\overline{\omega}}\, ,
\hspace{1cm} 
A = {\textstyle\frac{1}{4\pi}}i \log{\omega/\overline{\omega}}\, .
\end{equation}

The $T$ transformation generates out of the D-7-brane a configuration
with a different constant value for $\hat{C}^{(0)}$. Although this is
the only difference with the original D-7-brane solution, this
constant value induces $q$-charge through the Witten effect. The
presence of both $p$ and $q$ charges induces $r$-charge which here
seems not to be independent.


\subsection{7-Branes with a  Compact Transverse Dimension}

We want to transform 7-branes under T~duality and therefore we need to
consider the corresponding solutions with a compact transverse
dimension.

If one of the transverse coordinates, say $x_{2}^{1}\equiv y$ is
compact $y\sim y+2\pi \ell$ then the function $H_{D7}$ that solves
Eq.~(\ref{eq:harmonic}) in $\mathbb{R}\times S^{1}$ takes a different
form (we set to zero the additive constant for simplicity):

\begin{equation}
H_{D7} =  \frac{h_{D_{7}}}{2\ell}|x_{2}^{2}| 
+h_{D7}  \log{\sqrt{1 -2 e^{-|x_{2}^{2}|} \cos{y/\ell} 
+e^{-2|x_{2}^{2}|}} }\, .
\end{equation}

Usually, only the zero-mode in the Fourier expansion of this function
is considered when performing T~duality transformations because the
only T~duality rules known (Buscher's \cite{kn:Bu}) apply only to
solutions independent of the compact coordinate (at least the metric
has to be). This is a strong limitation which only recently started to
be appreciated \cite{kn:GHM}. Nevertheless, the behavior of this
zero-mode seems to be well understood and we will focus on it. In our
case, then, we will take, for a single D-7-brane ($h_{D7}=1/(2\pi
e^{-\hat{\varphi}_{0}})$) and for $\ell=1/2\pi$ ($y\sim y+1$)

\begin{equation}
H_{D7} = \frac{h_{D_{7}}}{2\ell}|x_{2}^{2}|\, .  
\end{equation}

\noindent Restricting ourselves to the region $x_{2}^{2}>0$ for
simplicity we find for the complex scalar $\hat{\lambda}$ the
expression

\begin{equation}
\left\{
\begin{array}{rcl}
\hat{\lambda}_{(p=-1)_{0}} & = &
{\textstyle\frac{1}{2}} \omega\, .\\
& & \\
\hat{\lambda}_{(p=+1)_{0}} & = &
-{\textstyle\frac{1}{2}} \overline{\omega}\, .\\
\end{array}
\right.
\hspace{1cm}
\omega=y+ix_{2}^{2}
\end{equation}

Somewhat surprisingly, the solution does depend on the compact
coordinate $y$.  The metric does not, but, after an $SL(2,\mathbb{R})$
transformation, the string metric will depend on $y$ while the
Einstein metric will not.

Again, it is convenient to rewrite $\hat{\lambda}$ as follows:

\begin{equation}
\hat{\lambda}_{(p=1)_{0}} = 
{\textstyle\frac{1}{2}} e^{-\hat{\varphi}_{0}}
\left( 
\begin{array}{c}
z\\
-\bar{z} \\
\end{array}
\right)\, ,
\hspace{1cm}
z=y+ix^{2}_{2}\, . 
\end{equation}

\begin{figure}[!ht]
\begin{center}
\leavevmode
\epsfxsize= 12cm
\epsfysize= 10cm
\epsffile{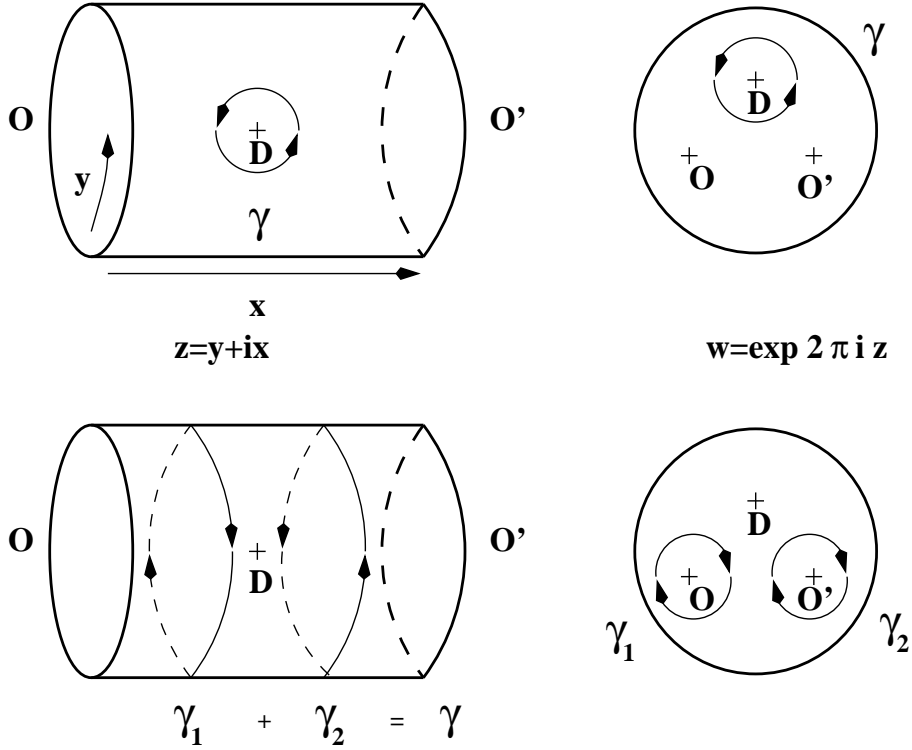}
\caption{\footnotesize If we place a 7-brane on a cylinder, 
  one has to take into account that automatically 7-branes are created
  at the boundaries. This can be easily seen by conformally
  transforming the cylinder into a punctured sphere. Consistency of
  the monodromy implies that the total sum of the charges in the sphere
  is nil. \normalsize}
\label{fig:cylinder}
\end{center}
\end{figure}

Let us now start by analyzing the monodromy of the positive charge
solution zeromode (the holomorphic one). The above function is regular
everywhere: The D-7-brane has been smeared out. The only non-trivial
cycle to study is the one along $y$, and one finds that the zeromode
is shifted by $1/2$. This is not an $SL(2,\mathbb{Z})$
transformation. To understand this result it is convenient to map the
cylinder into the Riemann sphere with two punctures by means of the
conformal transformation $1/w=e^{2\pi i z}$. $w$ is the coordinate in
the patch around infinity. Going around the origin in the $w$ plane is
the same as going around the cylinder's $S^{1}$ parametrized by $y$ in
the negative sense. The complex scalar zeromode becomes

\begin{equation}
\hat{\lambda}_{(p=1)_{0}} 
=-{\textstyle\frac{1}{2}}\frac{1}{2\pi i}\log{w}\, ,
\end{equation}

\noindent which obviously corresponds to a D-7-brane with charge
$-1/2$ placed at infinity in the Riemann sphere, i.e.~at infinity in
the cylinder (we are considering only the positive $x_{2}^{2}$ part of
the cylinder). Something analogous happens at minus infinity. Then,
the presence of a D-7-brane on a cylinder induces the presence of
other D-7-branes at infinity. The D-7-branes at infinity have to have
integer charge and thus we can only place a D-7-brane of charge
$(p=2)$ to have a consistent picture. The situation is depicted in
Figure~\ref{fig:cylinder} The monodromies along the compact coordinate
measure the 7-brane charges at infinity and are, therefore
$SL(2,\mathbb{Z})$ matrices as discussed in the previous section (now
with $\xi=y$).  These are precisely the monodromy matrices that appear
in our massive 9-dimensional type~II supergravity theory.

In the supergravity theory, the monodromy matrices are determined by
the mass matrix $m$ and, comparing with the results of the previous
Section, this is identical to the $pq$-7-brane charge matrix $m$:

\begin{equation}
m={\textstyle\frac{1}{2}} \left(
\begin{array}{cc}
m^{1}       & m^{2}+m^{3} \\
& \\
m^{2}-m^{3} &  -m_{1}     \\
\end{array}
\right)={\cal Q}= \left(
\begin{array}{cc}
\frac{\delta}{2}r & \beta q             \\
\gamma p          & -\frac{\delta}{2} r \\
\end{array}
\right)\, .
\end{equation}

This is the sought for relation between the background of 7-branes
and the mass parameters of the massive 9-dimensional type~II
supergravity theory.


\section{KK-7A- and KK-8A-branes and T~Duality}
\label{sec-KK789}

\begin{figure}[!ht] 
\begin{center} 
\leavevmode 
\epsfxsize= 10cm
\epsfysize= 17.5cm
\epsffile{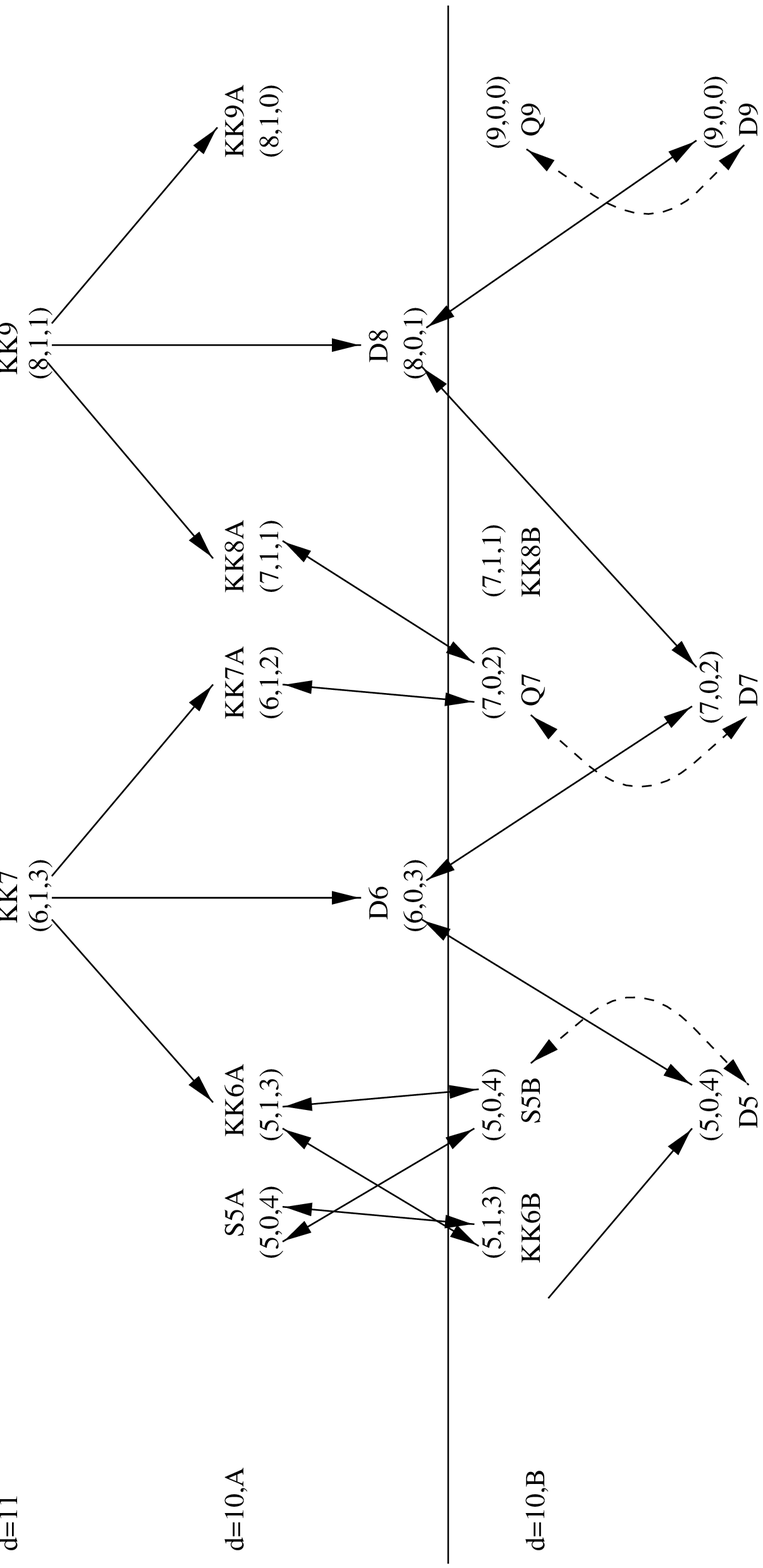} 
\caption{\footnotesize  This figure is a magnified and
  more detailed piece of Figure~\ref{fig:dualbran} in which a general
  picture of all the known extended objects of M/string theory and
  their duality relations is given. Only well-established relations
  are shown, and so no duality connections between the conjectured
  KK-8B-brane and other objects are drawn. In the triplets $(m,n,p)$
  $m$ stands for the number of standard spacelike dimensions of the
  object, $n$ for the number of special isometric directions ($z$) and
  $p$ for the number of standard transverse dimensions. The double
  arrows indicate on which directions T~duality acts.\normalsize}
\label{fig:kkduals} 
\end{center} 
\end{figure}

In this Section we are going to check explicitly the dualities between
extended objects underlying the generalized T~duality between the
type~IIA and type~IIB theories. We will find some of the objects whose
existence we conjectured in the Introduction. We will essentially
prove the connections shown in Figure~\ref{fig:kkduals}.

It is convenient to start with the 11-dimensional Kaluza-Klein
monopole which we refer to as KK-7M-brane. This is a 7-dimensional,
purely gravitational object, but one of the spacelike worldvolume
directions, with coordinate $z$ is compactified on a circle. Its
metric is given by

\begin{equation}
\label{eq:KK7M}
\begin{array}{c}
KK7M\\
(6,1,3)\\
\end{array}
\,\,\,
\left\{
\begin{array}{rcl}
d\hat{\hat{s}}{}^{2} & = & \eta_{ij}dy^{i}dy^{j}
-H^{-1}\left(dz^{2}+A_{m}dx^{m}\right)^{2} 
-Hd\vec{x}^{\ 2}_{3}\, ,\\
& & \\
2\partial_{[m}A_{n]} & = & \epsilon_{mnp}\partial_{p}H\, ,\\
\end{array}
\right.
\end{equation}

\noindent where $\vec{x}_{3}=(x^{m})=(x^{1},x^{2},x^{3})$ and
$i=0,1,\ldots,6$. The standard solution corresponds to the choice

\begin{equation}
H = 1+\frac{h}{|\vec{x}_{3}|}\, .
\end{equation}

We can reduce this solution in three different ways. First, we can
reduce in the isometry direction, $z$. It is well-known that the
resulting object is the D-6-brane. Reducing on one of the standard
spacelike worldvolume directions (double dimensional reduction)
trivially gives the KK-6A-brane, which is nothing but the
10-dimensional KK monopole.

Finally, we can reduce it on a transverse coordinate, $x^{3}$. We
obtain

\begin{equation}
\begin{array}{c}
KK7A\\
(6,1,2)\\
\end{array}
\,\,\,
\left\{
\begin{array}{rcl}
d\hat{s}^{2}_{IIA} & = & 
\left(\frac{H}{H^{2}+A^{2}}\right)^{-1/2}
\left[\eta_{ij}dy^{i}dy^{j}
-\frac{H}{H^{2}+A^{2}}dz^{2}
-Hd\omega d\overline{\omega}\right]\, ,\\
& & \\
e^{\hat{\phi}} & = & \left(\frac{H}{H^{2}+A^{2}}\right)^{-3/4}\, ,\\
& & \\
\hat{C}^{(1)}{}_{\underline{z}} & = & \frac{A}{H^{2}+A^{2}}\, ,\\
& & \\
\partial_{\omega}A & = & i\partial_{\omega}H\, ,\\
\end{array}
\right.
\end{equation}

\noindent where $\omega=x^{1}+ix^{2}$ and $A=A_{3}$ and the last
equation is simply $2\partial_{[m}A_{n]} =
\epsilon_{mnp}\partial_{p}H$ with the assumption that $H$ does not
depend on $x^{3}$ and in the $A_{1}=A_{2}=0$ gauge.  In complex
notation the last equation then reads $\partial_{\omega}\left(
A_{3}-iH\right) =0$, which has as a particular solution

\begin{equation}
H = {\textstyle\frac{h}{2}} \log{\omega\overline{\omega}}\, ,
\hspace{1cm} 
A = {\textstyle\frac{h}{2}}i \log{\omega/\overline{\omega}}\, .
\end{equation}

This kind of solutions has been previously considered in
Refs.~\cite{kn:BREJS1,kn:BREJS2,kn:LPTX}. To relate it with type~IIB
solutions, we further reduce it in the isometry direction $z$. The
resulting solution is a 9-dimensional ``Q-6-brane'':

\begin{equation}
\label{eq:Q69}
\begin{array}{c}
Q6_{9}\\
(6,0,2)\\
\end{array}
\,\,\,
\left\{
\begin{array}{rcl}
ds^{2}_{II} & = & 
\left(H^{2}+A^{2}\right)^{1/2}
\left[H^{-1/2}\eta_{ij}dy^{i}dy^{j}
-H^{1/2}d\omega d\overline{\omega}\right]\, ,\\
& & \\
e^{\phi} & = & \left(\frac{H}{H^{2}+A^{2}}\right)^{-1}\, ,\\
& & \\
C^{(0)} & = & \frac{A}{H^{2}+A^{2}}\, ,\\
& & \\
\partial_{\omega}A & = & i\partial_{\omega}H\, .\\
\end{array}
\right.
\end{equation}

This is a solution of our massive 9-dimensional type~II theory with
$m^{i=0}$. We are going to show it through duality arguments.

Notice that we have obtained two different solutions by reducing first
on $z$ and then on $x^{3}$ and in the inverse order. The difference is
a rotation in internal space $z,x^{3}$ and, by T~duality to an
S~duality transformation in the type~IIB side, as we are going to see.

We can now uplift this solution using the type~IIB rules and adding
the coordinate $y$. We obtain the Q-7-brane solution
Eq.~(\ref{eq:q=1Q7}):

\begin{equation}
\label{eq:Q7}
\begin{array}{c}
Q7\\
(7,0,2)\\
\end{array}
\,\,\,
\left\{
\begin{array}{rcl}
d\hat{s}^{2}_{IIB} & = & 
\left(H^{2}+A^{2}\right)^{1/2}
\left[H^{-1/2}\left( \eta_{ij}dy^{i}dy^{j} -dy^{2}\right)
-H^{1/2}d\omega d\overline{\omega}\right]\, ,\\
& & \\
\hat{\lambda} & = & -1/(-A +iH)\, .\\
\end{array}
\right.
\end{equation}

This solution is the S~dual of the standard D-7-brane solution. In
fact, performing the $SL(2,\mathbb{Z})$ transformation $S$ and
substituting the explicit expressions for $H$ and $A$ we get

\begin{equation}
\label{eq:D7}
\begin{array}{c}
D7\\
(7,0,2)\\
\end{array}
\,\,\,
\left\{
\begin{array}{rcl}
d\hat{s}^{2}_{IIB} & = & 
H^{-1/2}\left( \eta_{ij}dy^{i}dy^{j} -dy^{2}\right)
-H^{1/2}d\omega d\overline{\omega}\, ,\\
& & \\
\hat{\lambda} & = & -\frac{h}{i}\log{\overline{\omega}}\, ,\\
\end{array}
\right.
\end{equation}

\noindent which is the (positive charge)  D-7-brane solution of
 Eq.~(\ref{eq:p=17brane}) if we set $h=1/2\pi$.

We could have reduced the KK-7A-brane on another transverse direction
$x^{2}$. Equivalently, we could have simultaneously reduced the
KK-7M-brane on $x^{2}$ and $x^{3}$. We immediately face a problem: if
$H$ is a harmonic function that only depends on $x^{1}$, then
$A_{1}=0$ but $A_{2}$ and/or $A_{3}$ depend on $x^{3}$ and/or $x^{2}$.

The situation is identical to that of the reduction of the Q-7-brane
on a transverse coordinate. There, it was impossible to eliminate the
dependence on that coordinate and generalized dimensional reduction
was necessary. Here, only through generalized dimensional reduction of
11-dimensional supergravity one can find the 9-dimensional solution
and the T~dual. The T~dual must have a special isometric direction and
7 standard spacelike worldvolume coordinates. Such a configuration is
what we call a KK-8B-brane. The generalized dimensional reduction of
11-dimensional supergravity must give the same 9-dimensional theory as
the reduction of type~IIB in presence of KK-8B-branes.

We could have reduced the KK-7A-brane on a standard worldvolume
direction $y^{i}$, getting

\begin{equation}
\label{eq:KK79}
\begin{array}{c}
KK7_{9}\\
(6,1,1)\\
\end{array}
\,\,\,
\left\{
\begin{array}{rcl}
ds^{2}_{II} & = & 
H\left[H^{-1/2}\eta_{ij}dy^{i}dy^{j}
-H^{1/2}dx^{2}\right] -H^{-3/2}dz^{2} \, ,\\
& & \\
e^{\phi} & = & H^{1/8}\, ,\\
& & \\
k & = & H^{1/4}\, .\\
\end{array}
\right.
\end{equation}

This {\it not} a solution of our 9-dimensional massive type~IIB
theory. It would be a solution of another massive 9-dimensional
type~II theory with ``Killing vectors'' in its Lagrangian. Only after
the elimination by reduction of the special isometric direction will we
get a solution to some massive supergravity. Anyway, if we uplift this
configuration to ten dimensions using the standard type~IIB rules we
get

\begin{equation}
\label{eq:unk}
\begin{array}{c}
Unknown\\
(6,2,1)\\
\end{array}
\,\,\,
d\hat{s}^{2}_{IIB} = 
H\left[H^{-1/2}\eta_{ij}dy^{i}dy^{j} -H^{1/2}dx^{2}\right]
-H^{-1/2}dy^{2} -H^{-3/2}dz^{2}\, .
\end{equation}

This purely gravitational configuration is similar to the KK-9M-brane
but with 2 isometric directions instead of just one.  Its presence as
a 10-dimensional type~IIB background will give an 8-dimensional fully
covariant massive type~II theory. 

Objects of this kind can be useful in considering massive theories in
lower dimensions, which are out of the scope of this paper and so we
will not discuss them any further.

We have already checked the left hand side of
Figure~\ref{fig:kkduals}.  It is convenient now to start from the
KK-9M-brane, recently constructed and studied in
Ref.~\cite{kn:BS}\footnote{In that reference it is called
``M-9-brane''. We prefer the name KK-9M-brane because it stresses the
fact that it has a special isometric direction as the usual KK
monopole.}.  This purely gravitational field configuration is not a
solution of the standard 11-dimensional supergravity, but it is a
solution of the massive 11-dimensional supergravity constructed in
Ref.~\cite{kn:BLO} which we have just generalized in a manifestly
$SL(2,\mathbb{R})$-covariant way. Its defining property is that it has
a special isometric direction ($z$) and reduction in this direction
gives the D-8-brane.

Choosing $\epsilon=-1$, the metric of the KK-9M-brane is

\begin{equation}
\label{eq:KK9M}
\begin{array}{c}
KK9M\\
(8,1,1)\\
\end{array}
\,\,\,
\left\{
\begin{array}{rcl}
d\hat{\hat{s}}{}^{2} & = & H^{1/3}\eta_{ij}dy^{i}dy^{j}
-H^{-5/3}dz^{2} -H^{4/3}dx^{2}\, ,\\
& & \\
H & = & c +Qx\, ,\\
\end{array}
\right.
\end{equation}

\noindent where now $i=0,1,\ldots,8$.

If we reduce the KK-9M-brane in the isometry direction ($z$) we get
the D-8-brane

\begin{equation}
\begin{array}{c}
D8\\
(8,0,1)\\
\end{array}
\,\,\,
\left\{
\begin{array}{rcl}
d\hat{s}^{2}_{IIA} & = & 
H^{-1/2}\eta_{ij}dy^{i}dy^{j} -H^{1/2}dx^{2}\, ,\\
& & \\
e^{\hat{\phi}} & = & H^{-5/4}\, ,\\
\end{array}
\right.
\end{equation}

\noindent  which is a solution of Romans' massive type~IIA
supergravity \cite{kn:Ro2}.

Reducing further in one of the spacelike worldvolume directions
($y^{8}$) we get the 9-dimensional D-7-brane 

\begin{equation}
\label{eq:D79}
\begin{array}{c}
D7_{9}\\
(7,0,2)\\
\end{array}
\,\,\,
\left\{
\begin{array}{rcl}
ds^{2}_{II} & = & 
H^{-1/2}\eta_{ij}dy^{i}dy^{j} -H^{1/2}dx^{2}\, ,\\
& & \\
e^{\phi} & = & H^{-9/8}\, ,\\
& & \\
k & = & H^{-1/4}\, .\\
\end{array}
\right.
\end{equation}

Uplifting to 10 dimensional using the type~IIA rules we get the
D-7-brane is also the solution we obtained by compactifying in a
transverse dimension the D-7-brane. This establishes T~duality between
the D-8- and the D-7-brane \cite{kn:BRGPT}.

If we reduce first the KK-9M-brane on a standard worldvolume direction 
we get the following field configuration

\begin{equation}
\label{eq:KK8A}
\begin{array}{c}
KK8A\\
(7,1,1)\\
\end{array}
\,\,\,
\left\{
\begin{array}{rcl}
d\hat{s}^{2}_{IIA} & = & 
H\left[H^{-1/2}\eta_{ij}dy^{i}dy^{j}
-H^{1/2}dx^{2}\right] -H^{-3/2}dz^{2}\, ,\\
& & \\
e^{\hat{\phi}} & = & H^{1/4}\, ,\\
\end{array}
\right.
\end{equation}

\noindent which we call KK-8A-brane. This is not a solution of any
standard 10-dimensional supergravity. Instead, it is a solution of the
massive type~IIA supergravity that one finds by reduction of the
massive 11-dimensional supergravity of Ref.~\cite{kn:BLO} in a
direction different from the isometric one. This theory is related by
a rotation in internal space with Romans' massive supergravity
\cite{kn:Ro2}.

Reducing further in the isometry direction ($z$), we get
the 9-dimensional Q-7-brane

\begin{equation}
\label{eq:Q79}
\begin{array}{c}
Q7_{9}\\
(7,0,1)\\
\end{array}
\,\,\,
\left\{
\begin{array}{rcl}
ds^{2}_{II} & = & 
H\left[H^{-1/2}\eta_{ij}dy^{i}dy^{j} -H^{1/2}dx^{2}\right]\, ,\\
& & \\
e^{\phi} & = & H^{5/8}\, ,\\
& & \\
k & = & H^{-3/4}\, .\\
\end{array}
\right.
\end{equation}

We observe again that we have obtained two different 9-dimensional
results which must be related by a rotation in the 2-dimensional
internal space and, therefore, by an S~duality transformation in the
T~dual type~IIB theory. Thus, not surprisingly, if we uplift the
9-dimensional Q-7-brane to ten dimensions using the standard type~IIB
rules we get

\begin{equation}
\label{eq:Q7bare}
\begin{array}{c}
Q7^{\rm bare}\\
(7,0,2)\\
\end{array}
\,\,\,
\left\{
\begin{array}{rcl}
d\hat{s}^{2\ {\rm b}}_{IIB} & = & 
H \left[H^{-1/2}\left( \eta_{ij}dy^{i}dy^{j} -dy^{2}\right)
-H^{1/2}d\omega d\overline{\omega}\right]\, ,\\
& & \\
\hat{\lambda}^{\rm b} & = & +iH^{-1}\, .\\
\end{array}
\right.
\end{equation}

This is nothing but the {\it bare} field configuration of the
Q-7-brane Eq.~(\ref{eq:Q7}). Using the generalized rules for
uplifting, the dependence on the internal coordinate is fully
recovered. This establishes T~duality between the Q-7-brane and the
KK-8A-brane under the generalized Buscher T~duality rules of
Appendix~\ref{sec-tdual}.


\section{Conclusion}
\label{sec-conclusion}

We have successfully completed the program put forward in the
Introduction. However, there are still some missing pieces in the
general picture. In particular, we have found explicitly the T~duals
of the S~duals of the D-7-brane (the KK-8A-brane, which comes from the
reduction of the KK-9M-brane recently constructed in
Ref.~\cite{kn:BS}) but we have not built the corresponding 9-form
potential, related by T~duality to the 8-form potential of the
Q-7-brane which we have not constructed either.

If the picture we have proposed is correct and the 9-form potential of
the KK-8A-brane is a purely gravitational object (as suggested by Hull
in Ref.~\cite{kn:Hu2}) there might an analogous solutions in each
string theory. Thus, there ought to be KK-8-branes of the type~IIB theory
(KK-8B-branes) and we can ask ourselves what the effect of having one
of these objects in the background would be. Clearly, we could obtain
a massive 9-dimensional type~II theory with one more mass parameter!
To what theory would it correspond in the type~IIA/M side?

The answer is simple: If we reduce the massive 11-dimensional theory
of Section~\ref{sec-toro} directly to nine dimensions we can use
Scherk \& Schwarz's original generalized dimensional reduction to
produce an extra 9-dimensional mass parameter. As we have seen, the
reduction of the 11-dimensional KK monopole (KK-7M-brane) over two
transverse dimensions can only be performed using this technique.
Eventually one could construct a 9-dimensional type~II theory with
four mass parameters. It is reasonable to expect that they are
organized in a multiplet of the 9-dimensional duality group
($GL(,2\mathbb{R})$.

However, we only know how to produce one mass parameter using this
technique if we compactify at least two dimensions. Thus, we do not
know what the 10-dimensional theory would be like.

The next question would be to ask what would happen if we placed more
KK-9M-branes in eleven dimensions or KK-8-branes in ten (or
KK-$(d-2)$-branes in $d$ dimensions). Clearly this should give us new
massive theories in dimensions lower than nine and should be seen in
the T~dual picture as the result of a generalized dimensional
reduction (i.e.~putting Q-$(d-3)$-branes in the background).

The general picture we have obtained seems to agree with the
suggestion of Ref.~\cite{kn:KKM} of the existence of a general massive
theory of which all other should be particular cases. In fact, the
general massive 4-dimensional type~II theory (massive $N=8$
supergravity) has to be consistent with U~duality, which acts on the
mass parameters. These should then fit into a multiplet of $E_{7}$. On
the other hand, since the mass parameters are in a sense potentials
associated to branes, the maximally massive $N=8$ supergravity should
be considered {\it the} $N=8$ supergravity theory and {\it the}
low-energy limit of type~II string theories. The presence of
KK-$(d-2)$-branes in the corresponding higher-dimensional theories
which lead to the maximally massive $N=8$ supergravity is unavoidable.

\begin{figure}[!ht]
\begin{center}
\leavevmode
\epsfxsize= 13cm
\epsfysize= 18cm
\epsffile{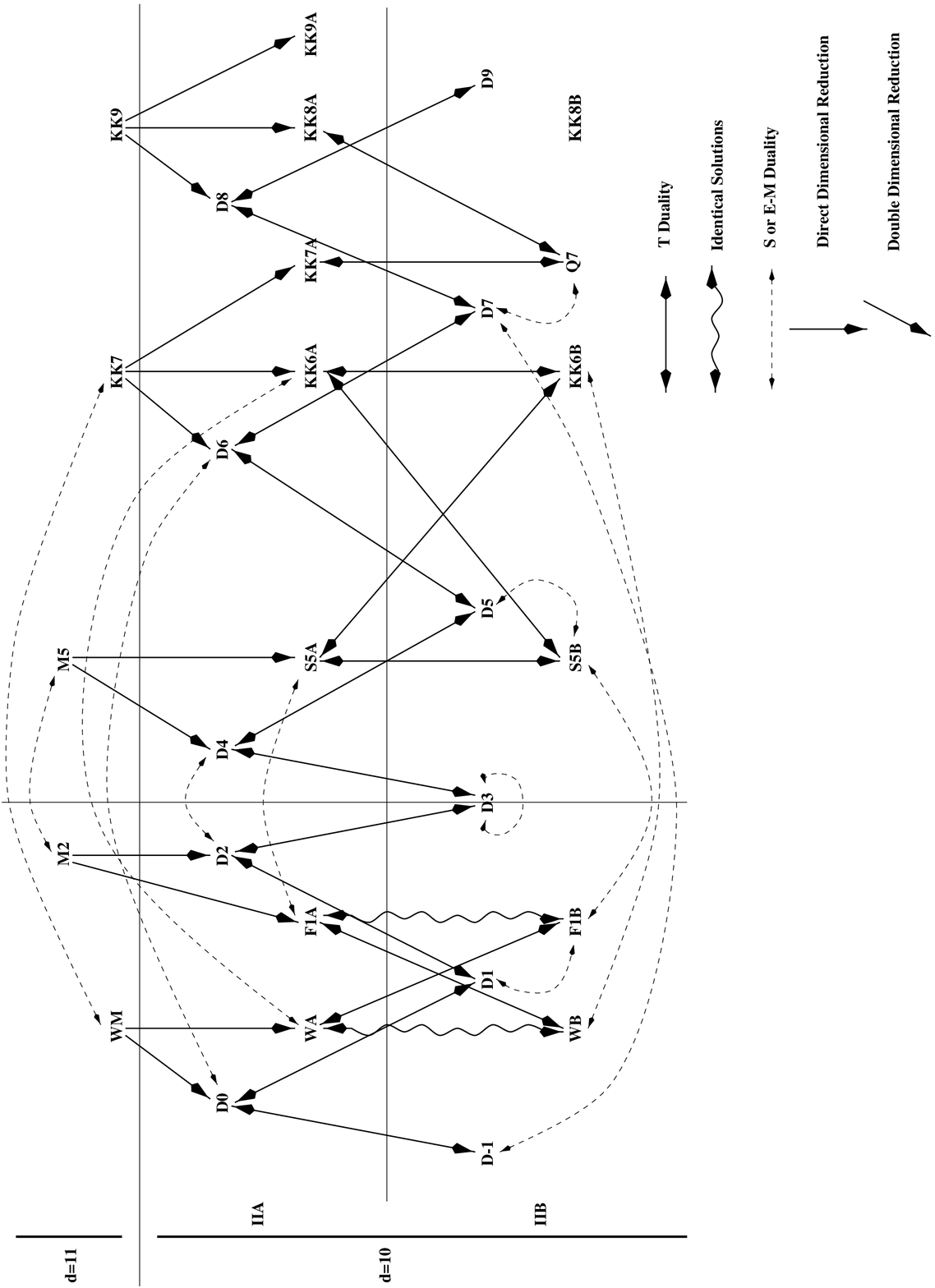}
\caption{\footnotesize Duality relations between 
  classical solutions of 10- and 11-dimensional supergravity theories
  describing string/M-theory solitons: p-branes, M-branes, D-branes,
  gravitational waves, Kaluza-Klein monopoles and other KK-type
  solutions. Lines with two arrows denote T~duality
  relations. Dashed lines denote S~duality relations. Lines with
  a single arrow denote relations of dimensional reduction, either
  vertical (direct dimensional reduction) or diagonal (double
  dimensional reduction).\normalsize}
\label{fig:dualbran}
\end{center}
\end{figure}

In Figure~\ref{fig:dualbran} we represent the present knowledge about
classical solutions of 11- and 10-dimensional supergravity theories
describing string/M-theory solitons. Missing from this {\it
bestiarium} are still objects such as orientifolds which we do not
know as classical solutions.

Finally, we would also like to comment on the possible relation
between the superalgebras of supergravity theories in presence of
KK-$(d-2)$-branes and the superalgebras studied by Bars in twelve and
higher dimensions \cite{kn:Ba}. These superalgebras can accommodate
naturally vectors that break Poincar\'e covariance. Being global
algebras, a proof is not easy, but one is nevertheless tempted to
identify those vectors with the Killing vectors of our supergravity
theories. The presence of a preferred vector would therefore be the
signal of the presence of KK-$(d-2)$-branes in the background.

\section*{Note Added in Proof}

Soon after the appearence of this paper, another paper
\cite{art:tonin} appeared which, using other techniques, reached the
same conclusions as the ones presented in
Sec.~(\ref{sec-pq-7-branes}).


\section*{Acknowledgments}

We would like to thank Prof.~C.~G\'omez for many useful conversations.
T.O.~is indebted to M.M.~Fern\'andez for her support.  The work of
P.M.~has been partially supported by the European Union under contract
number ERBFMBI-CT96-0616. The work of T.O.~is supported by the
European Union TMR program FMRX-CT96-0012 {\sl Integrability,
Non-perturbative Effects, and Symmetry in Quantum Field Theory} and by
the Spanish grant AEN96-1655.

\appendix

\section{9-Dimensional Einstein Fields Vs.
10-Dimensional Type~IIA String Fields}
\label{sec-9E-10S-IIA}

In the main body of the paper we went directly from 11 to 9 dimensions
and thus we need to repeat the reduction from 11 to 10 dimensions
\cite{kn:Wi,kn:BHO} to be able to relate 9- with 10-dimensional
fields.

As usual, we assume now that all fields are independent of the
spacelike coordinate $z =x^{\underline{10}}$ and we rewrite the fields
and action in a ten-dimensional form.  The dimensional reduction of
11-dimensional supergravity Eq.~(\ref{eq:11daction}) gives rise to the
fields of the ten-dimensional $N=2A,d=10$ supergravity theory

\begin{equation}
\left\{
\hat{g}_{\hat{\mu}\hat{\nu}},
\hat{B}_{\hat{\mu}\hat{\nu}},
\hat{\phi},
\hat{C}^{(3)}{}_{\hat{\mu}\hat{\nu}\hat{\rho}},
\hat{C}^{(1)}{}_{\hat{\mu}},
\right\}\, .
\end{equation}

The metric, the two-form and the dilaton are NS-NS fields and the
three-form and the vector are RR fields. We are going to use for RR
forms the conventions proposed in Refs.~\cite{kn:GHT,kn:BCT,kn:BLO}.

The fields of the 11-dimensional theory can be expressed in terms of the
10-dimensional ones as follows:

\begin{equation}
\begin{array}{rclrcl}
\hat{\hat{g}}_{\hat{\mu}\hat{\nu}}
&
=
&
e^{-\frac{2}{3}\hat{\phi}}\hat{g}_{\hat{\mu}\hat{\nu}}
- e^{\frac{4}{3}\hat{\phi}}\hat{C}^{(1)}{}_{\hat{\mu}}
\hat{C}^{(1)}{}_{\hat{\nu}}\, ,\hspace{1cm}
&
\hat{\hat{C}}_{\hat{\mu}\hat{\nu}\hat{\rho}}
&
=
&
\hat{C}^{(3)}{}_{\hat{\mu}\hat{\nu}\hat{\rho}}\, ,
\\
& & & & &
\\
\hat{\hat{g}}_{\hat{\mu}\underline{z}}
&
=
&
-e^{\frac{4}{3}\hat{\phi}}\hat{C}^{(1)}{}_{\hat{\mu}}\, ,
&
\hat{\hat{C}}_{\hat{\mu}\hat{\nu}\underline{z}}
&
=
&
\hat{B}_{\hat{\mu}\hat{\nu}}\, ,
\\
& & & & &
\\
\hat{\hat{g}}_{\underline{z}\underline{z}}
&
=
&
-e^{\frac{4}{3}\hat{\phi}}\, .
& & &
\\
\end{array}
\end{equation}

\noindent For the Elfbeins we have

\begin{equation}
\begin{array}{rcl}
\left( \hat{\hat{e}}_{\hat{\hat{\mu}}}{}^{\hat{\hat{a}}} \right) & = &
\left(
\begin{array}{cc}
e^{-\frac{1}{3}\hat{\phi}} \hat{e}_{\hat{\mu}}{}^{\hat{a}}
&
e^{\frac{2}{3}\hat{\phi}} \hat{C}^{(1)}{}_{\hat{\mu}}
\\
&
\\
0
&
e^{\frac{2}{3}\hat{\phi}}
\\
\end{array}
\right)
\, , \\
& & \\
\left( \hat{\hat{e}}_{\hat{\hat{a}}}{}^{\hat{\hat{\mu}}} \right) & = &
\left(
\begin{array}{cc}
e^{\frac{1}{3}\hat{\phi}} \hat{e}_{\hat{a}}{}^{\hat{\mu}}
&
-e^{\frac{1}{3}\hat{\phi}} \hat{C}^{(1)}{}_{\hat{a}}
\\
&
\\
0
&
e^{-\frac{2}{3}\hat{\phi}}
\\
\end{array}
\right)\, . \\
\end{array}
\label{eq:basis}
\end{equation}

\noindent Conversely, the 10-dimensional fields can be expressed
in terms of the 11-dimensional ones by:

\begin{equation}
\begin{array}{rclrcl}
\hat{g}_{\hat{\mu}\hat{\nu}}
&
=
&
\left( -\hat{\hat{g}}_{\underline{z}\underline{z}} \right)^{\frac{1}{2}}
\left (\hat{\hat{g}}_{\hat{\mu}\hat{\nu}}
-\hat{\hat{g}}_{\hat{\mu}\underline{z}}
\hat{\hat{g}}_{\hat{\nu}\underline{z}}
/\hat{\hat{g}}_{\underline{z}\underline{z}} \right)\, ,\hspace{1cm}
&
\hat{C}^{(3)}{}_{\hat{\mu}\hat{\nu}\hat{\rho}}
&
=
&
\hat{\hat{C}}_{\hat{\mu}\hat{\nu}\hat{\rho}}\, ,
\\
& & & & &
\\
\hat{C}^{(1)}{}_{\hat{\mu}}
&
=
&
\hat{\hat{g}}_{\hat{\mu}\underline{z}}
/\hat{\hat{g}}_{\underline{z}\underline{z}}\, ,
&
\hat{B}_{\hat{\mu}\hat{\nu}}
&
=
&
\hat{\hat{C}}_{\hat{\mu}\hat{\nu}\underline{z}}\, ,
\\
& & & & &
\\
\hat{\phi}
&
=
&
\frac{3}{4}\log{\left( -\hat{\hat{g}}_{\underline{z}\underline{z}}
\right)}\, .
& & &
\\
\end{array}
\label{eq:10ddef}
\end{equation}

After some standard calculations that we omit we find the bosonic part
of the $N=2A,d=10$ supergravity action in ten dimensions in the string
frame:

\begin{equation}
\begin{array}{rcl}
\hat{S} & = &
\int d^{10}x\
\sqrt{|\hat{g}|} \left\{ e^{-2\hat{\phi}}
\left[ \hat{R} -4\left( \partial\hat{\phi} \right)^{2}
+{\textstyle\frac{1}{2\cdot 3!}} \hat{H}^{2}\right]\right.\\
& & \\
& & 
-\left[
{\textstyle\frac{1}{4}} \left( \hat{G}^{(2)} \right)^{2}
+{\textstyle\frac{1}{2\cdot 4!}}\left(\hat{G}^{(4)}\right)^{2}
\right]
-{\textstyle\frac{1}{144}} \frac{1}{\sqrt{|\hat{g}|}}\
\hat{\epsilon}\partial\hat{C}^{(3)}\partial\hat{C}^{(3)}\hat{B}
\biggr \}\, .\\
\end{array}
\label{eq:IIAaction}
\end{equation}

\noindent where the fields strengths are defined as follows:

\begin{equation}
\left\{
\begin{array}{rcl}
\hat{H} & = & 3\partial \hat{B}\, ,\\
& & \\
\hat{G}^{(2)} & = & 2\partial \hat{C}^{(1)}\, ,\\
& & \\
\hat{G}^{(4)} & = & 4\left(\partial \hat{C}^{(3)}
-3\partial \hat{B} \hat{C}^{(1)}\right)\, ,\\
\end{array}
\right.
\end{equation}

\noindent and they are invariant under the gauge transformations

\begin{equation}
\left\{
\begin{array}{rcl}
\delta\hat{B} & = & \partial\hat{\Lambda}\, ,\\
& & \\
\delta\hat{C}^{(1)} & = & \partial\hat{\Lambda}^{(0)}\, ,\\
& & \\
\delta\hat{C}^{(3)} & = & 3\partial\hat{\Lambda}^{(2)}
+3\hat{B}\partial\hat{\Lambda}^{(0)}\, .\\
\end{array}
\right.
\end{equation}

Now, using these results together with the relation between 9- and
11-dimensional  fields obtained in Section~\ref{sec-toro} we get

\begin{equation}
\begin{array}{lcl}
{\cal M} & = & 
e^{\hat{\phi}}|\hat{g}_{\underline{xx}}|^{-1/2}
\left(
\begin{array}{lc}
e^{-2\hat{\phi}}|\hat{g}_{\underline{xx}}| 
+(\hat{C}^{(1)}{}_{\underline{x}})^{2}\,\,\,\, & 
\hat{C}^{(1)}{}_{\underline{x}} \\
& \\
\hat{C}^{(1)}{}_{\underline{x}} & 1 \\
\end{array}
\right)\, ,\\
& & \\
K & = & 
e^{\hat{\phi}/3}|\hat{g}_{\underline{xx}}|^{1/2}\, ,\\
& & \\
A_{(1)\ \mu} & = & 
\hat{B}_{\mu\underline{x}}\, ,\\
\end{array}
\end{equation}

\begin{equation}
\begin{array}{lcl}
\vec{A}_{(1)\ \mu} & = & 
\left( 
\begin{array}{c}
\hat{C}^{(1)}{}_{\mu} -\hat{C}^{(1)}{}_{\underline{x}}
\hat{g}_{\mu\underline{x}}/\hat{g}_{\underline{xx}}  \\
\\
-\hat{g}_{\mu\underline{x}}/\hat{g}_{\underline{xx}} \\ 
\end{array}
\right)\, ,\\
& & \\
\vec{A}_{(2)\ \mu\nu} & = & 
\left( 
\begin{array}{c}
\hat{C}^{(3)}{}_{\mu\nu\underline{x}}
-2\hat{B}_{[\mu|\underline{x}|} \hat{C}^{(1)}{}_{\nu]}
+2\hat{C}^{(1)}{}_{\underline{x}} \hat{B}_{[\mu|\underline{x}|}
\hat{g}_{\nu]\underline{x}}/\hat{g}_{\underline{xx}} \\
\\
\hat{B}_{\mu\nu} +2\hat{B}_{[\mu |\underline{x}|} 
\hat{g}_{\nu]\underline{x}}/\hat{g}_{\underline{xx}}\\ 
\end{array}
\right)\, ,\\
& & \\
A_{(3)\ \mu\nu\rho} & = & \hat{C}^{(3)}{}_{\mu\nu\rho}
-\frac{3}{2} \hat{g}_{[\mu|\underline{x}|} 
\hat{C}^{(3)}{}_{\nu\rho]\underline{x}}/\hat{g}_{\underline{xx}}
-\frac{3}{2}\hat{C}^{(1)}{}_{\underline{x}} 
\hat{g}_{[\mu|\underline{x}|} \hat{B}_{\nu\rho}/\hat{g}_{\underline{xx}}
\\
& & \\
& &
-\frac{3}{2}\hat{C}^{(1)}{}_{[\mu} \hat{B}_{\nu\rho]}\, ,\\
& & \\
g_{E\ \mu\nu} & = & e^{-4\hat{\phi}/7}
|\hat{g}_{\underline{xx}}|^{1/7}
\left[
\hat{g}_{\mu\nu}
-\hat{g}_{\mu\underline{x}}\hat{g}_{\nu\underline{x}}
/\hat{g}_{\underline{xx}}
\right]
\, .\\
\end{array}
\end{equation}


\section{9-Dimensional Einstein Fields Vs.
10-Dimensional Type~IIB String Fields}
\label{sec-9E-10S-IIB}

Using the results of Section~\ref{sec-generalized} we find

\begin{equation}
\begin{array}{lcl}
{\cal M} & = &
\Lambda^{-1}(y)\hat{\cal M}(\hat{x})(\Lambda^{-1})^{T}(y)
= \hat{\cal M}^{\rm b}= 
e^{\hat{\varphi}^{\rm b}}
\left(
\begin{array}{cc}
|\hat{\lambda}^{\rm b}|^{2} &    \hat{C}^{{\rm b}\ (0)}  \\
& \\
\hat{C}^{{\rm b}\ (0)}       &  1          \\
\end{array}
\right)\, ,\\
& & \\
K & = & e^{\hat{\varphi}/3} 
|\hat{\jmath}_{\underline{y}\underline{y}}|^{-2/3}
= e^{\hat{\varphi}^{\rm b}/3} 
|\hat{\jmath}^{\rm b}{}_{\underline{y}\underline{y}}|^{-2/3}\, ,\\
& & \\
A_{(1)\ \mu} & = & \hat{\jmath}_{\mu\underline{y}}/
\hat{\jmath}_{\underline{yy}}=
\hat{\jmath}^{\rm b}{}_{\mu\underline{y}}/
\hat{\jmath}^{\rm b}{}_{\underline{yy}}\, ,\\
& & \\
\vec{A}_{(1)\ \mu} & = & 
-\Lambda^{-1}(y)
\left( 
\begin{array}{l}
\hat{C}^{(2)}{}_{\mu\underline{y}} \\
\\
\hat{\cal B}_{\mu\underline{y}} \\ 
\end{array}
\right)=
\left( 
\begin{array}{l}
\hat{C}^{{\rm b}\ (2)}{}_{\mu\underline{y}} \\
\\
\hat{\cal B}^{\rm b}{}_{\mu\underline{y}} \\ 
\end{array}
\right)\, ,\\
& & \\
\vec{A}_{(2)\ \mu\nu} & = & 
\Lambda^{-1}(y)
\left( 
\begin{array}{l}
\hat{C}^{(2)}{}_{\mu\nu} \\
\\
\hat{\cal B}_{\mu\nu} \\ 
\end{array}
\right)=
\left( 
\begin{array}{l}
\hat{C}^{{\rm b}\ (2)}{}_{\mu\nu} \\
\\
\hat{\cal B}^{\rm b}{}_{\mu\nu} \\ 
\end{array}
\right)\, ,\\
& & \\
A_{(3)\ \mu\nu\rho} & = & 
-\hat{C}^{(4)}{}_{\mu\nu\rho\underline{y}}
-\frac{3}{2}\hat{\cal B}_{[\mu\nu}
\hat{C}^{(2)}{}_{\rho]\underline{y}}\
-\frac{3}{2}\hat{\cal B}_{[\mu|\underline{y}|}
\hat{C}^{(2)}{}_{\nu\rho]}\, ,\\
& & \\
g_{E\ \mu\nu} & = & e^{-4\hat{\varphi}/7}
|\hat{\jmath}_{\underline{yy}}|^{1/7}
\left[\hat{\jmath}_{\mu\nu}-
\hat{\jmath}_{\mu\underline{y}}\hat{\jmath}_{\nu\underline{y}}/
\hat{\jmath}_{\underline{yy}}\right]\, .\\
\end{array}
\end{equation}


\section{Generalized Buscher T~Duality Rules}
\label{sec-tdual}

Now we just have to compare the results of
Appendix~\ref{sec-9E-10S-IIB} and Appendix~\ref{sec-9E-10S-IIA} to
identify the 10-dimensional fields of the type~IIA and~IIB theories.
This identification produces for us the searched for generalization of
Buscher's T~duality rules \cite{kn:Bu}. These rules generalize the
standard type~II T~duality rules of Ref.~\cite{kn:BHO} in the same way
as those of Ref.~\cite{kn:BRGPT}: The rules have exactly the same form
as the massless ones if we replace the type~IIB fields by the {\it
bare type~IIB} fields.

The only deficiency of these rules is with respect to the S~duals of
D-7-branes: It is necessary to dualize their 8-form potentials which
transform independently of $\hat{\lambda}^{\rm b}$. 

Thus, indicating by a superscript $b$ the bare type~IIB fields the
T~duality rules take the form\footnote{These rules apply to RR
$n$-forms for any $n$. For the values of $n$ that do not appear in the
main body of this paper, one simply has to use the general expression
for the RR field strengths and gauge transformations given in
Ref.~\cite{kn:BLO} inspired by those of Refs.~\cite{kn:GHT,kn:BCT}.}:


{\bf From IIA to IIB:}

\begin{equation}
\begin{array}{l}
\begin{array}{lcllcl}
\hat{\jmath}^{\rm b}{}_{\mu\nu} & = & \hat{g}_{\mu\nu} 
-\left(\hat{g}_{\mu\underline{x}}\hat{g}_{\nu\underline{x}} 
-\hat{B}_{\mu\underline{x}}\hat{B}_{\nu\underline{x}}\right)/
\hat{g}_{\underline{x}\underline{x}}\, , 
\hspace{.1cm}&
\hat{\jmath}^{\rm b}{}_{\mu\underline{y}} & = & 
\hat{B}_{\mu\underline{x}}/
\hat{g}_{\underline{x}\underline{x}}\, , \\
& & & & & \\
\hat{\cal B}^{\rm b}{}_{\mu\nu} & = & \hat{B}_{\mu\nu} 
+2\hat{g}_{[\mu|\underline{x}}\hat{B}_{\nu]\underline{x}}/
\hat{g}_{\underline{x}\underline{x}}\, , &
\hat{\cal B}^{\rm b}{}_{\mu\underline{y}} & = & 
\hat{g}_{\mu\underline{x}}/
\hat{g}_{\underline{x}\underline{x}}\, , \\
& & & & & \\
\hat{\varphi}^{\rm b} & = & \hat{\phi} 
-\frac{1}{2}\log{|\hat{g}_{\underline{x}\underline{x}}|}\, , &
\hat{\jmath}^{\rm b}{}_{\underline{y}\underline{y}} & = & 
1/\hat{g}_{\underline{x}\underline{x}}\, ,\\
\end{array}
\\
\\
\begin{array}{lcl}
\hat{C}^{{\rm b}\ (2n)}{}_{\mu_{1}\ldots\mu_{2n}} & = & 
\hat{C}^{(2n+1)}{}_{\mu_{1}\ldots\mu_{2n}\underline{x}}
+2n \hat{B}_{[\mu_{1}|\underline{x}|}\hat{C}^{(2n-1)}
{}_{\mu_{2}\ldots\mu_{2n}]}\\
& & \\
& &
-2n (2n -1) \hat{B}_{[\mu_{1}|\underline{x}|}\hat{g}_{\mu_{2}|\underline{x}|}
\hat{C}^{(2n-1)}{}_{\mu_{3}\ldots\mu_{2n}]\underline{x}}/
\hat{g}_{\underline{x}\underline{x}}\, , \\
& & \\
\hat{C}^{{\rm b}\ (2n)}{}_{\mu_{1}\ldots\mu_{2n-1}\underline{y}} & = & 
-\hat{C}^{(2n-1)}{}_{\mu_{1}\ldots\mu_{2n-1}} \\
& & \\
& &
+(2n -1) \hat{g}_{[\mu_{1}|\underline{x}|}
\hat{C}^{(2n-1)}{}_{\mu_{2}\ldots\mu_{2n-1}]\underline{x}}/
\hat{g}_{\underline{x}\underline{x}}\, .\\
\end{array}\\
\end{array}
\end{equation}


{\bf From IIB to IIA:}

\begin{equation}
\begin{array}{l}
\begin{array}{lcllcl}
\hat{g}_{\mu\nu} & = & \hat{\jmath}^{\rm b}{}_{\mu\nu} 
-\left(\hat{\jmath}^{\rm b}{}_{\mu\underline{y}}
\hat{\jmath}^{\rm b}{}_{\nu\underline{y}} 
-\hat{\cal B}^{\rm b}{}_{\mu\underline{y}}
\hat{\cal B}^{\rm b}{}_{\nu\underline{y}}\right)/
\hat{\jmath}^{\rm b}{}_{\underline{y}\underline{y}}\, , 
\hspace{.1cm}&
\hat{g}_{\mu\underline{x}} & = & 
\hat{\cal B}^{\rm b}{}_{\mu\underline{y}}/
\hat{\jmath}^{\rm b}{}_{\underline{y}\underline{y}}\, , \\
& & & & & \\
\hat{B}_{\mu\nu} & = & 
\hat{\cal B}^{\rm b}{}_{\mu\nu} 
+2\hat{\jmath}^{\rm b}{}_{[\mu|\underline{y}|}
\hat{\cal B}^{\rm b}{}_{\nu]\underline{y}}/
\hat{\jmath}^{\rm b}{}_{\underline{y}\underline{y}}\, , &
\hat{B}_{\mu\underline{x}} & = & 
\hat{\jmath}^{\rm b}{}_{\mu\underline{y}}/
\hat{\jmath}^{\rm b}{}_{\underline{y}\underline{y}}\, , \\
& & & & & \\
\hat{\phi} & = & \hat{\varphi}^{\rm b}{} 
-\frac{1}{2}\log{|\hat{\jmath}^{\rm b}
{}_{\underline{y}\underline{y}}|}\, , &
\hat{g}_{\underline{x}\underline{x}} & = & 
1/\hat{\jmath}^{\rm b}{}_{\underline{y}\underline{y}}\, ,\\
\end{array}
\\
\\
\begin{array}{lcl}
\hat{C}^{(2n+1)}{}_{\mu_{1}\ldots\mu_{2n+1}} & = & 
-\hat{C}^{{\rm b}\ (2n+2)}{}_{\mu_{1}\ldots\mu_{2n+1}\underline{y}}
\\
& & \\
& &
+(2n+1) \hat{\cal B}^{\rm b}{}_{[\mu_{1}|\underline{y}|}
\hat{C}^{{\rm b}\ (2n)}{}_{\mu_{2}\ldots\mu_{2n+1}]}\\
& & \\
& &
-2n (2n +1) \hat{\cal B}^{\rm b}{}_{[\mu_{1}|\underline{y}|}
\hat{\jmath}^{\rm b}{}_{\mu_{2}|\underline{y}}
\hat{C}^{{\rm b}\ (2n)}{}_{\mu_{3}\ldots\mu_{2n+1}]\underline{y}}/
\hat{\jmath}^{\rm b}{}_{\underline{y}\underline{y}}\, , \\
& & \\
\hat{C}^{(2n+1)}{}_{\mu_{1}\ldots\mu_{2n}\underline{x}} & = & 
\hat{C}^{{\rm b}\ (2n)}{}_{\mu_{1}\ldots\mu_{2n}} \\
& & \\
& &
+2n \hat{\jmath}^{\rm b}{}_{[\mu_{1}|\underline{y}|}
\hat{C}^{{\rm b}\ (2n)}{}_{\mu_{2}\ldots\mu_{2n}]\underline{y}}/
\hat{\jmath}^{\rm b}{}_{\underline{y}\underline{y}}\, .\\
\end{array}\\
\end{array}
\end{equation}

The relation between the {\it bare} fields and the real fields is
given in Section~\ref{sec-generalized}.



\begin{thebibliography}{50}


\bibitem{kn:Wi} E.~Witten,
                {\sl String Theory Dynamics in Various Dimensions},
                {\it Nucl.~Phys.}~{\bf B443} (1995) 85-126.

\bibitem{kn:M} J.M.~Maldacena,
               Ph.D.~Thesis, Princeton University,
               {\sl Black Holes in String Theory},
               {\tt hep-th/9607235}.

\bibitem{kn:Bu} T.~Buscher,
                {\sl Quantum Corrections and Extended Supersymmetry
                in New Sigma Models},
                {\it Phys.~Lett.}~{\bf 159B} (1985) 127;
                {\sl A Symmetry of the String Background Field
                Equations},
                {\it ibid} {\bf 194B} (1987) 59;
                {\sl Path Integral derivation of Quantum Duality in
                Non-Linear Sigma Models},
                {\it ibid} {\bf 201B} (1988) 466.

\bibitem{kn:MS} J.~Maharana and J.H.~Schwarz,
                {\sl Non-Compact Symmetries in String Theory},
                {\it Nucl.~Phys.}~{\bf B390}, (1993) 3-32.

\bibitem{kn:BKO2} E.~Bergshoeff, R.~Kallosh and T.~Ort\'{\i}n,
                  {\sl Duality Versus Supersymmetry and
                  Compactification},
                  {\it Phys.~Rev.}~{\bf D51}, (1995) 3009--3016.

\bibitem{kn:BHO} E.~Bergshoeff, C.M.~Hull and T.~Ort\'{\i}n,
                 {\sl Duality in the Type~II Superstring Effective
                 Action},
                 {\it Nucl.~Phys.}~{\bf B451} (1995) 547-578.


\bibitem{kn:SS} J.~Scherk and J.H.~Schwarz,
                {\sl How to Get Masses from Extra Dimensions},
                {\it Nucl.~Phys.}~{\bf B153} (1979) 61-88.

\bibitem{kn:BRGPT} E.~Bergshoeff, M.~de Roo, M.B.~Green,
                   G.~Papadopoulos and P.K.~Townsend,
                   {\sl Duality of Type~II 7-Branes and 8-Branes},
                   {\it Nucl.~Phys.}~{\bf B470} (1996) 113-135.

\bibitem{kn:Ro2} L.J.~Romans,
                 {\sl Massive $N=2a$ Supergravity in Ten Dimensions},
                 {\it Phys.~Lett.}~{\bf 169B} (1986) 374.

\bibitem{kn:PW} J.~Polchinski and E.~Witten,
                {\sl Evidence for Heterotic - Type I String Duality},
                {\it Nucl.~Phys.}~{\bf B460} (1996) 525.

\bibitem{kn:LLP2} I.V.~Lavrinenko, H.~Lu and C.N.~Pope,
                  {\sl Fiber Bundles and Generalized Dimensional Reduction},
                  Report CTP-TAMU-43-97 and
                  {\tt hep-th/9710243}.

\bibitem{kn:BJO} E.~Bergshoeff, B.~Janssen and T.~Ort\'{\i}n,
                 {\sl Solution-Generating Transformations and the
                 String Effective Action},
                 {\it Class.~Quantum Grav.}~{\bf 13} (1996) 321-343.

\bibitem{kn:Be} E.~Bergshoeff,
                {\sl Duality Symmetries and the Type II String 
                Effective Action},
                Presented at ICTP {\it Trieste
                Conference on Physical and Mathematical
                Implications of Mirror Symmetry in String
                Theory}, Trieste, Italy, Jun 5-9, 1995.  
                {\it Nucl.~Phys.~B~Proc.~Suppl.}~{\bf 46} (1996) 39-48.

\bibitem{kn:As} P.S.~Aspinwall, 
                {\sl Some Relationships Between Dualities in String
                Theory}, 
                Presented at ICTP {\it Trieste
                Conference on Physical and Mathematical
                Implications of Mirror Symmetry in String
                Theory}, Trieste, Italy, Jun 5-9, 1995.  
                {\it Nucl.~Phys.~B~Proc.~Suppl.}~{\bf 46} (1996) 30-38.

\bibitem{kn:BDHS} K.~Bautier, S.~Deser, M.~Henneaux and
                  D.~Seminara,
                  {\sl No Cosmological D = 11 Supergravity},
                  {\it Phys.~Lett.}~{\bf 406} (1997) 49.
                  \\
                  S.~Deser,
                  {\sl Uniqueness of d=11 Supergravity},
                  {\tt hep-th/9712064}.
                  \\
                  S.~Deser,
                  {\sl D=11 Supergravity Revisited},
                  {\tt hep-th/9805205}.

\bibitem{kn:L} Y.~Lozano,
               {\sl Eleven Dimensions from the Massive D-2-Brane},
               {\it Phys.~Lett.}~{\bf B414} (1997) 52.

\bibitem{kn:O} T.~Ort\'{\i}n,
               {\sl A Note on the D-2-Brane of the Massive 
               Type~IIA Theory and Gauged Sigma Models},
               {\it Phys.~Lett.}~{\bf B415} (1997) 39-44.

\bibitem{kn:BLO} E.~Bergshoeff, Y.~Lozano and T.~Ort\'{\i}n,
                 {\sl Massive Branes},
                 Report UG-8/97, QMW-PH-97-28, CERN-TH/97-229, 
                 IFT-UAM/CSIC-97-2 and
                 {\tt hep-th/9712115}.

\bibitem{kn:BJO2} E.~Bergshoeff, B.~Janssen and T.~Ort\'{\i}n,
                  {\sl Kaluza-Klein Monopoles and Sigma-Models},
                  {\it Phys.~Lett.}~{\bf B410} (1997) 131-141.


\bibitem{kn:BEL} E.~Bergshoeff, E.~Eyras and Y.~Lozano,
                 {\sl The Massive Kaluza-Klein Monopole},
                 {\tt hep-th/9802199}

\bibitem{kn:BS} E.~Bergshoeff and J.P.~van der Schaar,
                {\sl On the M-9-Brane},
                Report UG-10/98 and 
                {\tt hep-th/9806069}.

\bibitem{kn:St} E.C.G.~Stueckelberg,
                {\it Helv.~Phys.~Acta} {\bf 11} (1938) 225.
%

\bibitem{kn:BBO} E.~Bergshoeff, H.-J.~Boonstra and T.~Ort\'{\i}n,
                 {\sl $S$~Duality and Dyonic $p$-Brane Solutions
                 in Type~II String Theory},
                 {\it Phys.~Rev.}~{\bf D53} 7206-7212.

\bibitem{kn:JHS} J.H.~Schwarz, 
                 {\sl Covariant Field Equations of Chiral $N=2,d=10$
                 Supergravity}, 
                 {\it Nucl.~Phys.}~{\bf B226} (1983) 269-288.

\bibitem{kn:GHT} M.B.~Green, C.M.~Hull and P.K.~Townsend,
                 {\sl D-p-brane Wess-Zumino Actions, T-Duality and the
                 Cosmological Constant},
                 {\it Phys.~Lett.}~{\bf B382} (1996) 65.

\bibitem{kn:BCT} E.~Bergshoeff, P.M.~Cowdall and P.K.~Townsend, 
                 {\sl Massive IIA Supergravity from the Topologically 
                 Massive D-2-Brane},
                 Report DAMTP-R-97-20 and
                 {\tt hep-th/9706094}.

\bibitem{kn:HT} C.M.~Hull and P.K.~Townsend,
                {\sl Unity of Superstring Dualities},
                {\it Nucl.~Phys.}~{\bf B438} (1995) 109-137.

\bibitem{kn:Nak} M.~Nakahara,
                 {\sl Geometry, Topology and Physics},
                 IOP Publishing Ltd., London 1990. 

\bibitem{kn:CJS} E.~Cremmer, B.~Julia and J.~Scherk,
                 {\sl Supergravity Theory in 11 Dimensions},
                 {\it Phys.~Lett.}~{\bf 76B} (1978) 409.

\bibitem{kn:DLP} J.~Dai, R.G.~Leigh and J.~Polchinski,
                 {\sl New Connections Between String Theories},
                 {\it Mod.~Phys.~Lett.}~{\bf A4} (1989) 2073.

\bibitem{kn:DHS} M.~Dine, P.~Huet and N.~Seiberg, 
                 {\sl Large and Small Radius in String Theory},
                 {\it Nucl.~Phys.}~{\bf B322} (1989) 301.

\bibitem{kn:O2} T.~Ort\'{\i}n,
                {\sl Extremality Versus Supersymmetry in Stringy
                Black Holes},
                {\it Phys.~Lett.}~{\bf B422} (1998) 93-100.

\bibitem{kn:KhO1} R.R.~Khuri and T.~Ort\'{\i}n,
                  {\sl Supersymmetric Black Holes in $N=8$ Supergravity},
                  {\it Nucl.~Phys.}~{\bf B467}), (1996) 355-382.

\bibitem{kn:Ba} I.~Bars,
                {\sl S-Theory},
                {\it Phys.~Rev.}~{\bf D55} (1997) 2373-2381.

\bibitem{kn:Ba2} I.~Bars,
                 {\sl Black Hole Entropy Reveals a 12th ``Dimension''},
                 {\it Phys.~Rev.}~{\bf D55} 3633-3641.

\bibitem{kn:JHS2} J.H.~Schwarz,
                  {\sl An $SL(2,\mathbb{Z})$ Multiplet of Type~IIB Strings},
                  {\it Phys.~Lett.}~{\bf B360} (1995) 13-18, {\bf ERRATUM} 
                  {\it ibid.}~{\bf B364} (1995) 252.

\bibitem{kn:LR} J.X.~Lu and S.~Roy,
                {\sl An $SL(2,\mathbb{Z})$ Multiplet of Type~IIB 
                Super-Five-Branes},
                Report CTP-TAMU-06/98, SINP-TNP/98-05 and
                {\tt hep-th/9802080}.

\bibitem{kn:GHZ}  M.R.~Gaberdiel, T.~Hauer and  B.~Zwiebach,
                  {\sl Open string - string junction transitions},
                  Report DAMTP-98-6, MIT-CTP-2712 and
                  {\tt hep-th/9801205}.
                    
\bibitem{kn:GHM} R.~Gregory, J.A.~Harvey and G.~Moore,
                 {\sl Unwinding Strings and T-duality of
                 Kaluza-Klein and H-Monopoles},
                 Report DTP/97/31, EFI-97-26, YCTP-P15-97 and
                 {\tt hep-th/9708086}.

\bibitem{kn:BREJS1} E.~Bergshoeff, M.~de Roo, E.~Eyras, B.~Janssen, 
                     J.P.~van der Schaar ,
                     {\sl Multiple Intersections of D-branes and M-branes},
                     {\it Nucl.~Phys.}~{\bf B494} (1997) 119-143.
   
\bibitem{kn:BREJS2} E.~Bergshoeff, M.~de Roo, E.~Eyras, B.~Janssen, 
                     J.P.~van der Schaar,
                     {\sl Intersections Involving Waves and Monopoles 
                     in Eleven Dimensions},
                     {\it Class.~Quant.~Grav.}~{\bf 14} (1997) 2757-2769.

\bibitem{kn:LPTX} H.~Lu, C.~N.~Pope, T.R.~Tran, K.-W.~Xu,
                  {\sl Classification of p-branes, NUTs, Waves 
                  and Intersections}
                  {\it Nucl.~Phys.}~{\bf B511} (1998) 98-154.

\bibitem{kn:Hu2} C.M.~Hull,
                 {\sl Gravitational Duality, Branes and Charges},
                 Report QMW-97-16, NI97028-NQF and
                 {\tt hep-th/9705162}.

\bibitem{kn:KKM} N.~Kaloper, R.R.~Khuri and R.C.~Myers,
                 {\sl On Generalized Axion Reductions},
                 Report UCSB NSF-ITP-98-023, SU-ITP-98-08, McGill/98-04,
                 QMW-PH-98-08 and
                 {\tt hep-th/9803006}.

\bibitem{art:tonin} G.~Dall'Agata, K.~Lechner, M.~Tonin,
                    {\sl $D=10$, $N=IIB$ Supergravity: Lorentz-invariant
                    actions and duality},
                    J. High Energy Phys. 9807:17,1998.
                    {\tt hep-th/9806140}.
\end{thebibliography}
\end{document}